\documentclass[12pt,a4paper]{article}
\usepackage[utf8]{inputenc}
\usepackage{epsf}
\usepackage{latexsym,amssymb,euscript}
\usepackage[dvips]{graphicx}
\usepackage[numbers,sort&compress]{natbib}
\usepackage{amsmath}
\usepackage{nicefrac}
\usepackage{slashed}
\usepackage{booktabs}
\usepackage[linktocpage]{hyperref}
\usepackage{braket}
\usepackage{chngcntr}
\usepackage{bbold}
\usepackage{multicol}
\usepackage{graphics}
\usepackage{graphicx}
\usepackage{mciteplus}
\usepackage{caption}
\usepackage{subcaption}
\usepackage{pdfpages}
\usepackage[titletoc]{appendix}
\usepackage[normalem]{ulem}
\graphicspath{{./figures/}}
\hypersetup{
 linktocpage = false,
 urlcolor = citegreen,
 colorlinks = true,
 linkcolor = citegreen,
 anchorcolor = linkred,
 citecolor = urlblue,
 pdfstartview = {XYZ null null 1.25} 
           }
\usepackage[left=2cm, right=2cm]{geometry}
\usepackage{pstricks}
\usepackage{color}
\usepackage{xcolor}
\definecolor{urlblue}{rgb}{0.2,0.4,0.7}
\definecolor{citegreen}{rgb}{0,0.4,0.2}
\definecolor{linkred}{rgb}{0.9,0.2,0.1}
\usepackage{float}
\usepackage{academicons}
\definecolor{orcidlogocol}{HTML}{A6CE39}
\usepackage{changepage}
\usepackage{fancyhdr}
\newcommand{\drv}{{\rm d}}

\newcommand{\as}{\alpha_s}
\newcommand{\LQCD}{\Lambda_{\rm QCD}}
\newcommand{\MSb}{\overline{\rm MS}}

\newcommand{\LL}{{\rm LL/LO}}

\newcommand{\NLLp}{{\rm NLL/NLO^+}}
\newcommand{\NLLpp}{{\rm NLL/NLO^{(+)}}}
\newcommand{\HENLOp}{{\rm HE}\mbox{-}{\rm NLO^+}}

\newcommand{\ClLL}{{\cal C}_l^\LL}

\newcommand{\ClNLLp}{{\cal C}_l^\NLLp}

\newcommand{\ClHENLOp}{{\cal C}_l^{{\rm HE}\text{-}{\rm NLO}^+}}

\newcommand{\DY}{\Delta Y}

\newcommand{\vqTTa}{\langle {\vec q}_T^{\;2} \rangle}

\newcommand{\E}{{\cal E}}

\newcommand{\HQ}{{\cal H}_Q}

\newcommand{\Jpsi}{J/\psi}

\newcommand{\BCs}{B_c(^1S_0)}
\newcommand{\Bss}{B_c(^3S_1)}

\newcommand{\TQc}{T_{4c}}

\newcommand{\PQc}{P_{5c}}

\newcommand{{\HFNRevo}}{\tt HF-NRevo}

\newcommand{{\Jethad}}{\tt JETHAD}
\newcommand{{\symJethad}}{\tt symJETHAD}
\newcommand{{\psymJethad}}{\tt (sym)JETHAD}
\newcommand{{\Hell}}{\tt HELL}
\newcommand{{\RadISH}}{\tt RadISH}
\newcommand{{\Pegasus}}{\tt QCD-PEGASUS}
\newcommand{{\HOPPET}}{\tt HOPPET}
\newcommand{{\QCDNUM}}{\tt QCDNUM}
\newcommand{{\APFEL}}{\tt APFEL}
\newcommand{{\APFELpp}}{\tt APFEL++}
\newcommand{{\APFELppp}}{\tt APFEL(++)}
\newcommand{{\EKO}}{\tt EKO}
\newcommand{{\FeynCalc}}{\tt FeynCalc}

\newcommand{\eref}[1]{~\eqref{#1}}

\newcommand{\orcidFGC}{\href{https://orcid.org/0000-0003-3299-2203}{\includegraphics[scale=0.1]{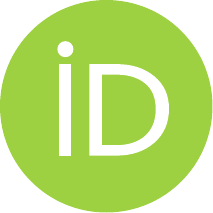}}}

\begin{document}

\begin{titlepage}

{
\begin{adjustwidth}{-1cm}{-1cm}
\begin{center}
  {\Large \bf Toward Precision Fragmentation of $\Omega_{3Q}$ Baryons: \vskip.035cm The {\tt OMG3Q1.1} Framework}
\end{center}
\end{adjustwidth}
}

\vskip 0.75cm

\centerline{
Francesco~Giovanni~Celiberto$^{\;1\;\dagger}$ {\orcidFGC}
}

\vskip .4cm

\centerline{${}^1$ {\sl Universidad de Alcal\'a (UAH), Departamento de Física y Matemáticas,}}
\centerline{\sl E-28805 Alcal\'a de Henares, Madrid, Spain}
\vskip 1.15cm

\begin{abstract}
\vspace{0.25cm}
\headrule 
\vspace{0.50cm}
Recent experimental advances in the baryon sector, including the observation of doubly charmed states, have renewed interest in the production mechanisms of increasingly heavy hadronic systems, calling for precision and uncertainty-controlled descriptions. 
We present the {\tt OMG3Q1.1} framework for the fragmentation of same-flavor all-heavy $\Omega_{3Q}$ baryons in high-energy hadronic collisions. 
The construction combines diquark-inspired inputs for constituent-heavy-quark and gluon channels with threshold-aware DGLAP evolution within the {\HFNRevo} scheme. 
A replica-based strategy consistently quantifies perturbative missing-higher-order effects (F-MHOUs) and nonperturbative wave-function uncertainties (F-NPWFs), yielding the first uncertainty-resolved fragmentation-function set for the $\Omega_{3Q}$ sector. 
The resulting {\tt LHAPDF6} grids are employed to investigate semi-inclusive $\Omega_{3Q}$ plus jet production at the HL-LHC and future FCC within the {\tt (sym)JETHAD} environment. 
The {\tt OMG3Q1.1} framework establishes a precision-oriented baseline for rare triply heavy baryons and provides a foundation for future studies of the heavy-flavor baryon landscape.
\vspace{0.50cm} 
\headrule
\vspace{0.75cm}
{
 \setlength{\parindent}{0pt}
 \textsc{Keywords}: \vspace{0.15cm} \\ 
 Exotic matter \\
 Omega sector \\
 Rare baryons \\
 Heavy flavor \\
 Precision QCD \\
 Hadronic structure \\
 Fragmentation \\
 Resummation \\
 {\tt OMG3Q1.1} FF release
\vspace{0.65cm} 
\headrule
}
\end{abstract}

\vspace{-0.00cm}
\vfill
$^{\dagger}${\it e-mail}:
\href{mailto:francesco.celiberto@uah.es}{francesco.celiberto@uah.es}

\end{titlepage}

\tableofcontents
\clearpage

\section{Introduction}
\label{sec:intro}

Heavy-flavored hadrons provide a privileged laboratory for investigating the interplay between perturbative and nonperturbative Quantum Chromodynamics (QCD). 
Within this sector, same-flavor all-heavy baryons such as $\Omega_{3c}$ and $\Omega_{3b}$ offer an especially clean environment, as their valence structure contains no light degrees of freedom and is therefore directly governed by heavy-quark dynamics~\cite{Bjorken:1985ei,Fleck:1989mb,Martynenko:2007je,Martynenko:2013eoa,Karliner:2014gca,Yoshida:2015tia,Ebert:2002ig,Roberts:2007ni,Chen:2011mb,Padmanath:2013zfa,Brown:2014ena,Meinel:2012qz}. 
Their spectroscopy and structural properties have been extensively investigated through constituent-quark and potential models, lattice QCD, effective approaches, and nonrelativistic calculations~\cite{Mathur:2018rwu,Francis:2018jyb,Eichten:1994gt,Godfrey:1985xj,Padmanath:2013zfa,Faustov:2021qqf,Cakir:2026fzd,Llanes-Estrada:2011gwu,Wei:2016jyk,Yang:2019lsg,Gomez-Rocha:2023jfr,Najjar:2024deh}, with recent developments also exploring quantum-computing implementations of heavy-baryon dynamics~\cite{deArenaza:2024dhe}. 
The increasing discovery potential offered by the High-Luminosity Large Hadron Collider (HL-LHC)~\cite{Apollinari:2015wtw}, the Future Circular Collider (FCC)~\cite{FCC:2025lpp,FCC:2025uan,FCC:2025jtd}, and complementary $e^+e^-$ facilities such as Belle~II~\cite{Belle-II:2010dht} further motivates dedicated studies of these rare systems.

This motivation is reinforced by recent experimental progress in the multicharm sector. 
The discovery of all-charm structures in double-$\Jpsi$ and $\Jpsi+\psi(2S)$ final states~\cite{CMS:2023owd}, together with spin-parity information favoring a compact $J^{PC}=2^{++}$ interpretation for the leading structure~\cite{CMS:2025fpt}, has established fully heavy spectroscopy as an experimentally accessible frontier. 
In parallel, the observation of the doubly charmed baryons $\Xi_{cc}^{++}$~\cite{LHCb:2017iph} and $\Xi_{cc}^{+}$~\cite{LHCb:2026pxn}, followed by the recent report of the $\Omega_{cc}^{+}$ state~\cite{Wang:2026OmegaCC}, completes the ground-state doubly charmed sector and demonstrates the growing experimental reach toward complex heavy-baryon systems. 
Although triply heavy baryons are expected to be considerably rarer, these advances strengthen the case for dedicated searches and for testing quark-diquark descriptions of baryonic structure. 
Connections with heavier multiquark configurations further place $\Omega_{3Q}$ states at the interface between conventional heavy-baryon spectroscopy and the exotic-hadron sector~\cite{Gershtein:2000nx,Ali:2017jda,LHCb:2020bwg,Karliner:2020vsi}.

At sufficiently large transverse momentum, parton fragmentation provides a natural leading-power description of multiply heavy hadron production, separating short-distance partonic dynamics from the long-distance formation of the observed bound state~\cite{Braaten:1994bz,Braaten:1993rw,Braaten:1993mp,Braaten:1994xb,Braaten:1993jn,Kiselev:1994pu}. 
For triply heavy baryons, diquark-inspired approaches provide an economical description of this mechanism~\cite{Anselmino:1992vg,Ebert:1995fp,MoosaviNejad:2016qdx,Chang:2006eu}, and dedicated calculations of heavy-quark- and gluon-initiated fragmentation functions (FFs) have been developed at different perturbative accuracies~\cite{Adamov:1997yk,Yang:2002gh,GomshiNobary:2004mq,MoosaviNejad:2017bda,MoosaviNejad:2017rvi,Delpasand:2019xpk}. 
Their use in collider phenomenology requires a consistent treatment of DGLAP evolution and heavy-parton thresholds~\cite{Mele:1990cw,Cacciari:1993mq,Buza:1996wv,Cacciari:2001cw,Mitov:2006wy}, whose technical implementation is discussed in the following sections.

Recent progress in threshold-aware fragmentation has led to several variable-flavor number-scheme (VFNS) determinations for heavy and exotic hadrons, including the {\tt TQHL1.1}, {\tt TQ4Q1.1}, {\tt TQ4Q2.0}, {\tt PQ5Q1.1}, and {\tt NRFF1.0} families~\cite{Celiberto:2024beg,Celiberto:2025dfe,Celiberto:2025ziy,Celiberto:2026kks,Celiberto:2026idp,Celiberto:2025ipt,Celiberto:2026rdk,Celiberto:2025euy,Celiberto:2026ajo}. 
These rely on the Heavy-Flavor NonRelativistic evolution ({\HFNRevo}) scheme~\cite{Celiberto:2025euy,Celiberto:2024mex,Celiberto:2024bxu,Celiberto:2024rxa,Celiberto:2025xvy,Celiberto:2026rzi,Celiberto:2026zss}, which connects model-based heavy-hadron FF inputs with DGLAP evolution. 
Within this program, {\tt OMG3Q1.0}~\cite{Celiberto:2025ogy} provided the first public VFNS FFs for same-flavor all-heavy $\Omega_{3Q}$ baryons, establishing their fragmentation-based description but remaining limited to central values without systematic uncertainty assessment.

The present work introduces {\tt OMG3Q1.1}, which extends and supersedes the original determination by promoting uncertainty quantification to an integral component of the FF construction. 
Perturbative fragmentation missing-higher-order uncertainties (F-MHOUs) are assessed through controlled variations of the scales entering the initial FF construction and evolution, while fragmentation nonperturbative wave-function uncertainties (F-NPWFs) probe the sensitivity to the transverse-momentum structure governing bound-state formation. 
Both effects are organized within a replica-based architecture, allowing their separate or combined propagation from the FF level to physical observables. 
In this way, {\tt OMG3Q1.1} moves beyond the central-value description of {\tt OMG3Q1.0} and provides an uncertainty-aware framework for leading-power fragmentation into same-flavor all-heavy baryons.

As a phenomenological application, we study semi-inclusive $\Omega_{3Q}$ plus light-jet production at the HL-LHC and FCC. 
Predictions are obtained within the $\NLLp$ hybrid-factorization framework, combining NLL high-energy resummation with NLO collinear ingredients, and are evaluated through the {\Jethad} platform and its symbolic module {\symJethad}~\cite{Celiberto:2020wpk,Celiberto:2022rfj,Celiberto:2023fzz,Celiberto:2024mrq,Celiberto:2024swu,Celiberto:2025csa,Celiberto:2026ooh}. 
This setup permits us to examine simultaneously the perturbative stability of the high-energy calculation and the propagation of the newly introduced fragmentation uncertainties to collider observables.

The manuscript is organized as follows. 
Section~\ref{sec:FFs} reviews the main ingredients of heavy-flavor fragmentation and introduces the theoretical framework adopted for $\Omega_{3Q}$ baryons. 
Section~\ref{sec:FFs-architecture} presents the construction of the {\tt OMG3Q1.1} FFs. 
The $\NLLp$ hybrid-factorization formalism is outlined in Section~\ref{sec:hybrid-factorization}.
Phenomenological results for $\Omega_{3Q}$ plus jet production at the HL-LHC and FCC are discussed in Section~\ref{sec:phenomenology}. 
Our conclusions and outlook are presented in Section~\ref{sec:conclusions}.

\section{Fragmentation functions of all-heavy $\Omega$ baryons}
\label{sec:FFs}

We start by reviewing a number of relevant aspects of heavy-flavor fragmentation, with emphasis on heavy-light hadrons, quarkonia, and exotic states (Section~\ref{ssec:FFs-intro}). 
Next, we discuss the diquark-inspired framework employed for all-heavy baryons (Section~\ref{ssec:FFs-diquark}). 
The section concludes with an overview of the general structure of the initial-scale $\Omega_{3Q}$ FFs (Section~\ref{ssec:FFs-O3Q-general}).

\subsection{Heavy-flavor fragmentation in a VFNS}
\label{ssec:FFs-intro}

Heavy-flavor fragmentation differs structurally from its light-hadron counterpart because heavy-quark masses introduce perturbative scales already at the lowest Fock level. 
Initial-scale heavy-hadron FFs therefore combine short-distance partonic dynamics with nonperturbative bound-state formation, providing the natural starting point for their subsequent evolution within a variable-flavor number scheme (VFNS).

For singly heavy states, such as $D$ and $B$ mesons or $\Lambda_{c,b}$ baryons, this structure can be represented as a two-stage mechanism~\cite{Cacciari:1996wr,Cacciari:1993mq,Helenius:2023wkn,Generet:2023vte,Jaffe:1993ie,Kniehl:2005mk,Helenius:2018uul}. 
A heavy quark $Q$ is first generated perturbatively, with the corresponding short-distance coefficient (SDC) known up to NLO~\cite{Mele:1990yq,Mele:1990cw} and NNLO accuracy~\cite{Rijken:1996vr,Mitov:2006wy,Blumlein:2006rr,Melnikov:2004bm,Mitov:2004du,Biello:2024zti}, and subsequently hadronizes through long-distance dynamics described by phenomenological models~\cite{Kartvelishvili:1977pi,Bowler:1981sb,Peterson:1982ak,Andersson:1983jt,Collins:1984ms,Colangelo:1992kh} or effective approaches~\cite{Georgi:1990um,Eichten:1989zv,Grinstein:1992ss,Neubert:1993mb,Jaffe:1993ie}. 
The corresponding initial-scale FF for a singly heavy hadron $\HQ$ can be written as~\cite{Cacciari:1996wr,Cacciari:1997du}
\begin{equation}
\label{FFs_HF_initial}
 D_{i}^{\HQ} (z, \mu_{F,0}) =
 \int_z^1 \frac{\drv \xi}{\xi} D_i^Q (\xi, \mu_{F,0}) \,
 D_{\rm [np]}^{\HQ} \left( \frac{z}{\xi} \right) \;,
\end{equation}
where $D_i^Q$ denotes the perturbative SDC and $D_{\rm [np]}^{\HQ}$ the nonperturbative hadronization component. 
These inputs are then evolved through timelike DGLAP equations.

For quarkonia and multiply heavy states, the same separation between short- and long-distance dynamics persists, although the lowest Fock configuration acquires a richer structure. 
In quarkonium production, this separation is systematically realized through NRQCD factorization~\cite{Caswell:1985ui,Thacker:1990bm,Bodwin:1994jh,Cho:1995vh,Cho:1995ce,Leibovich:1996pa,Bodwin:2005hm}; see Refs.~\cite{Grinstein:1998xb,Kramer:2001hh,QuarkoniumWorkingGroup:2004kpm,Lansberg:2005aw,Lansberg:2019adr} for reviews. 
At sufficiently large transverse momentum, single-parton fragmentation provides the leading-power contribution, whereas direct production of the heavy Fock state can be interpreted as a higher-power fixed-flavor number-scheme (FFNS) mechanism~\cite{Alekhin:2009ni,Kang:2014tta,Boer:2023zit,Celiberto:2025euy,Celiberto:2024mex,Celiberto:2024bxu,Celiberto:2024rxa,Celiberto:2025xvy,Celiberto:2026rzi,Celiberto:2026zss}.

Initial-scale heavy-quark and gluon FFs into $S$-wave vector quarkonia were first obtained at LO~\cite{Braaten:1993rw,Braaten:1993mp} and subsequently extended to NLO~\cite{Zheng:2019gnb}. 
They provided the basis for the {\tt ZCW19$^+$}~\cite{Celiberto:2022dyf,Celiberto:2023fzz} and {\tt ZCFW22}~\cite{Celiberto:2022keu,Celiberto:2024omj} VFNS determinations. 
The latter also yielded $\BCs$-to-$B$ production rates below $0.1\%$, consistently with LHCb measurements~\cite{LHCb:2014iah,Celiberto:2024omj}, supporting the applicability of the fragmentation picture in the high-$p_T$ regime.

This strategy has progressively been extended to multiquark systems. 
NRQCD descriptions of the di-$\Jpsi$ structures observed at the LHC~\cite{LHCb:2020bwg,ATLAS:2023bft,CMS:2023owd} and their interpretation as compact all-charm tetraquarks~\cite{Zhang:2020hoh,Zhu:2020xni} motivated the first NRQCD gluon-fragmentation input for color-singlet $S$-wave $\TQc$ states~\cite{Feng:2020riv}. 
Subsequent VFNS developments produced the {\tt TQHL1.0}, {\tt TQ4Q1.1}, and {\tt TQHL1.1} families~\cite{Celiberto:2023rzw,Celiberto:2024mrq,Celiberto:2024beg}, incorporating additional heavy-quark inputs~\cite{Bai:2024ezn} and enabling applications to all-heavy tetraquark phenomenology~\cite{Celiberto:2025ziy,Celiberto:2025dfe,Celiberto:2026zed}, electroweak and Higgs decays~\cite{Ma:2025ryo}, intrinsic-charm-sensitive observables~\cite{Celiberto:2025vra}, and gluon fragmentation~\cite{Nakhaei:2025zty}.

More recent developments include the {\tt NRFF1.0} determination for pseudoscalar quarkonia~\cite{Celiberto:2025euy} and the uncertainty-aware {\tt TQ4Q2.0} framework for all-heavy tetraquarks~\cite{Celiberto:2026kks}. 
Within the baryonic and higher-multiquark sectors, the {\tt PQ5Q1.0} and {\tt OMG3Q1.0} sets were introduced in Refs.~\cite{Celiberto:2025ipt,Celiberto:2025ogy,Celiberto:2025csa}, while the subsequent {\tt PQ5Q1.1} release~\cite{Celiberto:2026rdk} incorporated multimodal inputs and a replica-based uncertainty treatment. 
These developments provide the methodological basis for the uncertainty-aware $\Omega_{3Q}$ FF construction presented below.

\subsection{The diquark-inspired fragmentation framework}
\label{ssec:FFs-diquark}

The quark-diquark picture offers an effective description of baryon structure in which two quarks are treated as a correlated subsystem interacting with a third constituent quark~\cite{Gell-Mann:1964ewy}. Owing to its simplicity and phenomenological success, this framework has been extensively employed in hadron spectroscopy and in studies of baryon production and decay, particularly in the heavy-flavor sector~\cite{Maiani:2004vq,Jaffe:2003sg,Guo:2013xga,DeSanctis:2016zph}. Within this approach, diquarks are usually modeled as scalar ($J=0$) or axial-vector ($J=1$) states, with their nonperturbative internal structure encoded through effective form factors.

Scalar diquarks are characterized by a single form factor and a comparatively simple spin structure, whereas axial-vector configurations require a richer set of form factors and spin correlations. Both have been widely used in the description of baryon formation and in spectator-model studies of polarized and unpolarized partonic distributions inside the nucleon~\cite{Bacchetta:2008af,Bacchetta:2010si,Bacchetta:2020vty,Bacchetta:2024fci,Chakrabarti:2023djs,Banu:2024ywv}.

Applications of the quark-diquark formalism to fragmentation processes date back to early studies of light and heavy baryons~\cite{Nzar:1995wb,Ma:2001ri,Yang:2002gh,Falk:1993gb,Adamov:1997yk,MoosaviNejad:2017rvi,Delpasand:2019xpk}. 
Similar ideas have also been extended to exotic-hadron phenomenology, including pentaquark configurations~\cite{Maiani:2015vwa} and the spectroscopy of doubly and all-heavy tetraquarks in relativistic quasipotential approaches based on diquark-anti\-diquark dynamics~\cite{Faustov:2020qfm,Faustov:2021hjs,Faustov:2022mvs}.

In this work we follow the scalar-diquark implementation adopted in several fragmentation studies~\cite{Adamov:1997yk,Martynenko:1996bt,MoosaviNejad:2017rvi,Delpasand:2019xpk}. 
Besides reducing the complexity of the spin algebra, this choice minimizes the number of independent form factors and leads to a tractable analytical framework. 
Scalar diquarks have been successfully employed in perturbative-QCD calculations of baryon FFs~\cite{Adamov:1997yk} and have been shown to provide phenomenologically viable descriptions while preserving analytical control~\cite{Martynenko:1996bt,GomshiNobary:2007xk}.

Axial-vector diquarks, by contrast, require a more elaborate treatment of spin correlations and additional form factors. 
They are nevertheless indispensable for describing spin-$3/2$ baryons and polarization effects.
In particular, a realistic treatment of polarized $\Omega_{3c}^{*}$ production would require a vector-diquark configuration.

The scalar-diquark approximation employed here should therefore be regarded as an effective starting point for a perturbative-QCD description of fragmentation into unpolarized all-heavy baryons. Extensions incorporating spin-1 diquarks could provide a more complete account of polarization and spin-dependent dynamics.

Our use of scalar diquarks in the description of unpolarized $\Omega_{3c}$ and $\Omega_{3b}$ production is consistent with the symmetry requirements of the baryonic wave function when the process is interpreted within a two-step fragmentation mechanism. 
In this picture, a heavy quark first fragments into a color-antitriplet heavy diquark, which subsequently evolves into the physical triply heavy baryon through nonperturbative QCD dynamics.

Although a scalar diquark does not by itself reproduce the full symmetrization pattern of a $J=1/2$ state composed of three identical fermions, the missing structure can be effectively absorbed into the nonperturbative component of the FF.
Conceptually, this decoupling of perturbative and nonperturbative regimes is analogous to the NRQCD framework for quarkonium. 
In that case, SDCs govern the creation of color-octet $[Q\bar{Q}]$ states, which are then projected onto physical color-singlets by means of LDMEs.

\subsection{Anatomy of the $\Omega_{3Q}$ fragmentation functions}
\label{ssec:FFs-O3Q-general}

The initial-scale FFs employed throughout this work are taken from existing perturbative calculations of color-singlet all-heavy baryons. 
For the heavy-quark channel we adopt the LO and NLO results of Refs.~\cite{MoosaviNejad:2017bda,MoosaviNejad:2017rvi}, while for the gluon channel we use the NLO determination of Ref.~\cite{Delpasand:2019xpk}. 
These studies describe the baryon as a color-singlet state and model the correlated heavy-quark pair as a spin-0 scalar diquark.

Two complementary fragmentation pictures are commonly considered. 
In the direct mechanism, the fragmenting heavy quark evolves into the baryon through a genuine three-body process involving all constituent quarks. The alternative diquark scenario, adopted here, assumes that two quarks first form a compact scalar diquark, which subsequently combines with the third quark to generate the final baryon. 
Besides reducing the complexity of the calculation, this approximation has become the standard framework for describing unpolarized $J=1/2$ triply heavy baryons.

The indirect mechanism discussed in Refs.~\cite{MoosaviNejad:2017rvi,Delpasand:2019xpk} can be interpreted as the diquark realization of the two-stage fragmentation picture introduced in Section~\ref{ssec:FFs-intro}. 
Within this framework, perturbative dynamics produce an intermediate colored diquark, while the formation of the physical baryon is treated as a separate nonperturbative step. 
Since both quark- and gluon-induced FFs are available at NLO accuracy, the formalism provides a consistent starting point for VFNS studies.

A central ingredient of this construction is the Suzuki model~\cite{Suzuki:1985up}, which plays a role analogous to that of NRQCD in quarkonium production. Originally formulated to describe heavy-hadron fragmentation, the model provides a perturbative transition amplitude in which a constituent heavy quark fragments into a diquark state and a spectator, followed by the formation of the baryonic bound state. Nonperturbative effects enter through the wave function at the origin and an effective diquark form factor.

Although the Suzuki framework can be applied to both direct and diquark mechanisms, its advantages are most evident in the latter case. 
By replacing a full three-body transition with an effective two-body process, it leads to compact analytic expressions for the corresponding SDCs and greatly simplifies the treatment of spin, color, and kinematic structures. 
This feature is particularly valuable for systems such as $\Omega_{3Q}$, where a complete perturbative-QCD calculation becomes analytically cumbersome.

The analogy with NRQCD is instructive. 
In both approaches, short-distance dynamics are separated from long-distance hadronization through effective composite degrees of freedom, allowing a transparent factorization between perturbative production and bound-state formation. 
The Suzuki picture therefore offers a practical realization of perturbative/nonperturbative factorization for baryon fragmentation while retaining sensitivity to the underlying hadronic structure.

In the implementation of Refs.~\cite{MoosaviNejad:2017rvi,Delpasand:2019xpk}, the nonperturbative dynamics of the baryon are encoded through a light-cone distribution amplitude. Following the Lepage-Brodsky prescription~\cite{Lepage:1980fj}, this amplitude is approximated by a Dirac $\delta$ function in the momentum-fraction variables, forcing the constituents to move collinearly inside the baryon. 
The resulting simplification yields compact analytic expressions in which the nonperturbative input is effectively absorbed into the fragmentation amplitude. 
Similar strategies have successfully been employed in studies of all-heavy-light tetraquarks~\cite{Nejad:2021mmp,Celiberto:2023rzw} and pentaquark-charmonium systems~\cite{Farashaeian:2024son,Farashaeian:2024cpd,Celiberto:2025ipt}.

A key ingredient of the initial-scale FF construction is the nonperturbative transition from the constituent diquark ${\cal D}$ to the physical baryon $\Omega_{3Q}$. 
Following Refs.~\cite{MoosaviNejad:2017rvi,Delpasand:2019xpk}, we describe this stage through the Peterson-Schlatter-Schmitt-Zerwas (PSSZ) parametrization~\cite{Peterson:1982ak}, a widely used model for heavy-flavor fragmentation and particularly suitable for hadrons containing multiple heavy constituents.
We write
\begin{equation}
\label{HFF_PSSZ}
 D_{\cal D}^{\Omega_{3Q}} (z) \equiv D_{\rm [np]}^{\rm [PSSZ]} (z) = {\cal N}_P \frac{z(1-z)^2}{\left[(1-z)^2 + \gamma_P z \right]^2} \;,
\end{equation}
with normalization factor
\begin{equation}
\begin{split}
\label{HFF_PSSZ_Nrm}
 {\cal N}_P &\,=\, 
  \left\{
 \frac{\gamma_P^2-6\gamma_P+4}{(4-\gamma_P)\sqrt{4\gamma_P-\gamma_P^2}}
  \right.
  \left[ 
 \arctan \frac{\gamma_P}{\sqrt{4 \gamma_P - \gamma_P^2}} + \arctan \frac{2-\gamma_P}{\sqrt{4 \gamma_P - \gamma_P^2}} 
  \right]
  \\ &\,
 + \,
  \left.
 \frac{1}{2} \ln \gamma_P + \frac{1}{4-\gamma_P} 
  \right\} \;.
\end{split}
\end{equation}
The normalization is obtained by summing over all hadrons containing a constituent heavy quark $Q$ quark~\cite{Cacciari:1997du}; see also Section~IV of Ref.~\cite{Peterson:1982ak}. 
As in the original formulation, the parameter $\gamma_P$ cannot be determined from first principles, since it encodes genuinely nonperturbative dynamics.

For singly heavy hadrons, such as $D$ or $B$ mesons with leading Fock states $|Q\bar q\rangle$, one expects $\gamma_P\simeq {\Lambda}/m_Q$, where ${\Lambda}$ denotes a hadronic scale of the order of the constituent-light-quark mass and $m_Q$ stands for the charm- or bottom-quark mass.
In this case, the average momentum fraction satisfies $\langle z\rangle = 1-\sqrt{\gamma_P}$, consistently with the expected heavy-quark scaling behavior of FFs~\cite{Suzuki:1977km,Bjorken:1977md,Kinoshita:1985mh,Cacciari:1996wr}.
For all-heavy baryons, on the other hand, the relevant scale is determined entirely by heavy constituents.
Following Refs.~\cite{Adamov:1997yk,MoosaviNejad:2017rvi,Delpasand:2019xpk}, we therefore adopt
\begin{equation}
\label{HFF_PSSZ_gamma_p}
 \gamma_P = \left[\frac{m_Q}{m_Q+m_{\bar Q}}\right]^2 = \frac{1}{4} \;.
\end{equation}
This choice is consistent with previous studies of all-heavy baryons~\cite{MoosaviNejad:2017rvi,Delpasand:2019xpk}, where the Peterson form is used to model the hadronization of the intermediate diquark. 
The parametrization naturally complements the perturbative calculation by supplying a universal nonperturbative component compatible with the Suzuki picture and with collinear factorization.

The physical motivation is straightforward. 
The Peterson model was originally devised for heavy-quark fragmentation, where the produced hadron inherits a large fraction of the momentum of the parent parton. 
The same expectation applies to all-heavy baryons, whose valence structure contains only heavy constituents. 
Consequently, the FF is expected to peak at large $z$ and to be suppressed at small $z$, a behavior accurately reproduced by Eq.~\eqref{HFF_PSSZ}. 
The simple analytic form of the model, controlled by the single parameter $\gamma_P$, also facilitates its implementation within collinear convolution frameworks and DGLAP evolution.

The resulting initial-scale FF for a parton $i$ fragmenting into a $\Omega_{3Q}$ baryon can then be written as
\begin{equation}
\label{FFs_O3Q_initial}
 D_i^{\Omega_{3Q}} (z, \mu_{F,0}) = 
 \int_z^1 \frac{\drv \xi}{\xi} D_i^{\cal D} (\xi, \mu_{F,0}) \, D_{\cal D}^{\Omega_{3Q}} \left( \frac{z}{\xi} \right) \;.
\end{equation}
Equation~\eqref{FFs_O3Q_initial} mirrors the general heavy-hadron expression of Eq.~\eqref{FFs_HF_initial}, with the substitutions
\begin{subequations}
\begin{align}
\label{FFs_O3Q_vs_HF_initial_a}
 D_i^Q (\xi, \mu_{F,0})
  \quad &\longrightarrow \quad
 D_i^{\cal D} (\xi, \mu_{F,0})
  \;,
  \\[0.25cm]
\label{FFs_O3Q_vs_HF_initial_b}
 D_{\rm [np]}^{\HQ} \left( \frac{z}{\xi} \right)
  \quad &\longrightarrow \quad
 D_{\cal D}^{\Omega_{3Q}} \left( \frac{z}{\xi} \right)
  \;.
\end{align}
\end{subequations}
The perturbative component is therefore described by SDCs for the splitting of a massless parton into a diquark state ${\cal D}$, Eq.~\eqref{FFs_O3Q_vs_HF_initial_a}, while the nonperturbative stage is encoded in the diquark-to-baryon transition of Eq.~\eqref{FFs_O3Q_vs_HF_initial_b}. 
In this sense, the diquark ${\cal D}$ plays the role of the heavy constituent entering the lowest Fock state of the singly heavy hadron $\HQ$. 
As discussed above, the latter transition is modeled through the PSSZ form of Eq.~\eqref{HFF_PSSZ}, whereas the quark- and gluon-induced SDCs are presented in the next section.

\section{Construction of the {\tt OMG3Q1.1} functions}
\label{sec:FFs-architecture}

\begin{figure*}[!t]
\centering
\includegraphics[width=0.475\textwidth]{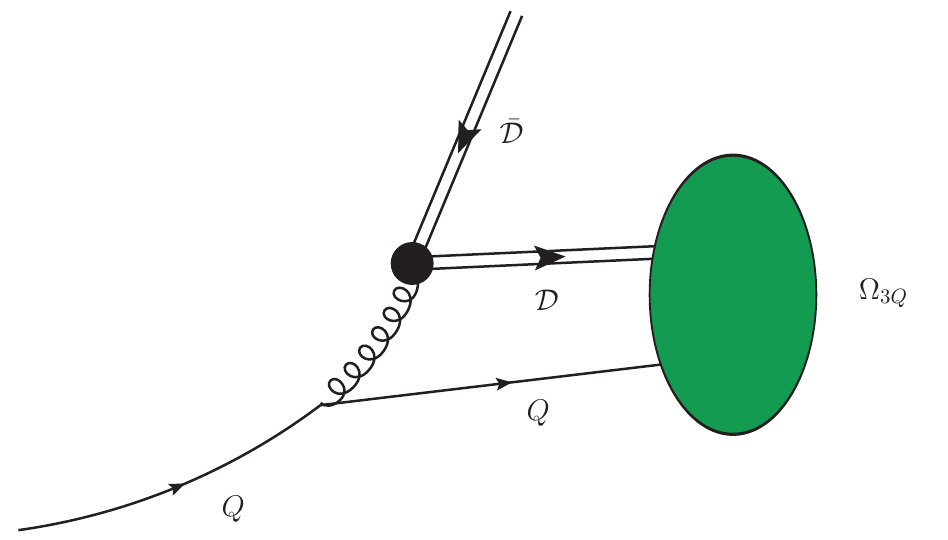}
\hspace{0.40cm}
\includegraphics[width=0.475\textwidth]{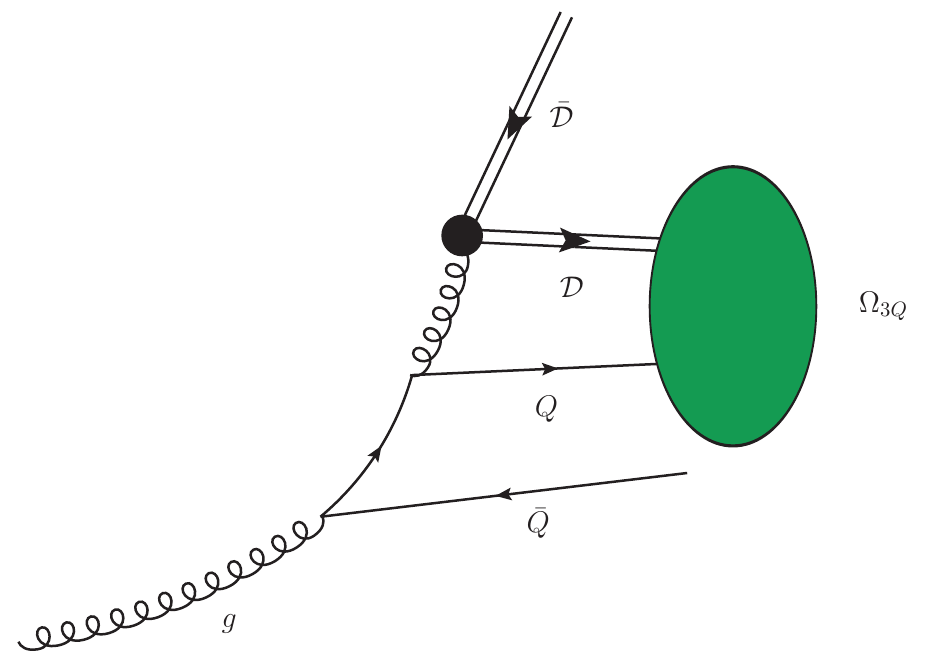}

\caption{Representative, leading-order contributions to the diquark-inspired proxy model for the initial fragmentation of constituent heavy quarks (left) and gluons (right) into color-singlet $S$-wave $\Omega_{3Q}$ baryons. 
Double lines represent ${\cal D}$ and $\bar{\cal D}$ diquark degrees of freedom, while green blobs account for the nonperturbative component of the fragmentation process. 
Black bullets denote effective couplings associated with the scalar-diquark form factor.}
\label{fig:O3Q_FF_diagrams}
\end{figure*}

In this section, we describe the procedure adopted to construct the {\tt OMG3Q1.1} VFNS fragment\-ation-function set for same-flavor all-heavy $\Omega$ baryons.
The determination starts from diquark-inspired initial-scale inputs for both constituent-heavy-quark and gluon fragmentation channels and incorporates threshold-aware timelike DGLAP evolution within the {\HFNRevo} framework~\cite{Celiberto:2025euy,Celiberto:2024mex,Celiberto:2024bxu,Celiberto:2024rxa,Celiberto:2025xvy,Celiberto:2026rzi,Celiberto:2026zss}.

All ingredients entering the {\tt OMG3Q1.1} construction have been derived with {\symJethad}, the \textsc{Mathematica}~\cite{Mathematica_V14-2} symbolic component of the {\Jethad} environment~\cite{Celiberto:2020wpk,Celiberto:2022rfj,Celiberto:2023fzz,Celiberto:2024mrq,Celiberto:2024swu,Celiberto:2025csa,Celiberto:2026ooh}, developed for analytical calculations in hadron structure and precision QCD studies.

The constituent-quark and gluon initial conditions are presented in Sections~\ref{ssec:FFs-Q} and~\ref{ssec:FFs-g}, respectively.
The resulting timelike DGLAP evolution and the main properties of the {\tt OMG3Q1.1} set are discussed in Section~\ref{ssec:FFs-OMG3Q11}.

\subsection{Constituent-quark initial-scale input}
\label{ssec:FFs-Q}

The left panel of Fig.~\ref{fig:O3Q_FF_diagrams} illustrates the diquark-inspired fragmentation mechanism considered in this work. Heavy diquark and antidiquark states, ${\cal D}\equiv |QQ\rangle$ and $\bar{\cal D}\equiv |\bar Q\bar Q\rangle$, are represented by double lines. 
The green blob denotes the nonperturbative transition function $D_{\cal D}^{\Omega_{3Q}}$, whereas black bullet vertices correspond to scalar-diquark form-factor couplings.

Although omitted from the analytical expressions reported below, the calculation includes the scalar-diquark form factor associated with the vertex shown in Fig.~\ref{fig:O3Q_FF_diagrams}. Following Ref.~\cite{MoosaviNejad:2017rvi}, the diquark is modeled through a simple-pole form in transverse-momentum space with a characteristic scale of approximately $5$~GeV. 
For compactness, its contribution is absorbed into the normalization of the perturbative amplitude.

The diagram in the left panel of Fig.~\ref{fig:O3Q_FF_diagrams} corresponds to the splitting process $[Q \to (Q,{\cal D})+\bar{\cal D}]$. 
Charge-conjugation symmetry yields the corresponding antiquark channel $[\bar Q \to (\bar Q,\bar{\cal D})+{\cal D}]$. 
In the present study, we assume identical formation mechanisms for all-heavy baryons and antibaryons and focus on charge-averaged observables. 
Under this assumption, the $Q$- and $\bar Q$-fragmentation channels are treated symmetrically (see Ref.~\cite{Bertone:2018ecm} for comparison with the light-hadron case).

Using {\symJethad}~\cite{Celiberto:2020wpk,Celiberto:2022rfj,Celiberto:2023fzz,Celiberto:2024mrq,Celiberto:2024swu,Celiberto:2025csa,Celiberto:2026ooh} together with {\FeynCalc}~\cite{Mertig:1990an,Shtabovenko:2016sxi,Shtabovenko:2020gxv}, we independently derived the LO and NLO heavy-quark SDCs and verified agreement with the results of Refs.~\cite{MoosaviNejad:2017bda,MoosaviNejad:2017rvi}.
Keeping the accuracy at LO, one has
\begin{equation}
\label{FFs_O3Q_Q}
 D_Q^{\cal D} (z, \mu_{F,0}) \,=\,
 \as^2
 \drv_Q^{\rm [LO]} (z; \mu_{F,0}) \\[0.10cm]
 \,+\,
 \as^3 \, \drv_Q^{\rm [NLO]} (z; \mu_{F,0})
 \;.
\end{equation}
where $\as \equiv \as(5m_Q)$ and
\begin{equation}
\label{FFs_O3Q_Q_d_LO}
 \drv_Q^{\rm [LO]} (z; \mu_{F,0}) =
 {\cal N}_Q(z)
 \frac{{\cal S}_Q^{\rm [LO]}(z; {\cal R}_{q_T/Q})}{{\cal T}_Q^{\rm [LO]}(z; {\cal R}_{q_T/Q})}
 \;,
\end{equation}
with
\begin{equation}
\label{R_qTQ}
  {\cal R}_{q_T/Q}^2 \equiv \vqTTa/m_Q^2
 \;,
\end{equation}
and
\begin{equation}
\label{FFs_O3Q_Q_d_norm}
 {\cal N}_Q(z) = \frac{\pi^2}{\sqrt{6}} \, f_{\cal B}^2 C_F^2 \, z^3(1-z)^3
 \;.
\end{equation}
Furthermore, one writes
\begin{equation}
\begin{split}
\label{FFs_O3Q_Q_d_num_LO}
  {\cal S}_Q^{\rm [LO]}(z; {\cal R}_{q_T/Q}) 
  \,&=\, 
  4(41z^2 - 80z + 96 + 256/z^2) \\
  \,&+\,
  {\cal R}_{q_T/Q}^2 (8z^3 + 5z^2 - 16z + 16) \\
  \,&+\,
  {\cal R}_{q_T/Q}^4 \, 4z^2
  \;,
\end{split}
\end{equation}
and
\begin{equation}
\label{FFs_O3Q_Q_d_den_LO}
 {\cal T}_Q^{\rm [LO]}(z; {\cal R}_{q_T/Q}) 
  \,=\, 
  [{\cal R}_{q_T/Q}^2 + (4-3z)^2]^2
  \,
  ({\cal R}_{q_T/Q}^2 \, z^2 + z^2 - 16 z + 16)^2
 \;.
\end{equation}
In Eq.~\eqref{FFs_O3Q_Q_d_norm}, the quantity $C_F=(N_c^2-1)/(2N_c)$ corresponds to the fundamental Casimir factor of $SU(3)_c$, whereas the parameter $f_{\cal B}$ stands for the decay constant of the rare baryon.
We set $f_{\cal B}=0.25$~GeV, following standard choices adopted in the literature~\cite{ParticleDataGroup:2020ssz}. 

The NLO heavy-quark SDC employed in this work was originally derived in Ref.~\cite{MoosaviNejad:2016qdx} and later adapted to the triply heavy baryon case in Ref.~\cite{MoosaviNejad:2017rvi}. One has
\begin{equation}
\label{FFs_O3Q_Q_d_NLO}
 \drv_Q^{\rm [NLO]} (z; \mu_{F,0}) =
 {\cal N}_c(z)
 \frac{{\cal S}_Q^{\rm [NLO]}(z; {\cal R}_{q_T/Q})}{{\cal T}_Q^{\rm [NLO]}(z; {\cal R}_{q_T/Q})}
 \;,
\end{equation}
where
\begin{equation}
\begin{split}
\label{FFs_O3Q_Q_d_num_NLO}
  {\cal S}_Q^{\rm [NLO]}&(z; {\cal R}_{q_T/Q}) 
  \,=\, 
  96 \pi z
\left[\right.
  {\cal R}_{q_T/Q}^{14} \, (17-20 z) z^{14} \\
\,&+\,
  {\cal R}_{q_T/Q}^{12} \, (240 z^4-1664 z^3 +
  3813 z^2-4130 z+1711) z^{12} \\
\,&+\,
  {\cal R}_{q_T/Q}^{10} \, (4368 z^6-39072 z^5+140635 z^4 \\
\,&-\,
  269144 z^3+300490 z^2 -
  189616 z+52243) z^{10} \\
\,&+\,
  {\cal R}_{q_T/Q}^8 \, (20976 z^8-208040 z^7 +
  870151 z^6-2070634 z^5+3255115 z^4 \\
\,&-\,
  3533740 z^3+2600201 z^2 -
  1202330 z+268205) z^8 \\
\,&+\,
  {\cal R}_{q_T/Q}^6 \, 4 (1-z)^2 (11256 z^8-75035 z^7 -
  151779 z^6+2535886 z^5 \\
\,&-\,
  8913460 z^4+15864355 z^3 -
  15844619 z^2+8395110 z-1823490) z^6 \\
\,&+\,
  {\cal R}_{q_T/Q}^4 \, 4 (1-z)^4 (12324 z^8-22830 z^7 -
  2045973 z^6 + 14641206 z^5 \\
\,&-\, 
  42277675 z^4 +
  65338590 z^3-57387186 z^2 +
  26330832 z-4933872) z^4 \\
  \,&+\,
  {\cal R}_{q_T/Q}^2 \, 48 (1-z)^6 (563 z^8+2272 z^7 -
  252314 z^6+1752016 z^5 \\
\,&-\,
  5068605 z^4 +
  7770780 z^3-6800544 z^2 +
  3505032 z-849528) z^2 \\
\,&+\,
  144 (1-z)^8 (41 z^8+438 z^7-36063 z^6 +
  258240 z^5-833328 z^4 \\
\,&+\, 
  1614600 z^3
  -1793016 z^2+676512 z-198288)
\left.\right]
  \;,
\end{split}
\end{equation}
and
\begin{equation}
\label{FFs_O3Q_Q_d_den_NLO}
\begin{split}
\hspace{-0.15cm}
 {\cal T}_Q^{\rm [NLO]}(z; {\cal R}_{q_T/Q}) 
  \,&=\, 
 [{\cal R}_{q_T/Q}^2 \, z^2 + 4 (3-2 z)^2]^2 \\
\,&\times\,
 [{\cal R}_{q_T/Q}^2 \, z^2 + (6-z)^2]^2  \\
\,&\times\,
[{\cal R}_{q_T/Q}^2 \, z^2 + (1-z)^2] \\
\,&\times\, 
[{\cal R}_{q_T/Q}^2 \, z^2 + 36(1-z)^2]^2  \\
\,&\times\,
[{\cal R}_{q_T/Q}^2 \, z^2 + z^2-35 z+36]
  \;.
\end{split}
\end{equation}
The initial scale of the $Q$-quark fragmentation channel is chosen as $\mu_{F,0}=5m_Q$, consistently with the kinematics of the LO process shown in the left panel of Fig.~\ref{fig:O3Q_FF_diagrams}.

Our heavy-quark FF differs from the construction of Ref.~\cite{MoosaviNejad:2017rvi} in two respects. 
First, the normalization factor ${\cal N}_Q(z)$ entering Eq.~\eqref{FFs_O3Q_Q_d_norm} is computed explicitly rather than fixed through an external normalization condition.
Second, we revisit the choice of the transverse-momentum parameter $\vqTTa$, which constitutes one of the key phenomenological ingredients of the Suzuki framework.

Within the Suzuki picture, the transverse momentum of the outgoing heavy quark is replaced by its average value, $\vqTTa$, effectively reducing the underlying TMD description to a collinear one. 
The parameter therefore controls the shape of the FF and must be determined phenomenologically. 
Larger values of $\vqTTa$ shift the distribution toward smaller $z$ and reduce its overall normalization~\cite{GomshiNobary:1994eq}.

The FF of Ref.~\cite{MoosaviNejad:2017rvi} was obtained using $\vqTTa=1~{\rm GeV}^2$, interpreted as an upper estimate of the average transverse momentum. 
Here, we adopt a different strategy, inspired by our previous studies of all-heavy tetraquarks and pentaquarks~\cite{Celiberto:2024mab,Celiberto:2025ipt,Celiberto:2026rdk}. 
In those analyses, $\vqTTa$ was fixed by requiring physically reasonable momentum-fraction distributions together with a balanced normalization of quark- and gluon-initiated channels. 
The resulting values, $\vqTTa_{\TQc}=70~{\rm GeV}^2$ and $\vqTTa_{\PQc}=90~{\rm GeV}^2$, were found to yield fragmentation patterns consistent with the kinematic region most relevant for collider phenomenology.

The same strategy is adopted here. A dedicated numerical scan of the initial-scale FFs leads to the reference choice
\begin{equation}
\label{eq:qT2_O3Q_central}
 \vqTTa_{\Omega_{3Q}} = 60~{\rm GeV}^2 \;,
\end{equation}
which places the maximum of the distribution in the phenomenologically relevant region, $\langle z \rangle \gtrsim 0.4$, while preserving a realistic balance between the heavy-quark and gluon fragmentation channels.

Although the value in Eq.~\eqref{eq:qT2_O3Q_central} lies well above the characteristic hadronic scale, its interpretation should be regarded as effective rather than directly connected to soft nonperturbative dynamics.
In all-heavy systems, the absence of light constituents prevents the emergence of a natural scale governed by $\LQCD$, while the presence of multiple heavy quarks is expected to generate a broader intrinsic momentum distribution inside the bound state.
Consequently, $\vqTTa$ should not be viewed as a direct measure of low-energy nonperturbative physics, but rather as an effective parameter encoding the momentum-space structure of the all-heavy Fock state.

This interpretation is consistent with the original heavy-flavor fragmentation picture~\cite{Suzuki:1977km,Bjorken:1977md,Kinoshita:1985mh,Peterson:1982ak}, where the large-$z$ behavior of the FF is primarily driven by the heavy-quark mass.
Unlike ordinary heavy hadrons, however, all-heavy baryons and exotic multiquark states do not contain a light constituent capable of fixing the position of the FF maximum through simple kinematic arguments.
The absence of a characteristic soft scale in their leading Fock configurations makes the shape of the FF more sensitive to the dynamics of multiquark binding.
As already observed for all-heavy tetraquarks and pentaquarks~\cite{Celiberto:2024mab,Celiberto:2025ipt}, a phenomenological determination of $\vqTTa$ therefore becomes necessary.

Within the Suzuki framework, $\vqTTa$ governs the transverse-momentum structure of the bound state and can be interpreted as an effective parameter entering the nonperturbative wave function.
Following the strategy adopted in our previous {\tt 1.0} determinations~\cite{Celiberto:2025ogy}, we identify $\vqTTa$ as the quantity controlling the nonperturbative-wave-function component (F-NPWF) of the FF.
In the {\tt OMG3Q1.1} framework, uncertainties associated with this contribution are estimated through systematic variations of $\vqTTa$ around the central value of Eq.~\eqref{eq:qT2_O3Q_central}, as discussed in Section~\ref{ssec:FFs-OMG3Q11}.

\subsection{Gluon initial-scale input}
\label{ssec:FFs-g}

The right panel of Fig.~\ref{fig:O3Q_FF_diagrams} illustrates the gluon-initiated fragmentation mechanism considered in this work. 
As for the heavy-quark channel, the final-state baryon is produced through the formation of an intermediate heavy diquark state, followed by its nonperturbative transition into the physical $\Omega_{3Q}$ baryon.

Although both channels represent the leading contributions within their respective topologies, they enter the perturbative expansion at different orders. 
The heavy-quark channel is known at both LO and NLO accuracy (Section~\ref{ssec:FFs-Q}), whereas the gluon channel starts at ${\cal O}(\alpha_s^3)$ and is currently available only at that order. 
This hierarchy originates from the absence of heavy-flavor quantum numbers in the initial gluon state, which requires the perturbative generation of the heavy constituents needed to form the final all-heavy baryon.

Throughout this work, we follow the convention adopted in Ref.~\cite{Delpasand:2019xpk} and refer to the ${\cal O}(\alpha_s^3)$ gluon contribution as the NLO initial-scale FF, consistently with the perturbative counting used for the heavy-quark channel.

Using {\symJethad}~\cite{Celiberto:2020wpk,Celiberto:2022rfj,Celiberto:2023fzz,Celiberto:2024mrq,Celiberto:2024swu,Celiberto:2025csa,Celiberto:2026ooh} together with {\FeynCalc}~\cite{Mertig:1990an,Shtabovenko:2016sxi,Shtabovenko:2020gxv}, we independently reconstructed the analytical expression of the gluon SDC and verified agreement with the result of Ref.~\cite{Delpasand:2019xpk}. 
We write
\begin{equation}
\label{FFs_O3Q_g}
 D_g^{\cal D} (z, \mu_{F,0}) \,=\, 
 \as^3 \, \drv_g^{\rm [NLO]} (z; \mu_{F,0}) \;.
\end{equation}
with $\as \equiv \as(6m_c)$ and
\begin{equation}
\label{FFs_O3Q_g_d}
 \drv_g^{\rm [NLO]} (z; \mu_{F,0}) =
 {\cal N}_g(z)
 \frac{{\cal S}_g^{\rm [NLO]}(z; {\cal R}_{q_T/Q})}{{\cal T}_g^{\rm [NLO]}(z; {\cal R}_{q_T/Q})}
 \;,
\end{equation}
with
\begin{equation}
\label{FFs_O3Q_g_d_norm}
 {\cal N}_g(z) = \frac{2 \pi^3}{3} \, f_{\cal B}^2 C_F^2 \, z^3(1-z)^2
 \;.
\end{equation}
Furthermore, one has
\begin{equation}
\begin{split}
\label{FFs_O3Q_g_d_num_NLO}
  {\cal S}_g^{\rm [NLO]}(z; {\cal R}_{q_T/Q}) 
  \,&=\, 
  8(16z^2-32z+15) \\
  \,&+\,
  {\cal R}_{q_T/Q}^2 \, 2z^2(4z^2-20z+17) \\ 
  \,&+\,
  {\cal R}_{q_T/Q}^4 \, z^4 \;,
\end{split}
\end{equation}
and
\begin{equation}
\label{FFs_O3Q_g_d_den_NLO}
\begin{split}
 {\cal T}_g^{\rm [NLO]}&(z; {\cal R}_{q_T/Q}) 
  \,=\, 
  (4 + z^2{\cal R}_{q_T/Q}^2)^5
 \;.
\end{split}
\end{equation}
As in Section~\ref{ssec:FFs-Q} (see Eq.\eqref{R_qTQ}), we define
${\cal R}_{q_T/Q}^2 = \vqTTa/m_Q^2$,
using the reference value $\vqTTa_{\Omega_{3Q}} = 60~{\rm GeV}^2$
introduced in Eq.~\eqref{eq:qT2_O3Q_central}. This choice defines the central member of the {\tt OMG3Q1.1} replica ensemble, while F-NPWF uncertainties are estimated through systematic variations around this reference configuration, as discussed in Section~\ref{ssec:FFs-OMG3Q11}.

Unlike the heavy-quark channel, the kinematics of the lowest-order gluon fragmentation process naturally set a higher initial scale. 
We therefore choose $\mu_{F,0} = 6m_Q$, which corresponds to the characteristic scale associated with the diagram shown in the right panel of Fig.~\ref{fig:O3Q_FF_diagrams}.

\subsection{The {\tt OMG3Q1.1} sets from {\HFNRevo}}
\label{ssec:FFs-OMG3Q11}

The {\tt OMG3Q1.1} release combines the initial-scale FFs derived in the previous sections with a threshold-aware DGLAP evolution performed within a VFNS framework. 
Unlike ordinary light-hadron fragmentation, the present case involves two independent parton sectors, namely the constituent heavy-quark and gluon channels, each characterized by its own activation scale. 
This structure reflects the perturbative production mechanisms of the lowest all-heavy baryonic configuration and requires a dedicated treatment of evolution thresholds.

For an $\Omega_{3Q}$ state, kinematics fixes the onset of heavy-quark fragmentation at $\mu_F=5m_Q$, whereas gluon fragmentation becomes active at the larger threshold $\mu_F=6m_Q$. 
The evolution between and beyond these scales is implemented through the {\HFNRevo} framework~\cite{Celiberto:2024mex,Celiberto:2024bxu,Celiberto:2024rxa,Celiberto:2025xvy,Celiberto:2026rzi,Celiberto:2026zss}, which was specifically developed to evolve heavy-hadron FFs originating from nonrelativistic inputs while preserving a consistent connection between threshold dynamics and VFNS evolution.

The {\HFNRevo} framework is built upon three complementary pillars: interpretation, evolution, and uncertainty quantification. 
The first provides a physical interpretation of low-scale heavy-hadron production in terms of effective fragmentation mechanisms, enabling a consistent connection between FFNS and VFNS descriptions. 
The second implements threshold-aware DGLAP evolution across multiple heavy-parton sectors, organized into two sequential stages, denoted as {\tt EDevo} and {\tt AOevo}.

The {\tt EDevo} stage provides a semianalytical, expanded treatment of the evolution between the lowest and highest initial fragmentation thresholds, with the relevant partonic channels activated according to their corresponding starting scales. 
This intermediate stage is particularly useful for retaining parton-dependent activation thresholds, including those associated with gluon and light-(anti)quark fragmentation channels, whose independent implementation can be less straightforward within standard all-order numerical DGLAP evolution codes.
At the highest threshold, the resulting FFs define a common evolution-ready boundary condition at the scale $Q_0$.

The subsequent {\tt AOevo} stage performs the standard coupled, all-order numerical DGLAP evolution from $Q_0$ toward higher factorization scales.
The third introduces a systematic treatment of theoretical uncertainties through controlled perturbative and nonperturbative variations, which in the present {\tt OMG3Q1.1} release are encoded through a replica-based structure.

Initially conceived for quarkonium phenomenology, {\HFNRevo} has progressively evolved into a general framework for heavy-flavor and multiquark fragmentation studies. 
Recent applications to fully heavy tetraquarks~\cite{Celiberto:2024mab,Celiberto:2024beg,Celiberto:2025ziy,Celiberto:2026kks} and early studies on all-heavy baryons~\cite{Celiberto:2025ogy} have demonstrated its capability to accommodate systems characterized by independent heavy-quark and gluon FF inputs defined at distinct initial scales.

In the specific $\Omega_{3Q}$ case, this structure is particularly transparent: independent initial-scale inputs are currently available only for the constituent-heavy-quark and gluon fragmentation channels, whereas all remaining partonic components are dynamically generated through the subsequent coupled DGLAP evolution.
Such a dual-input structure represents one of the defining features of heavy-hadron fragmentation and motivates the adoption of a dedicated evolution strategy.

The first stage of the evolution is confined to the interval between the heavy-quark and gluon thresholds. 
In this region only the constituent heavy-quark channel is active, so the corresponding FF evolves independently through the $P_{qq}$ splitting kernel. 
For $\Omega_{3Q}$ fragmentation, this {\tt EDevo} stage therefore extends from $\mu_F=5m_Q$ to $\mu_F=6m_Q$ and proceeds without quark-gluon mixing.
Owing to the absence of channel mixing, this step can be carried out semianalytically within the {\symJethad} environment~\cite{Celiberto:2020wpk,Celiberto:2022rfj,Celiberto:2023fzz,Celiberto:2024mrq,Celiberto:2024swu,Celiberto:2025csa,Celiberto:2026ooh}, providing the boundary condition for the subsequent coupled evolution above the gluon threshold.
More specifically, the constituent-heavy-quark input is evolved from $5m_Q$ up to $6m_Q$, where its evolved value is combined with the independently calculated gluon FF, whose initial condition becomes active precisely at that scale.

The FFs obtained at the gluon threshold define the \emph{evolution-ready} scale $Q_0$, corresponding to the highest active partonic threshold, $Q_0=6m_Q$ for the all-heavy baryons considered here. 
At this scale, the constituent-heavy-quark FF evolved through {\tt EDevo} and the independently calculated gluon FF provide the two nonzero boundary conditions for the subsequent {\tt AOevo} stage.
From this common starting point, all partonic species are evolved through the full NLO timelike DGLAP equations using {\tt APFEL++}~\cite{Bertone:2013vaa,Carrazza:2014gfa,Bertone:2017gds}. 
In this second stage, the evolution becomes fully coupled: quark-gluon mixing is restored through the corresponding splitting kernels, and the initially vanishing light- and nonconstituent heavy-quark FFs are dynamically generated as the factorization scale increases.
The {\tt AOevo} evolution is then continued from $Q_0$ up to the factorization scales relevant to the phenomenological analysis, and the resulting {\tt OMG3Q1.1} sets are released in {\tt LHAPDF} format.

Since no calculation is currently available for the fragmentation of nonconstituent quarks into all-heavy baryons, light- and nonconstituent heavy-quark channels are not assigned independent FF inputs at the initial scale. 
Instead, they are generated dynamically through the numerical {\tt AOevo} DGLAP evolution above $Q_0$. 
This approximation is expected to be well justified, as analogous studies of heavy-quarkonium fragmentation indicate a strong suppression of nonconstituent-quark channels relative to constituent heavy-quark and gluon contributions~\cite{Braaten:1993rw,Braaten:1993mp,Artoisenet:2014lpa,Zhang:2018mlo,Zheng:2021mqr,Zheng:2021ylc,Zheng:2019dfk}.

Should dedicated initial-scale calculations for these channels become available, they can be incorporated at their corresponding activation scales within the same threshold-aware construction. 
The {\tt EDevo} stage would then sequentially activate and evolve the enlarged set of independently initialized channels up to the highest threshold, thereby preserving their distinct physical starting scales before constructing the common boundary conditions required by {\tt AOevo}.

\begin{figure}[t]
\centering

 \includegraphics[scale=0.135,clip]{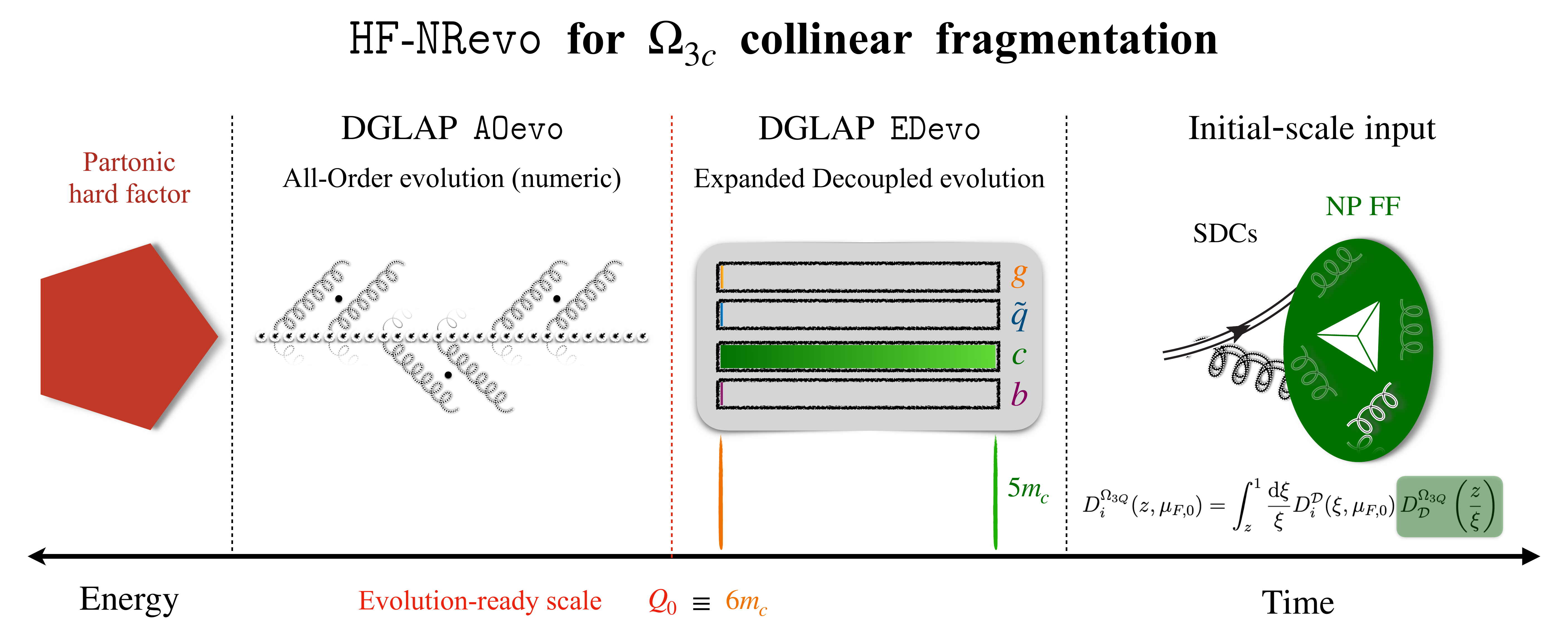}

 \vspace{0.55cm}

 \includegraphics[scale=0.135,clip]{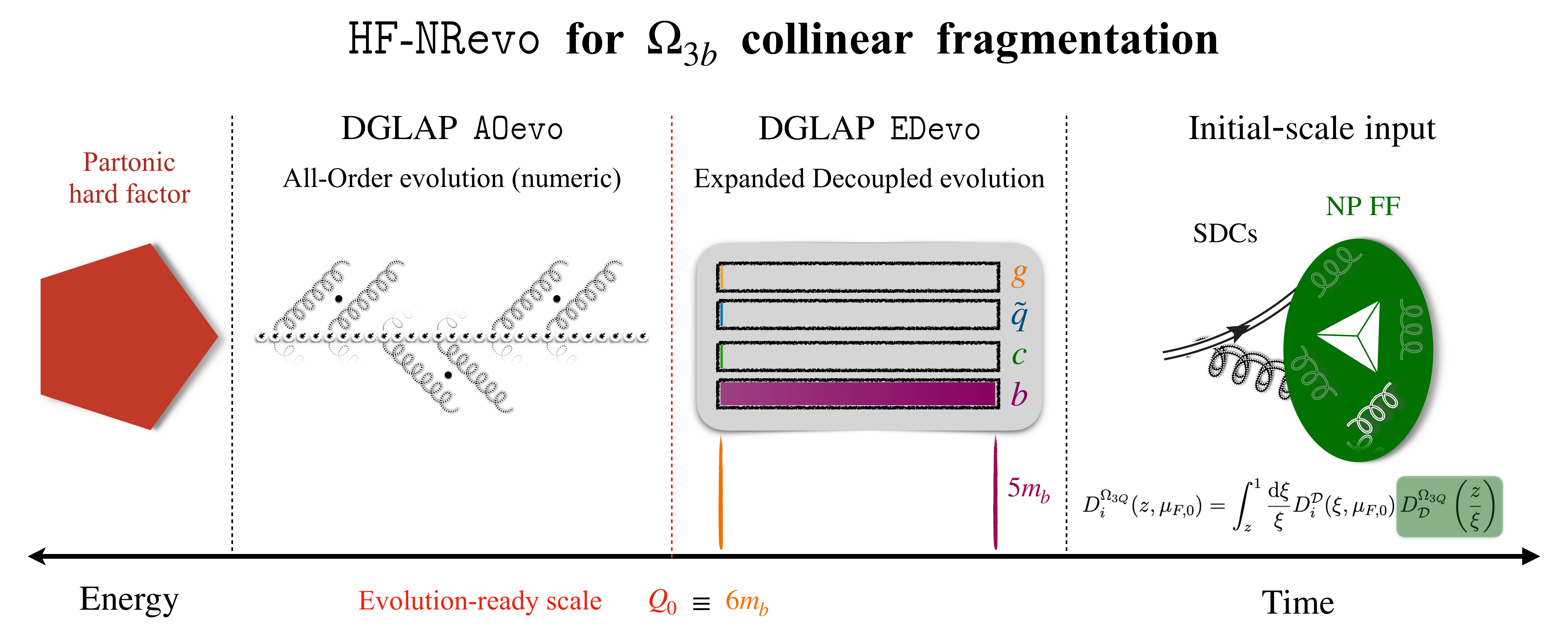}

\caption{
Two-stage evolution workflow for $\Omega_{3c}$ (upper) and $\Omega_{3b}$ (lower panels) collinear FFs within the {\HFNRevo} scheme.
The construction is based on a sequential DGLAP evolution: a first, semianalytical {\tt EDevo} stage extends from the lowest threshold, $5m_Q$, to the evolution-ready scale $Q_0=6m_Q$, where the independently calculated gluon input becomes active.
A fully numerical {\tt AOevo} stage then performs the coupled NLO DGLAP evolution from $Q_0$ toward higher factorization scales.
}
\label{fig:O3Q_cs_HFNRevo_sketch}
\end{figure}

The two-stage {\HFNRevo} construction described above is summarized in Fig.~\ref{fig:O3Q_cs_HFNRevo_sketch} for the $\Omega_{3c}$ and $\Omega_{3b}$ cases. 
The horizontal direction represents the evolution from the initial fragmentation thresholds toward increasingly larger factorization scales when read from right to left, thus following the actual construction of the FF sets. 
Starting from the constituent-heavy-quark input at $5m_Q$, the {\tt EDevo} stage evolves this channel up to the evolution-ready scale $Q_0=6m_Q$, where the independently calculated gluon input becomes active. 
The resulting heavy-quark and gluon FFs then provide the boundary conditions for the {\tt AOevo} stage, which performs the fully coupled NLO DGLAP evolution toward the hard scale and dynamically generates the remaining partonic components.

Conversely, reading the diagram from left to right provides the chronological fragmentation picture: a high-energy parton produced in the short-distance scattering undergoes partonic evolution toward lower scales and ultimately fragments into the observed $\Omega_{3Q}$ state. 
The same workflow applies to both the all-charm and all-bottom sectors, with the corresponding thresholds determined by the relevant heavy-quark mass.

A distinctive feature of the {\tt OMG3Q1.1} release is the incorporation of uncertainty quantification directly into the FF construction. Perturbative effects are probed through variations of the evolution-ready scale around its natural value, $Q_0 \to K_{\mu}Q_0$, with $1/2 \le K_{\mu} \le 2$. 
These variations define the fragmentation missing higher-order uncertainties (F-MHOUs) and provide a measure of the sensitivity to subleading perturbative contributions beyond the nominal accuracy.

Nonperturbative uncertainties are estimated through variations of the transverse-momentum parameter $\vqTTa$, which controls the shape of the wave-function sector entering the Suzuki-inspired construction. Although $\vqTTa$ does not represent a genuinely soft scale in all-heavy systems, it effectively parametrizes the internal momentum distribution of the bound state. 
The corresponding fragmentation nonperturbative wave-function (F-NPWF) variations therefore probe the dependence of the FFs on the underlying hadron-structure assumptions.

Both F-MHOUs and F-NPWF effects are encoded through a replica-based organization of the FF sets. 
Each replica corresponds to a specific combination of perturbative and nonperturbative parameters, allowing the resulting uncertainty envelope to be propagated directly to collider observables. 
Unlike the Monte Carlo replicas employed in modern PDF analyses, these replicas do not represent samples drawn from a probability distribution. 
Rather, they provide a structured exploration of physically motivated theory variations, combining the practical advantages of replica-based methodologies with a transparent interpretation of the underlying uncertainty sources. 
The complete correspondence between replica indices and parameter configurations is reported in Appendix~\hyperlink{app:A}{A}.

Figure~\ref{fig:FFs-z_O3Q_Q} displays the $z$ dependence of the {\tt OMG3Q1.1} FFs describing constituent-heavy-quark fragmentation into the all-heavy baryons $\Omega_{3c}$ (left panels) and $\Omega_{3b}$ (right panels). 
The central prediction is defined as the average over the full replica ensemble, while the shaded bands in the upper panels represent the total uncertainty envelope obtained from the replica set.
To illustrate the effects of DGLAP evolution, the FFs are shown at three representative factorization scales, $\mu_F=40$, $80$, and $160$~GeV. 
In both baryonic systems, the distributions exhibit a smooth scale dependence. 
As $\mu_F$ increases, the FFs become progressively broader, their maxima decrease, and part of the momentum is redistributed toward smaller-$z$ values. This behavior reflects the expected action of perturbative parton radiation and confirms the stability of the {\tt OMG3Q1.1} evolution framework across the explored scale range.

The FFs display a bell-shaped profile, with support concentrated in the intermediate-$z$ region and maxima located around $z\sim0.5$. 
This indicates that the produced baryon typically retains a substantial fraction of the momentum carried by the fragmenting heavy quark, leading to a relatively hard fragmentation pattern. 
While the overall shapes of the $\Omega_{3c}$ and $\Omega_{3b}$ FFs are similar, the bottom-initiated distributions are more suppressed in normalization and exhibit a slightly harder profile, with their maxima shifted toward larger values of $z$. 
Such a trend is consistent with the larger bottom-quark mass and the reduced momentum degradation expected during hadronization.

A useful comparison can be made with the recently released {\tt TQ4Q2.0} FFs for all-heavy tetraquarks~\cite{Celiberto:2026kks}. 
In that case, the distributions exhibit a pronounced enhancement toward small $z$, reflecting the structure of the underlying NRQCD short-distance coefficients entering the initial-scale inputs. 
By contrast, the {\tt OMG3Q1.1} FFs remain localized in the intermediate-$z$ region and display a smooth bell-shaped profile. 
This difference highlights the distinct physical origin of the two constructions. 
Whereas the tetraquark FFs retain a direct imprint of perturbative splitting dynamics, the present baryonic FFs are driven by a Suzuki-inspired framework in which bound-state formation plays a more prominent role in shaping the momentum-fraction distribution.

The uncertainty decomposition is shown in the two lower ancillary panels. 
The first one isolates the fragmentation missing higher-order uncertainties (F-MHOUs), obtained through correlated variations of the evolution-ready scale and of the scales entering the initial FF construction. 
The second panel displays the fragmentation nonperturbative wave-function (F-NPWF) uncertainty, generated by variations of the parameter controlling the bound-state wave-function profile. 
In both cases, uncertainties are reported as ratios to the corresponding central prediction.

A clear hierarchy emerges between the two uncertainty sources. 
F-MHOUs provide the dominant contribution over most of the phenomenologically relevant $z$ range, particularly in the intermediate- and large-$z$ regions where the FFs attain their maximum values. 
By contrast, F-NPWF effects remain very small throughout the bulk region and become appreciable only in a narrow region where the FFs attain very small values, causing ratio-based representations to amplify otherwise modest absolute differences. 
This pattern indicates that the perturbative construction and evolution of the FFs constitute the leading source of theoretical uncertainty, while the dependence on the specific wave-function modeling remains under good control.

Overall, the heavy-quark sector of the {\tt OMG3Q1.1} FFs provides a stable and physically transparent description of heavy-quark fragmentation into all-heavy baryons. 
Their regular shape, controlled scale dependence, and well-understood uncertainty structure support their use in precision phenomenological studies of rare-baryon production at present and future colliders. 
They also provide the baseline input for the uncertainty-aware collider phenomenology discussed in the following sections.

\begin{figure*}[!t]
\centering

   \hspace{-0.00cm}
   \includegraphics[scale=0.410,clip]{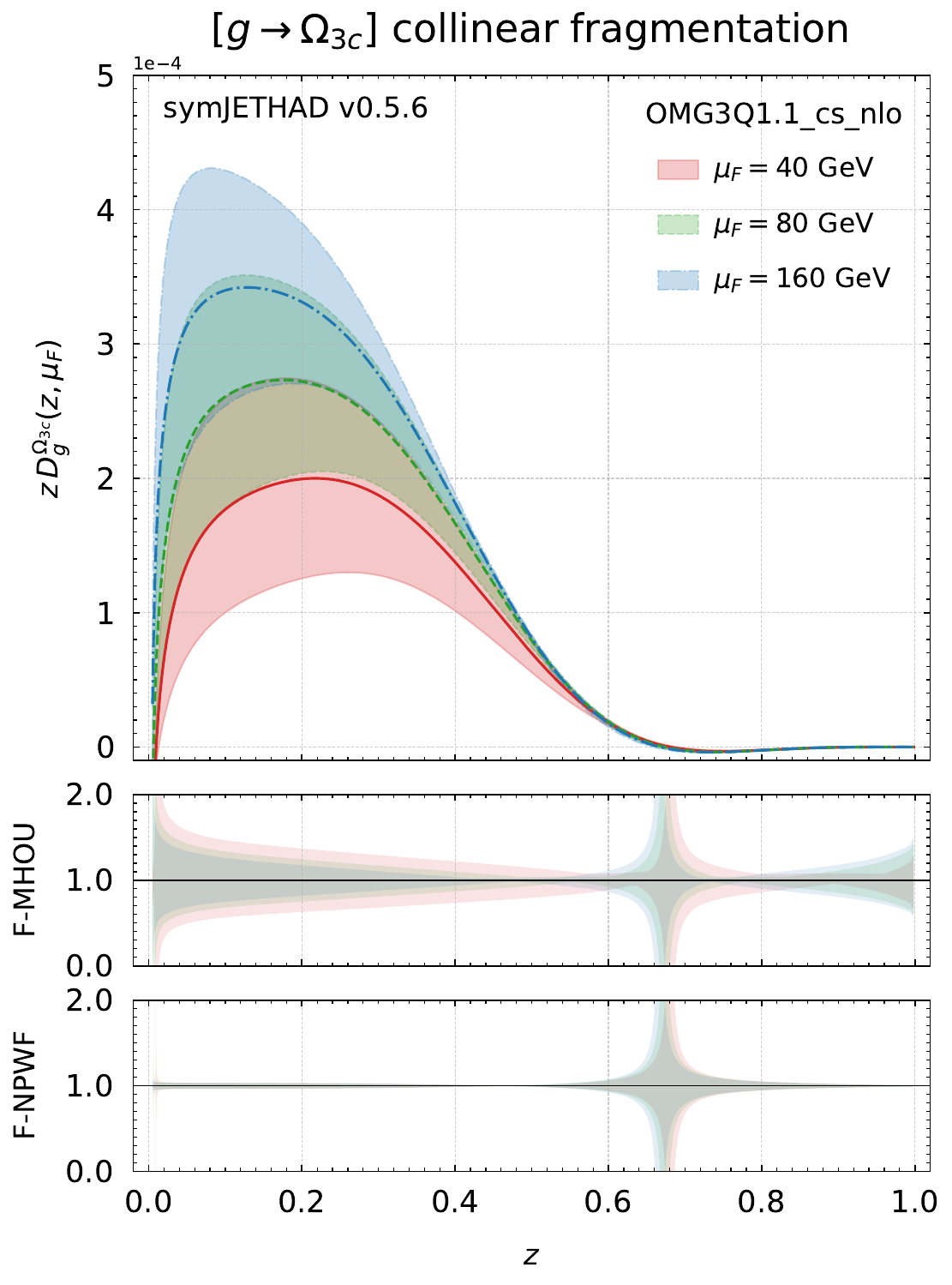}
   \hspace{0.90cm}
   \includegraphics[scale=0.410,clip]{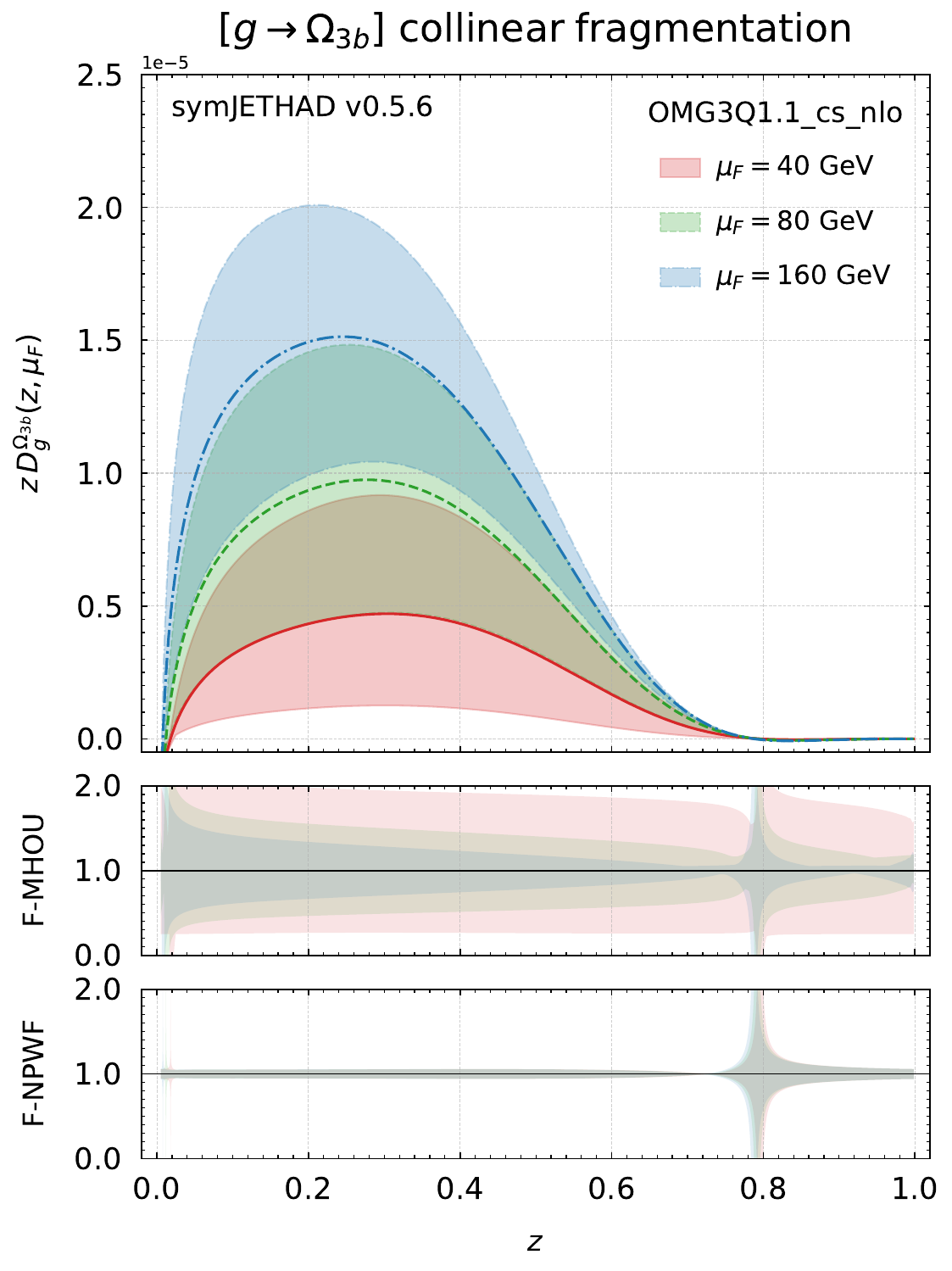}

\caption{
Momentum-fraction distributions of the {\tt OMG3Q1.1} FFs describing gluon fragmentation into $\Omega_{3c}$ (left) and $\Omega_{3b}$ (right) states. 
Results are reported for three representative factorization scales, $\mu_F=40$, $80$, and $160$~GeV. 
Shaded bands in the upper panels represent the total uncertainty obtained from the combination of F-MHOU and F-NPWF effects. 
The lower panels isolate the corresponding perturbative and nonperturbative uncertainty components. 
The replica structure adopted in the {\tt OMG3Q1.1} set is summarized in Table~\ref{tab:OMG3Q11_replicas} of Appendix~\protect\hyperlink{app:A}{A}.}
\label{fig:FFs-z_O3Q_g}
\end{figure*}

Figure~\ref{fig:FFs-z_O3Q_g} displays the $z$ dependence of the {\tt OMG3Q1.1} FFs describing gluon fragmentation into the all-heavy baryons $\Omega_{3c}$ (left panels) and $\Omega_{3b}$ (right panels). 
As in Fig.~\ref{fig:FFs-z_O3Q_Q}, the central prediction is obtained as the average over the full replica ensemble, while the shaded bands represent the combined uncertainty generated by F-MHOU and F-NPWF variations.

In both baryonic systems, the FFs exhibit smooth bell-shaped profiles with support concentrated in the small- and intermediate-$z$ region. 
The distributions reach their maximum around $z \sim 0.15$--$0.25$ for $\Omega_{3c}$ and at slightly larger values, $z \sim 0.20$--$0.35$ for $\Omega_{3b}$. 
Compared with the constituent-heavy-quark FFs, the gluon distributions are noticeably softer, indicating that the produced baryon typically carries a smaller fraction of the momentum of the fragmenting parton.

A distinctive feature of the gluon channel is its visible dependence on the factorization scale. 
As $\mu_F$ increases, the distributions gradually broaden and exhibit a moderate redistribution of strength toward smaller values of $z$. 
This behavior reflects the combined effect of the initial-scale gluon input and its subsequent DGLAP evolution. 
Compared with the constituent-heavy-quark channel, the gluon FF experiences a more noticeable modification of its shape, leading to a stronger scale dependence over the range explored here.

As in the constituent-heavy-quark case, a useful comparison can be made with the recently released {\tt TQ4Q2.0} FFs for all-heavy tetraquarks~\cite{Celiberto:2026kks}. 
However, the contrast is even more pronounced in the gluon sector. 
In the scalar $T_{4Q}(0^{++})$ channel, the gluon FF exhibits a marked enhancement toward small $z$, inherited from the NRQCD short-distance coefficients entering the initial-scale input. 
The {\tt OMG3Q1.1} gluon FFs instead remain concentrated in the intermediate-$z$ region and display a smooth bell-shaped profile. 
This difference reflects the distinct physical origin of the two constructions. 
While the tetraquark FFs retain a direct imprint of NRQCD fragmentation dynamics, the baryonic FFs originate from the Suzuki-inspired framework discussed above, where bound-state formation plays a more prominent role in shaping the momentum-fraction distribution.

A further noteworthy feature is the strong suppression of the gluon channel relative to constituent-heavy-quark fragmentation. 
As already discussed in the original {\tt OMG3Q1.0} analysis~\cite{Celiberto:2025ogy}, this hierarchy is not a peculiarity of the present implementation, but rather reflects a structural property of all-heavy baryon production. 
While fully heavy tetraquarks can be naturally generated through gluon-initiated cascades producing heavy-quark pairs, the formation of an $\Omega_{3Q}$ state requires the assembly of three valence heavy quarks into a color-singlet baryonic configuration. 
Such a final state is considerably less compatible with direct gluon fragmentation mechanisms, leading to a substantial suppression of the gluon FF. 
The persistence of this pattern in the uncertainty-aware {\tt OMG3Q1.1} framework further supports its interpretation as a genuine feature of the underlying production dynamics rather than a model-dependent artifact.

As for the heavy-quark FFs case (see Fig.~\ref{fig:FFs-z_O3Q_Q}), the uncertainty decomposition of gluon FFs is shown in the lower ancillary panels of Fig.~\ref{fig:FFs-z_O3Q_g}. 
The first one isolates the fragmentation missing higher-order uncertainties (F-MHOUs), obtained through correlated variations of the evolution-ready scale and of the scales entering the initial FF construction, while the second displays the fragmentation nonperturbative wave-function (F-NPWF) uncertainty associated with variations of the bound-state wave-function parameter. 
A clear hierarchy emerges between the two effects. 
F-MHOUs dominate the uncertainty budget across most of the phenomenologically relevant $z$ range, reflecting the strong sensitivity of the evolution-generated gluon channel to perturbative scale variations. 
By contrast, F-NPWF effects remain very small and become visible only in the vicinity of the node where the FF approaches zero, where ratio representations naturally amplify tiny absolute differences.

Overall, the gluon sector further confirms the internal consistency of the {\tt OMG3Q1.1} framework. 
Although strongly subleading with respect to constituent-heavy-quark fragmentation, consistently with the structural suppression discussed above, the gluon FFs display regular shapes, controlled scale dependence, and a well-understood uncertainty pattern. 
These features support the robustness of the underlying fragmentation picture and reinforce the use of the {\tt OMG3Q1.1} sets in precision phenomenological studies of all-heavy baryon production.

\begin{figure*}[!t]
\centering

   \hspace{-0.00cm}
   \includegraphics[scale=0.410,clip]{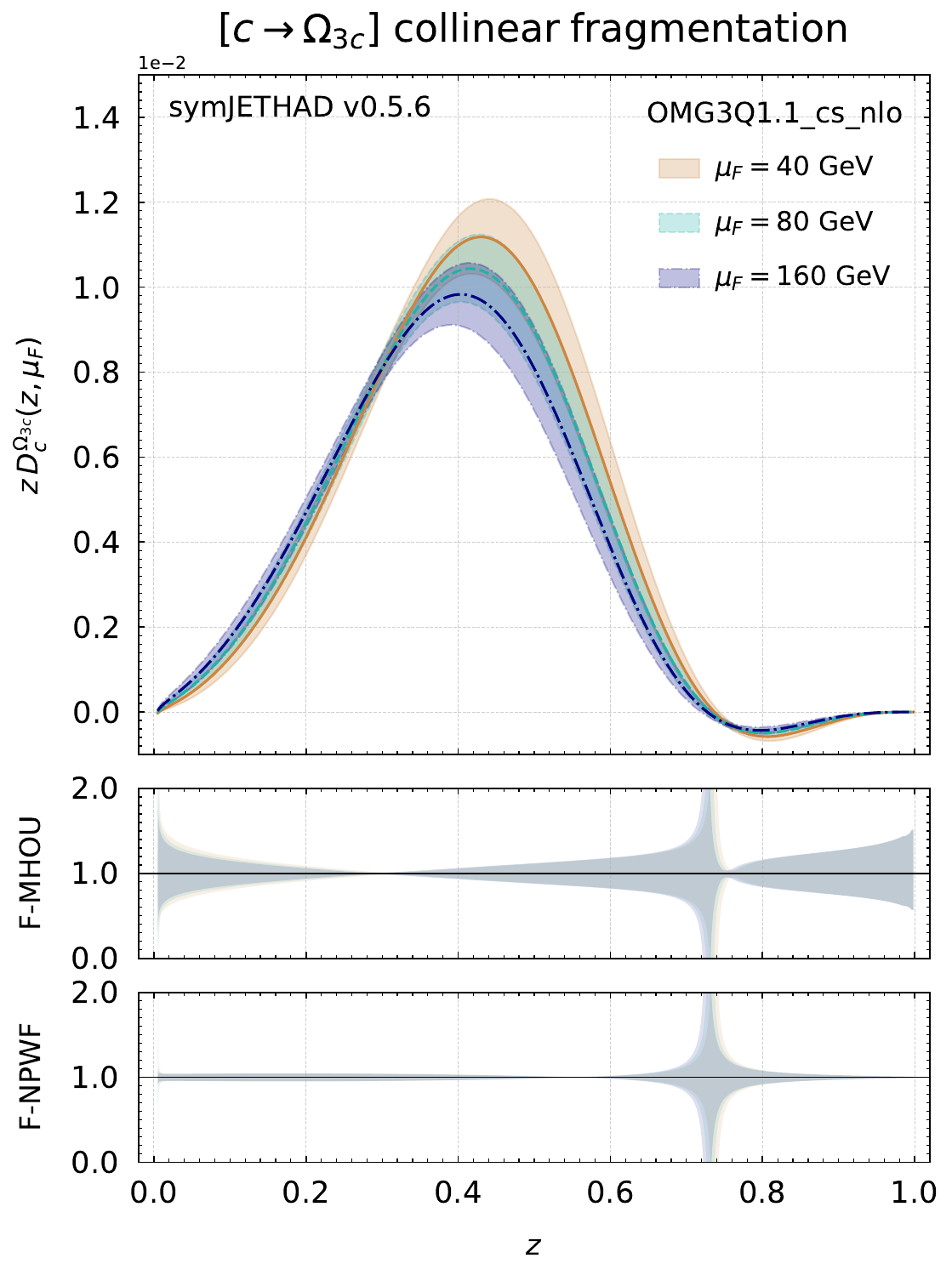}
   \hspace{0.90cm}
   \includegraphics[scale=0.410,clip]{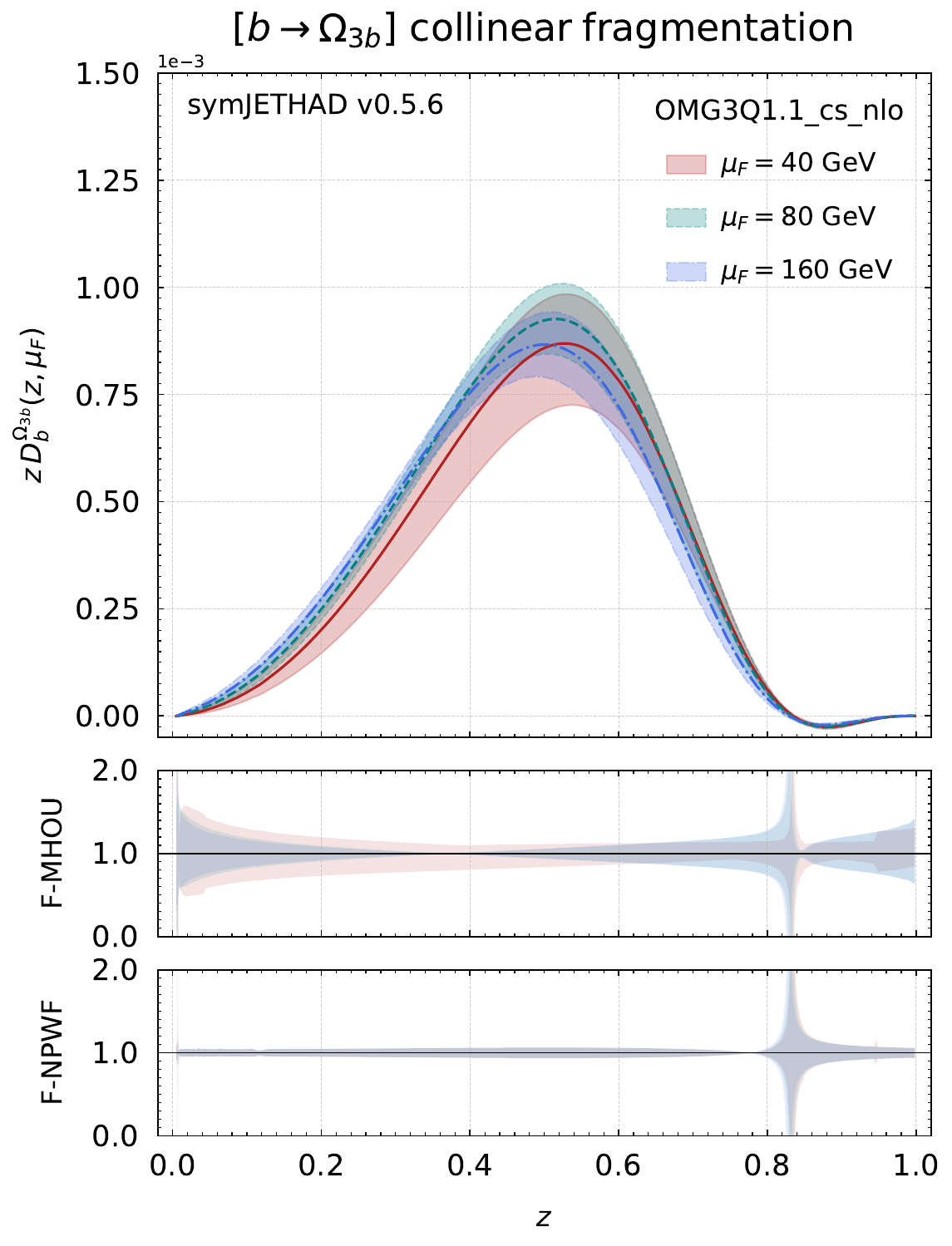}

\caption{
Momentum-fraction distributions of the {\tt OMG3Q1.1} FFs describing constituent-heavy-quark fragmentation into $\Omega_{3c}$ (left) and $\Omega_{3b}$ (right) states. 
Results are reported for three representative factorization scales, $\mu_F=40$, $80$, and $160$~GeV. 
Shaded bands in the upper panels represent the total uncertainty obtained from the combination of F-MHOU and F-NPWF effects. 
The lower panels isolate the corresponding perturbative and nonperturbative uncertainty components. 
The replica structure adopted in the {\tt OMG3Q1.1} set is summarized in Table~\ref{tab:OMG3Q11_replicas} of Appendix~\protect\hyperlink{app:A}{A}.}
\label{fig:FFs-z_O3Q_Q}
\end{figure*}

\begin{figure*}[!t]
\centering

   \hspace{-0.00cm}
   \includegraphics[scale=0.400,clip]{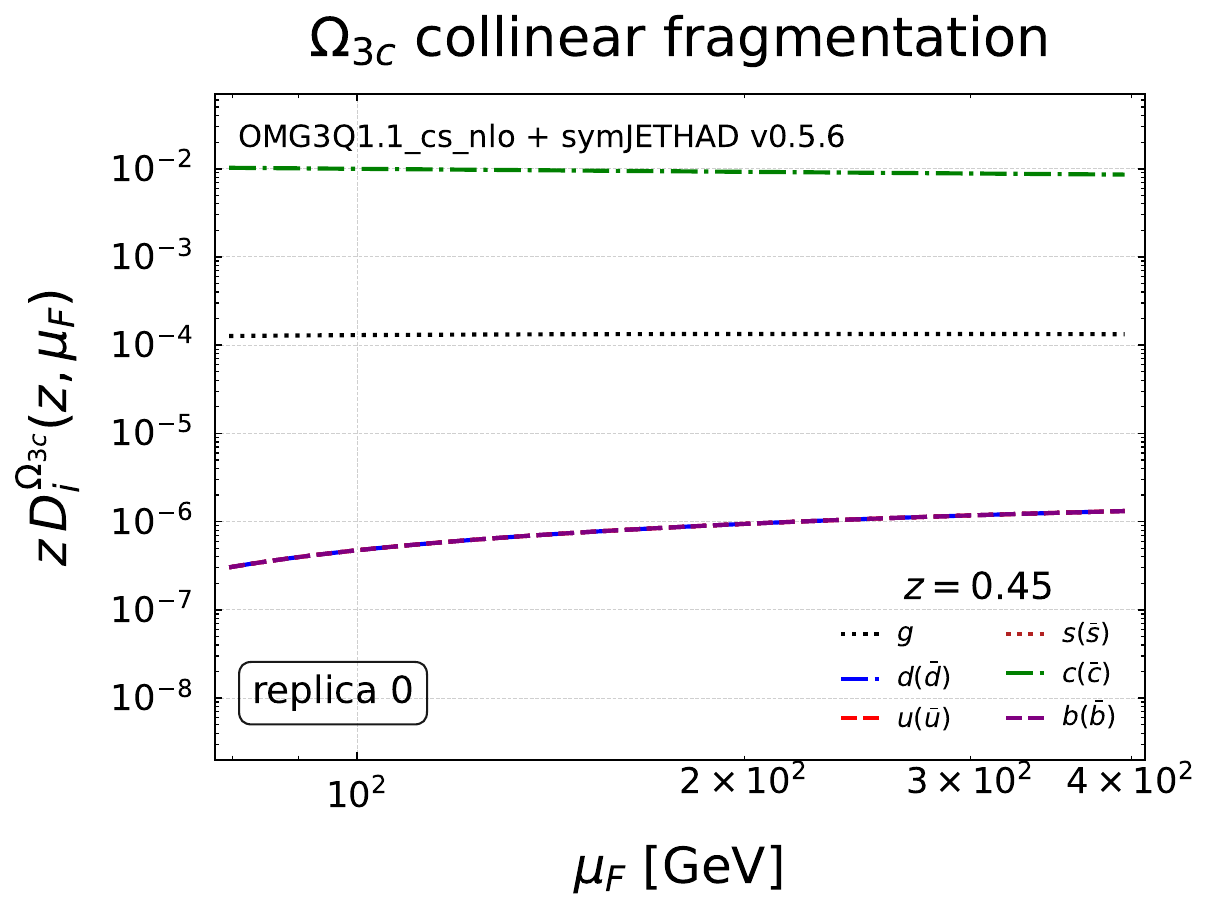}
   \hspace{-0.15cm}
   \includegraphics[scale=0.400,clip]{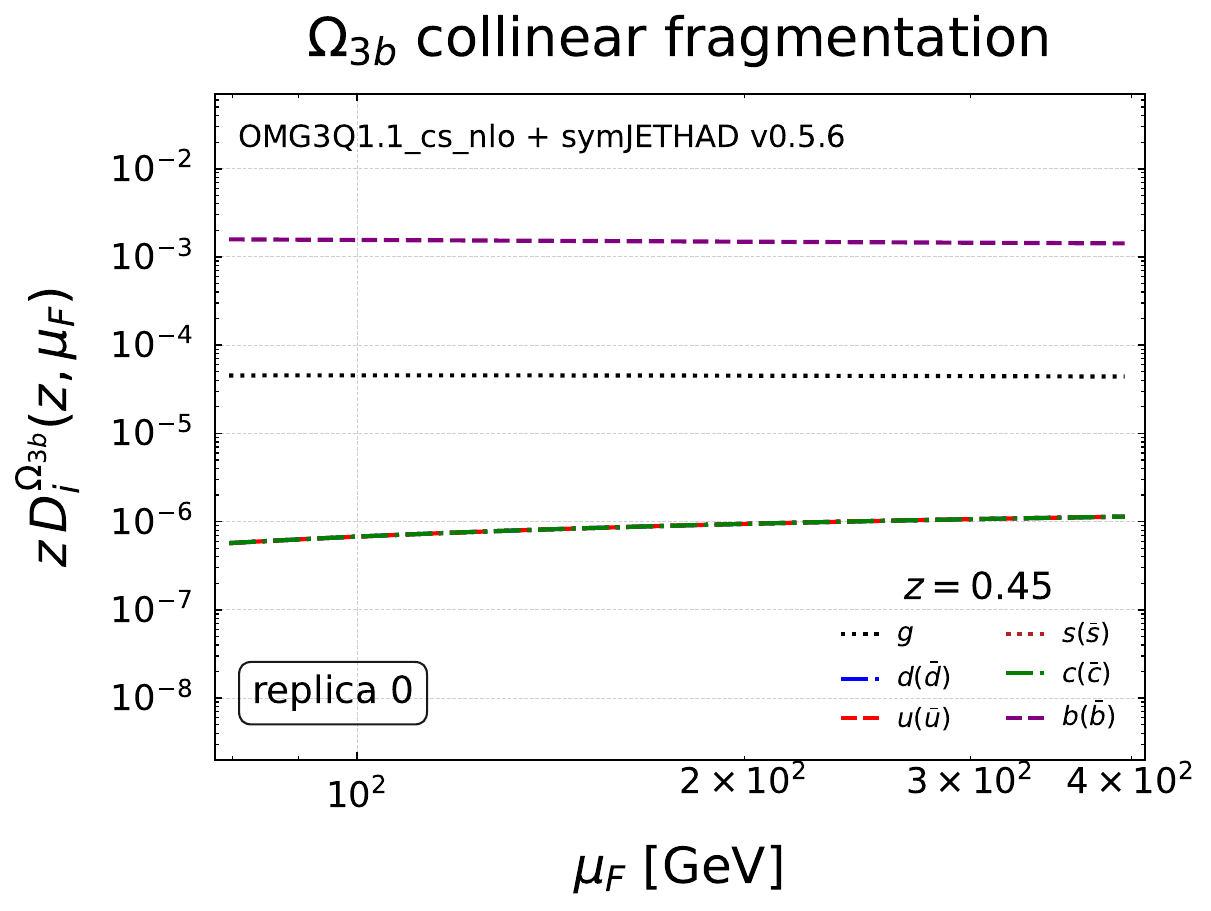}

\caption{Scale dependence of the {\tt OMG3Q1.1} FFs at fixed $z=0.45$ for $\Omega_{3c}$ (left) and $\Omega_{3b}$ (right) production. 
The distributions of all active parton species are shown as functions of the factorization scale $\mu_F$. 
Predictions correspond to the central replica (replica 0), with F-MHOU and F-NPWF variations switched off.
For details on the replica indexing and parameter mapping, see Table~\ref{tab:OMG3Q11_replicas} in Appendix~\protect\hyperlink{app:A}{A}.}
\label{fig:FFs-muF}
\end{figure*}

For completeness, we examine in Fig.~\ref{fig:FFs-muF} the evolution of the {\tt OMG3Q1.1} FFs as functions of the factorization scale for $\Omega_{3c}$ (left panel) and $\Omega_{3b}$ (right panel). 
The momentum fraction is fixed at $z=0.45$, a representative value within the bulk region of the FFs and close to the maxima of the constituent-heavy-quark distributions discussed above.

To isolate the genuine evolution effects, only the central {\tt OMG3Q1.1} replica is shown, with F-MHOU and F-NPWF variations switched off. 
Unlike the previous figures, where uncertainty bands summarize the behavior of the full replica ensemble, the present representation focuses on the scale evolution of each active partonic species and therefore provides a more direct view of the internal structure of the {\HFNRevo} framework.

The distributions reveal a well-defined hierarchy among partonic channels. 
For $\Omega_{3c}$ production, the constituent-charm contribution remains by far the dominant FF over the entire scale range explored. 
Likewise, in the $\Omega_{3b}$ case, the constituent-bottom channel clearly governs the fragmentation process. 
The gluon contribution occupies an intermediate position, while all remaining quark species stay strongly suppressed and are generated only through DGLAP evolution above the corresponding thresholds.

A noteworthy feature is the remarkable smoothness of the evolution pattern. 
The constituent-heavy-quark FFs exhibit only a mild dependence on $\mu_F$, indicating that their normalization and overall shape are largely determined by the initial-scale construction and remain stable under evolution. 
The gluon channel displays a more visible scale dependence, reflecting the combined action of its initial-scale input and subsequent partonic mixing. 
Nevertheless, its evolution remains regular and monotonic throughout the explored region, with no indication of pathological behavior or excessive sensitivity to threshold crossings.

The persistence of the constituent-heavy-quark dominance across a broad range of scales provides additional support for the physical picture underlying the {\tt OMG3Q1.1} construction. 
At the same time, the controlled evolution of the gluon sector confirms the consistency of the threshold-aware {\HFNRevo} strategy. 
Together, these features establish a robust foundation for the phenomenological predictions discussed in Sec.~\ref{sec:phenomenology}, where the interplay between constituent-heavy-quark and gluon fragmentation channels will be seen to play a central role in shaping collider observables.

\section{Semi-inclusive $\Omega_{3Q}$-jet production in high-energy QCD}
\label{sec:hybrid-factorization}

The first part of this section (Sec.~\ref{ssec:resummation}) reviews selected developments in semihard QCD phenomenology and high-energy resummation.
The second part (Sec.~\ref{ssec:NLL-cross-section}) introduces the $\NLLp$ hybrid factorization framework employed in this work and its application to semi-inclusive $\Omega_{3Q}$ plus jet production.

\subsection{semihard dynamics and theoretical background}
\label{ssec:resummation}

The production of heavy-flavored hadrons provides a powerful laboratory for exploring the semihard regime of QCD, where large energy logarithms can spoil the convergence of fixed-order perturbative calculations. 
In this domain, the Balitsky-Fadin-Kuraev-Lipatov (BFKL) formalism~\cite{Fadin:1975cb,Kuraev:1977fs,Balitsky:1978ic} systematically resums logarithmic contributions in the center-of-mass energy, including both leading-logarithmic (LL) terms $\alpha_s^n \ln^n(s)$ and NLL corrections $\alpha_s^{n+1}\ln^n(s)$.

More specifically, the semihard, or Regge-Gribov regime is characterized by perturbative transverse scales that remain well above the QCD scale while being much smaller than the collision energy~\cite{Gribov:1983ivg}. 
In this hierarchy, powers of energy logarithms compensate the smallness of the strong coupling and eventually spoil the convergence of a purely fixed-order expansion. 
The role of BFKL resummation is therefore to reorganize these enhanced contributions to all orders while retaining the hard transverse scales of the tagged final-state objects.

Within high-energy factorization, the reaction is organized into two process-dependent impact factors connected by the universal BFKL Green's function~\cite{Fadin:1998py,Ciafaloni:1998gs}. 
The impact factors describe the production of the tagged objects in the fragmentation regions of the two incoming protons and contain the corresponding collinear partonic information, whereas the Green's function controls the energy evolution across the rapidity interval separating them. 
The latter is independent of the nature of the tagged final states and resums the unresolved real and virtual radiation emitted in the central rapidity region. 
Since proton PDFs and, when an identified hadron is observed, collinear FFs are embedded in the external impact factors, this organization naturally realizes a hybrid description in which collinear and high-energy QCD dynamics act on distinct components of the factorized cross section.

BFKL resummation has been successfully applied to a broad class of semihard reactions, including Mueller-Navelet jets~\cite{Mueller:1986ey,Ducloue:2013hia,Colferai:2015zfa,Celiberto:2015yba,Celiberto:2015mpa,Celiberto:2016ygs,Celiberto:2017ius,Caporale:2018qnm,deLeon:2021ecb,Celiberto:2022gji,Baldenegro:2024ndr}, dihadron~\cite{Celiberto:2017ius,Celiberto:2016hae,Celiberto:2017ptm,Celiberto:2020rxb,Celiberto:2022rfj} and hadron-jet systems~\cite{Bolognino:2018oth,Bolognino:2019cac,Bolognino:2019yqj,Celiberto:2020wpk,Celiberto:2020rxb,Mohammed:2022gbk,Celiberto:2022kxx}, multijet production~\cite{GordoGomez:2018yjf,Celiberto:2017ius,Caporale:2016soq,Caporale:2016xku,Celiberto:2016vhn,Caporale:2016zkc}, forward-Higgs emissions~\cite{Hentschinski:2020tbi,Mohammed:2022gbk,Celiberto:2022fgx,Celiberto:2020tmb,Mohammed:2022gbk,Celiberto:2023rtu,Celiberto:2023uuk,Celiberto:2023eba,Celiberto:2023nym,Celiberto:2023rqp,Celiberto:2022zdg,Celiberto:2024bbv}, Drell-Yan processes~\cite{Celiberto:2018muu,Golec-Biernat:2018kem}, and heavy-flavored final states~\cite{Celiberto:2017nyx,Boussarie:2017oae,Bolognino:2019ouc,Bolognino:2019yls,Bolognino:2021mrc,Celiberto:2021dzy,Celiberto:2021fdp,Celiberto:2022dyf,Celiberto:2023fzz,Celiberto:2022grc,Bolognino:2022paj,Celiberto:2022keu,Celiberto:2022kza,Celiberto:2024omj}. 
Complementary studies of single-forward emissions have provided valuable constraints on small-$x$ gluon dynamics through unintegrated gluon distributions (UGDs) at HERA~\cite{Anikin:2011sa,Besse:2013muy,Bolognino:2018rhb,Bolognino:2018mlw,Bolognino:2019bko,Bolognino:2019pba,Celiberto:2019slj,Bolognino:2021bjd,Luszczak:2022fkf,Boroun:2023goy,Boroun:2023ldq} and at the forthcoming Electron-Ion Collider (EIC)~\cite{Bolognino:2021niq,Bolognino:2021gjm,Bolognino:2021bjd,Bolognino:2022uty,Bolognino:2022ndh}. 
These investigations have, in turn, stimulated the development of small-$x$ resummed PDFs~\cite{Ball:2017otu,Abdolmaleki:2018jln,Bonvini:2019wxf,Silvetti:2022hyc,Silvetti:2023suu,Rinaudo:2024hdb,Celiberto:2025nnq} and improved TMD parton densities~\cite{Bacchetta:2020vty,Bacchetta:2024fci,Celiberto:2021zww,Bacchetta:2021oht,Bacchetta:2021lvw,Bacchetta:2021twk,Bacchetta:2022esb,Bacchetta:2022crh,Bacchetta:2022nyv,Celiberto:2022omz,Bacchetta:2023zir}.

A particularly important development emerged from heavy-flavor phenomenology. Studies of ${\rm \Lambda}_c$~\cite{Celiberto:2021dzy} and $b$-hadron production~\cite{Celiberto:2021fdp} revealed a natural stabilization mechanism in semihard observables, driven by collinear VFNS fragmentation~\cite{Celiberto:2022grc}. 
Unlike light-particle emissions, which are often affected by large NLL corrections and threshold-enhanced effects~\cite{Bolognino:2018oth,Celiberto:2020wpk,Wang:2026nap}, heavy-flavored final states display improved perturbative stability.

This observation motivated the construction of VFNS fragmentation frameworks based on NRQCD inputs~\cite{Braaten:1993mp,Zheng:2019dfk,Braaten:1993rw,Zheng:2019gnb}, initially developed for vector quarkonia~\cite{Celiberto:2022dyf,Celiberto:2023fzz} and later extended to $\BCs$ and $\Bss$ mesons~\cite{Celiberto:2022keu,Celiberto:2024omj}. 
The same mechanism subsequently opened a phenomenological pathway toward exotic hadrons, enabling studies of doubly heavy tetraquarks~\cite{Celiberto:2023rzw,Celiberto:2024beg}, all-heavy tetraquarks~\cite{Gatto:2025kfl,Celiberto:2024mab,Celiberto:2024beg,Celiberto:2025dfe,Celiberto:2025ziy,Celiberto:2025vra,Celiberto:2026kks}, and all-charm pentaquarks~\cite{Celiberto:2025ipt,Celiberto:2026rdk}.

\begin{figure*}[!t]
\centering
\includegraphics[width=0.75\textwidth]{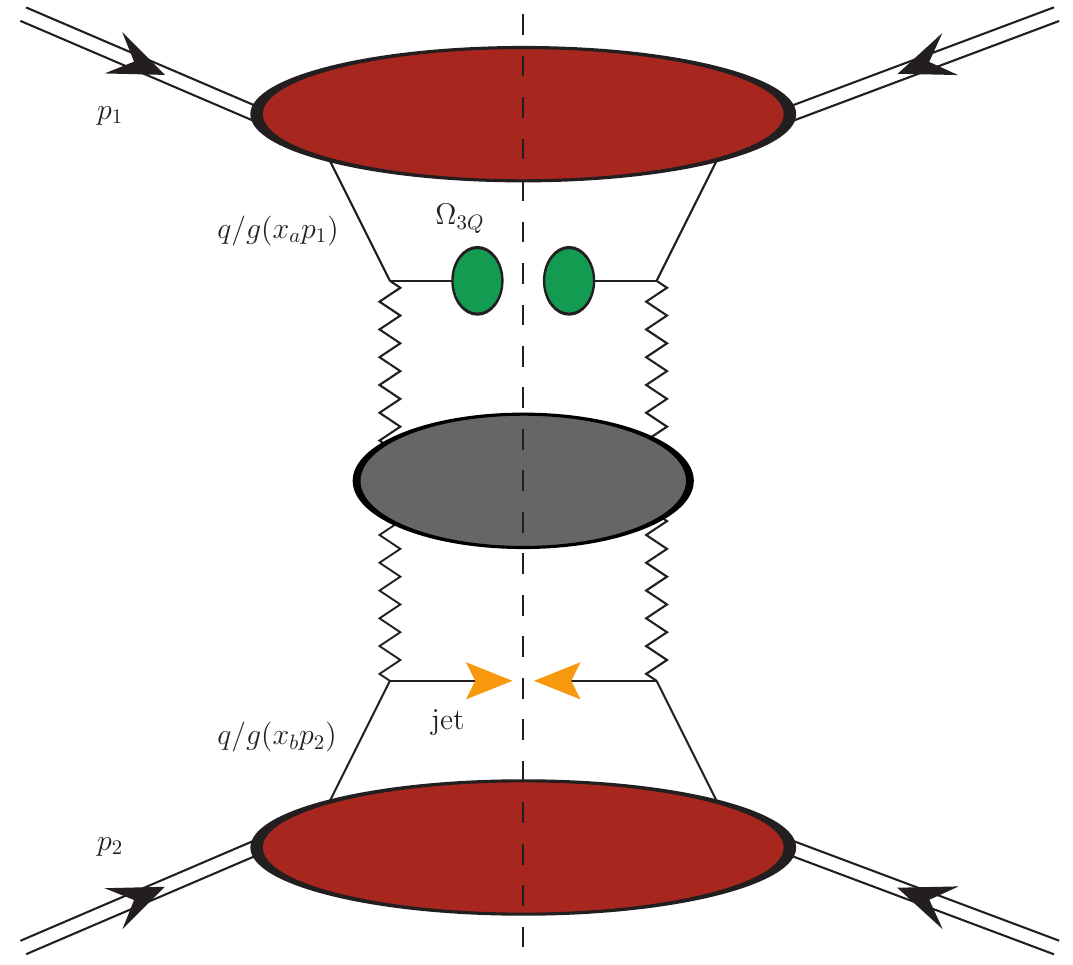}

\caption{Schematic illustration of the hybrid collinear/high-energy factorization employed for semi-inclusive $\Omega_{3Q}$-plus-jet production at hadron colliders.
Firebrick-colored blobs denote proton PDFs, whereas green ovals represent collinear FFs for the observed all-heavy $\Omega_{3Q}$ baryon.
The orange arrow identifies the associated jet.
The emission functions are connected through the universal BFKL Green's function (gray blob) via Reggeized-gluon exchanges.
The diagram was generated with {\tt JaxoDraw~2.0}~\cite{Binosi:2008ig}.}
\label{fig:reaction}
\end{figure*}

\subsection{NLL-resummed cross section}
\label{ssec:NLL-cross-section}

We consider the hadroproduction reaction depicted in Fig.~\ref{fig:reaction}
\begin{equation}
\label{process}
    {\rm p}(p_1) + {\rm p}(p_2) \, \rightarrow \, \Omega_{3Q}(q_1, y_1, \phi_1) + {\cal X} + {\rm jet}(q_2, y_2, \phi_2) \;,
\end{equation}
where the incoming protons, carrying momenta $p_{1,2}$, define the initial state of the collision. 
The symbol $\Omega_{3Q}$ denotes the observed all-heavy baryon, produced with momentum $q_1$, rapidity $y_1$, and azimuthal angle $\phi_1$. 
A jet is simultaneously tagged in the final state, characterized by momentum $q_2$, rapidity $y_2$, and azimuthal angle $\phi_2$. 
The system ${\cal X}$ collectively represents all additional particles that remain unobserved. 
Our analysis is performed in the semihard kinematic domain, where both final-state objects carry large transverse momenta, $|\vec q_{T_{1,2}}|$, and are separated by a wide rapidity interval, $\DY = y_1 - y_2$. 
This configuration guarantees the applicability of a VFNS description based on leading-power fragmentation, which is expected to dominate the production mechanism of the $\Omega_{3Q}$ states. 
The proton momenta are chosen as Sudakov light-cone vectors satisfying $p_1^2 = p_2^2 = 0$ and $(p_1 \cdot p_2)=s/2$, from which one obtains
\begin{equation}\label{sudakov}
q_{1,2} = x_{1,2} p_{1,2} - \frac{ q_{1,2\perp}^2}{x_{1,2} s}p_{2,1} + q_{1,2\perp} \ , \quad
\vec q_{T_{1,2}}^{\;2} \equiv -q_{1,2\perp}^2
\;.
\end{equation}
The longitudinal momentum fractions, $x_{1,2}$, are related to the rapidities of the observed final-state particles through
\begin{equation}\label{y-vs-x}
y_{1,2}=\pm\frac{1}{2}\ln\frac{x_{1,2}^2 s}
{\vec q_{T_{1,2}}^{\;2}}
\qquad \mbox{and} \qquad
\drv y_{1,2} = \pm \frac{\drv x_{1,2}}{x_{1,2}}
\;.
\end{equation}
One writes
\begin{equation}
\label{DeltaY}
 \DY = y_1 - y_2 = \ln \left( \frac{x_1 x_2}{|\vec q_{T_1}||\vec q_{T_2}|} s \right) \;.
\end{equation}

At LO in collinear QCD factorization, the differential cross section factorizes into the convolution of an on-shell partonic subprocess, the proton PDFs, and the FFs associated with the production of the observed rare baryon,
\begin{equation}
\label{sigma_collinear}
\begin{split}
\hspace{-0.25cm}
\frac{\drv\sigma^{\rm LO}_{\rm [coll.]}}{\drv x_1\drv x_2\drv ^2\vec q_{T_1}\drv ^2\vec q_{T_2}}
= \hspace{-0.25cm} 
\sum_{\alpha,\beta,\gamma}=q,{\bar q},g
\int_0^1 \hspace{-0.20cm} \drv x_a 
\int_0^1 \hspace{-0.20cm} \drv x_b\,
f_\alpha\left(x_a\right) f_\beta\left(x_b\right)
\int_{x_1}^1 \hspace{-0.15cm} \frac{\drv \xi}{\xi}
D^{\Omega_{3Q}}_{\gamma}\left(\frac{x_1}{\xi}\right) 
\frac{\drv {\hat\sigma}_{\alpha\beta\to\gamma\mathcal{X}}\left(\hat s\right)}
{\drv x_1\drv x_2\drv ^2\vec q_{T_1}\drv ^2\vec q_{T_2}}\;,
\end{split}
\end{equation}
Here, $\alpha$ and $\beta$ label the incoming partons, whereas $\gamma$ denotes the outgoing parton that fragments into the observed baryon.
The sums over $\alpha$ and $\beta$ and $\gamma$ run over all active partonic species, excluding the top quark, whose contribution to fragmentation is negligible. 
For brevity, the explicit dependence on the factorization scale $\mu_F$ has been suppressed in Eq.~\eqref{sigma_collinear}. 
The functions $f_{\alpha,\beta}(x_{a,b},\mu_F)$ denote the proton PDFs, whereas $D_{\gamma}^{\Omega_{3Q}}(x_1/\xi,\mu_F)$ describes the fragmentation of the outgoing parton $\gamma$ into the observed baryon. 
The variables $x_{a,b}$ represent the longitudinal momentum fractions carried by the incoming partons, while $\xi$ parametrizes the longitudinal momentum of the fragmenting parton relative to that of the observed baryon. 
The quantities $\drv\hat{\sigma}_{\alpha\beta\to\gamma\mathcal{X}}(\hat s)$ are the corresponding partonic hard-scattering coefficients, with $\hat s \equiv x_a x_b s$ denoting the squared center-of-mass energy of the underlying subprocess.

Equation~\eqref{sigma_collinear} provides a useful collinear reference, but the organization of the perturbative series changes in the semihard limit relevant to the present reaction. 
When the two tagged objects carry hard transverse momenta and are separated by a large rapidity interval, enhanced energy logarithms originate from additional partonic radiation emitted between the two fragmentation regions. 
In the hybrid-factorization description, these contributions are not assigned to a single fixed-order hard subprocess. 
Instead, the production of the $\Omega_{3Q}$ baryon and of the tagged jet is encoded in two separate emission functions, which are connected by the BFKL Green's function that resums the radiation spanning the rapidity gap.

The emission functions provide the interface between high-energy resummation and the collinear description of the external fragmentation regions. 
They embed the relevant collinear inputs, namely proton PDFs and, whenever an identified hadron is tagged, the corresponding FFs, together with the perturbative coefficients governing the production of the observed final state. 

In the present case, the $\Omega_{3Q}$ emission function therefore combines proton PDFs with the {\tt OMG3Q1.1} FFs, whereas the jet emission function involves the appropriate quark, antiquark, and gluon PDF contributions together with the perturbative jet-production vertex. 
The collinear inputs evolve according to DGLAP dynamics, while the BFKL Green's function independently governs the high-energy evolution across the rapidity interval separating the two tagged objects. 
The two evolution mechanisms thus act on distinct but complementary components of the hybrid-factorization framework.

At the level of the external emission vertices, all active gluon, quark, and antiquark species are included. 
The resulting flavor combinations comprise gluon-gluon, gluon-quark, quark-gluon, and quark-quark channels, together with the corresponding antiquark combinations and flavor crossings. 
At NLO, the baryon emission function further contains diagonal and mixed gluon-quark coefficient functions, so that the species extracted from the proton and the parton fragmenting into the observed $\Omega_{3Q}$ need not coincide. 
These flavor combinations characterize the partonic content of the impact factors and should not be interpreted as an independent decomposition of the resummed result into collinear [$2\to2$] subprocesses.

It is convenient to project the resulting cross section onto azimuthal harmonics by expanding it as a Fourier series in the relative azimuthal angle,
$\phi = \phi_1 - \phi_2 - \pi$,
so that
\begin{equation}
 \label{dsigma_Fourier}
 \frac{\drv \sigma}{\drv y_1 \drv y_2 \drv \vec q_{T_1} \drv \vec q_{T_2} \drv \phi_1 \drv \phi_2} =
 \frac{1}{(2\pi)^2} \left[{\cal C}_0 + 2 \sum_{l=1}^\infty \cos (l \phi)\,
 {\cal C}_l \right]\, .
\end{equation}
The first building block of the NLL resummation framework is the BFKL Green's function,
\begin{equation}
\label{G_BFKL_NLL}
 {\cal G}_{\rm NLL}(\DY,l,\nu,\mu_R) = e^{{\DY} \bar \alpha_s(\mu_R) \,
 \chi^{\rm NLO}(l,\nu)} \; ,
\end{equation}
where $\bar \alpha_s(\mu_R) \equiv \alpha_s(\mu_R) N_c/\pi$, and $\beta_0 = 11N_c/3 - 2 n_f/3$ denotes the leading coefficient of the QCD $\beta$ function. 
The quantity $\chi^{\rm NLO}(l,\nu)$ appearing in the exponent is the Mellin-space eigenvalue of the NLL BFKL kernel and embodies the all-order resummation of high-energy logarithms. 
Its explicit expression reads
\begin{eqnarray}
 \label{chi}
 \chi^{\rm NLO}(l,\nu) = \chi(l,\nu) + \bar\alpha_s \tilde{\chi}(l,\nu) \;,
\end{eqnarray}
where the LO kernel eigenvalues $\chi(l,\nu)$ read
\begin{eqnarray}
 \label{kernel_LO}
 \chi\left(l,\nu\right) = -2\gamma_{\rm E} - 2 \, {\rm Re} \left\{ \psi\left(\frac{1}{2} + \frac{l}{2} + i \nu \right) \right\} \,,
\end{eqnarray}
with $\gamma_{\rm E}$ being the Euler--Mascheroni constant and $\psi(z) \equiv \Gamma^\prime(z)/\Gamma(z)$ representing the digamma function. 
The NLO kernel correction, $\tilde{\chi}(l,\nu)$ in Eq.\eref{chi}, reads
\begin{equation}
\label{chi_NLO}
\tilde{\chi} \left(l,\nu\right) = \bar\chi(l,\nu)+\frac{\beta_0}{2 N_c}\chi(l,\nu)
\left( - \frac{1}{4}\chi(l,\nu) + \frac{5}{6} + \ln\frac{\mu_R}{\sqrt{|\vec q_{T_1}||\vec q_{T_2}|}} \right) \;,
\end{equation}
with $\bar\chi(l,\nu)$ being calculated in Ref.~\cite{Kotikov:2000pm} and explicitly reported in Appendix~\hyperlink{app:B}{B}.

The resummed cross section further requires the NLO baryon emission function in the representation of the LO BFKL eigenfunctions. 
Following the formalism of Ref.~\cite{Ivanov:2012iv}, particularly suitable for heavy-hadron production in the semihard regime, one obtains
\begin{equation}
\label{PIF}
\E_{\Omega_{3Q}}^{\rm NLO}(l,\nu,|\vec q_T|,x) =
\E_{\Omega_{3Q}}(\nu,|\vec q_T|,x) +
\alpha_s(\mu_R) \, \hat \E_{\Omega_{3Q}}(l,\nu,|\vec q_T|,x) \;.
\end{equation}
The corresponding LO function reads
\begin{equation}
\label{LOPIF}
\hspace{-0.30cm}
\E_{\Omega_{3Q}}(\nu,|\vec q_T|,x) 
= 2 \sqrt{\frac{C_F}{C_A}}
|\vec q_T|^{2i\nu-1}
\!\!\!\int_{x}^1\frac{\drv \xi}{\xi}
\left( \frac{\xi}{x} \right)
^{2 i\nu-1} 
 \!\left[\frac{C_A}{C_F}f_g(\xi)D_g^{\Omega_{3Q}}\left(\frac{x}{\xi}\right)
 +\!\!\!\sum_{\alpha=q,\bar q}\!f_\alpha(\xi)D_\alpha^{\Omega_{3Q}}\left(\frac{x}{\xi}\right)\right] \;,
\end{equation}
with the NLO correction, $\hat \E_{\Omega_{3Q}}(l,\nu,|\vec q_T|,x)$, taken from the identified-hadron NLO impact factor derived in Ref.~\cite{Ivanov:2012iv}, specifically Eq.~(4.58), and expressed in terms of the coefficient functions given in Eqs.~(4.59) to~(4.62). 
The structure of the complete NLO emission function in Eq.~\eqref{PIF} can be schematically represented as
\begin{equation}
{\cal E}_{\Omega_{3Q}}^{\rm NLO}
\sim
\sum_\alpha f_\alpha\otimes D_\alpha^{\Omega_{3Q}}
+
\frac{\alpha_s}{2\pi}
\sum_{\alpha,\beta}
f_\alpha\otimes c^h_{\alpha\beta}\otimes D_\beta^{\Omega_{3Q}}\;,
\end{equation}
where $\alpha$ and $\beta$ independently run over the active gluon, quark, and antiquark species, while $c^h_{\alpha\beta}$ denote the NLO coefficient functions. 
At LO, the emission function exhibits a purely flavor-diagonal convolution structure, $f_\alpha\otimes D_\alpha^{\Omega_{3Q}}$, in which the partonic species entering through the proton PDF coincides with the one fragmenting into the observed baryon.

At NLO, this structure is supplemented by the convolutions $f_\alpha\otimes c^h_{\alpha\beta}\otimes D_\beta^{\Omega_{3Q}}$, where the coefficient functions $c^h_{\alpha\beta}$ allow the PDF and FF flavor indices to be independent and therefore encode both diagonal and mixed partonic channels. 
These coefficient functions collect the real and virtual radiative contributions together with the collinear-subtraction terms. 
After PDF and FF renormalization, the infrared and ultraviolet singularities cancel, yielding a finite NLO emission function. 
The $\Omega_{3Q}$ FFs thus enter through the flavor-diagonal LO contribution as well as through the full set of NLO convolutions, providing the interface between the perturbative emission vertex and the fragmentation dynamics of the observed baryon.

The relative weighting of the gluon- and quark-initiated terms in Eq.~\eqref{LOPIF} originates from the color structure of the LO forward-parton impact factors. 
With the overall normalization convention adopted here, the quark contribution is normalized to the fundamental color factor $C_F$, whereas gluon emission carries the corresponding adjoint factor $C_A$, thus generating the relative weight $C_A/C_F$ multiplying the gluon PDF-FF convolution. 
These color factors therefore belong to the perturbative emission vertex and are independent of the nonperturbative structure encoded in the $\Omega_{3Q}$ FFs.

The resummed description is completed by the jet emission function,
\begin{equation}
\label{JIF}
\E_J^{\rm NLO}(l,\nu,|\vec q_T|,x) =
\E_J(\nu,|\vec q_T|,x) +
\alpha_s(\mu_R) \, \hat \E_J(l,\nu,|\vec q\,|,x) \;,
\end{equation}
whose LO expression is
\begin{equation}
 \label{LOJIF}
 \E_J(\nu,|\vec q_T|,x) = 2 \sqrt{\frac{C_F}{C_A}}
 |\vec q_T|^{2i\nu-1}\left[\frac{C_A}{C_F}f_g(x)
 +\sum_{\beta=q,\bar q}f_\beta(x)\right] \;.
\end{equation}
The NLO term $\hat \E_J(l,\nu,|\vec q_T|,x)$ is sensitive to the choice of jet definition. 
In this work, we follow the formalism developed in Ref.~\cite{Ivanov:2012ms}, adopting a formulation specifically designed for efficient numerical implementations. 
The corresponding jet-selection function is constructed within the small-cone approximation for cone-based algorithms (see Ref.~\cite{Colferai:2015zfa} for a detailed discussion). 
Consistently with recent CMS measurements of forward-jet production~\cite{Khachatryan:2016udy}, the jet radius is fixed to ${\cal R}_J = 0.5$.

The preceding construction makes explicit the factorized roles of the three ingredients: the two emission functions encode the process-dependent fragmentation-region dynamics, while the BFKL Green's function provides the universal high-energy evolution connecting them.

Combining the Green's function with the baryon and jet emission functions, one obtains the master formula for the $\NLLp$ azimuthal coefficients in the $\MSb$ renormalization scheme,

\begin{eqnarray}
\label{Cl_NLLp_MSb}
 \ClNLLp \!\! &=& \!\! 
 \frac{e^{\DY}}{s} 
 \int_{-\infty}^{+\infty} \drv \nu \, 
 {\cal G}_{\rm NLL}(\DY,l,\nu,\mu_R) \,
 \alpha_s^2(\mu_R) 
 \\ \nonumber
 \!\! &\times& \!\! \biggl\{\E_{\Omega_{3Q}}^{\rm NLO}(l,\nu,|\vec q_{T_1}|, x_1) \,
 [\E_J^{\rm NLO}(l,\nu,|\vec q_{T_2}|,x_2)]^*
 \\ \nonumber
 \!\! &+& \!\!
 \left.
 \alpha_s^2(\mu_R) \DY \frac{\beta_0}{4 \pi} \,
 \chi(l,\nu)
 \left[\ln\left(|\vec q_{T_1}| |\vec q_{T_2}|\right) + \frac{i}{2} \, \frac{\drv}{\drv \nu} \ln\frac{\E_{\Omega_{3Q}}}{\E_J^*}\right]
 \right\}
 \;.
\end{eqnarray}
The $\NLLp$ designation refers to the resummation of energy logarithms at full NLL accuracy combined with NLO impact factors. 
The superscript `$+$' indicates the inclusion of contributions beyond strict NLL accuracy, originating from products of NLO corrections to the emission functions. 

Within our notation, the labels $\LL$ and $\NLLpp$ refer exclusively to the logarithmic accuracy of the universal BFKL kernel. 
The $\LL$ approximation resums terms of the form $[\alpha_s \ln(s)]^n$ through the LO kernel, whereas $\NLLpp$ additionally incorporates $\alpha_s [\alpha_s \ln(s)]^n$ contributions generated by the NLO kernel. 
The resulting kernel is then convoluted with impact factors evaluated either at LO or NLO. 
This kernel-based classification provides a process-independent definition of logarithmic accuracy, avoiding ambiguities associated with the perturbative order of specific impact factors. 
Indeed, the leading power of $\alpha_s$ carried by the impact factors depends on the process under consideration, as exemplified by the difference between Higgs and jet/hadron production channels~\cite{Celiberto:2020tmb}.

For reference, we also consider the pure LL limit in the $\MSb$ scheme,
\begin{equation}
\label{Cl_LL_MSb}
 \ClLL = 
 \frac{e^{\DY}}{s} 
 \int_{-\infty}^{+\infty} \drv \nu \, 
 e^{{\cal G}_{\rm NLL}^{(0)}(\DY,l,\nu,\mu_R)} 
 \alpha_s^2(\mu_R) \, \E_{\Omega_{3Q}}(l,\nu,|\vec q_{T_1}|, x_1)[\E_J(l,\nu,|\vec q_{T_2}|,x_2)]^* \;.
\end{equation}

A meaningful assessment of high-energy resummation effects requires a comparison with predictions based on fixed-order perturbation theory. 
Since fully differential NLO calculations for inclusive hadron-plus-jet production are not currently available in a numerical implementation, we adopt a high-energy fixed-order approximation to quantify the impact of resummation. 
This procedure, first introduced in studies of di-jet~\cite{Celiberto:2015yba} and hadron-jet~\cite{Celiberto:2020wpk} correlations, consists in truncating the resummed high-energy series at NLO accuracy, thereby reproducing the asymptotic high-energy limit of a fixed-order NLO calculation. 
Specifically, the azimuthal coefficients of Eq.~(\ref{Cl_NLLp_MSb}) are expanded up to ${\cal O}(\alpha_s^3)$, leading to the high-energy fixed-order ($\HENLOp$) representation
\begin{align}
\label{Cl_HENLOp_MSb}
 \ClHENLOp &= 
 \frac{e^{\DY}}{s} 
 \int_{-\infty}^{+\infty} \drv \nu \, 
 \alpha_s^2(\mu_R) \,
 \left[ 1 + {\cal G}_{\rm NLL}^{(0)}(\DY,l,\nu,\mu_R) \right]
 \\ \nonumber
 &\times
 \E_{\Omega_{3Q}}^{\rm NLO}(l,\nu,|\vec q_{T_1}|, x_1)[\E_J^{\rm NLO}(l,\nu,|\vec q_{T_2}|,x_2)]^* \;,
\end{align}
with
\begin{equation}
\label{G_BFKL_0}
 {\cal G}_{\rm NLL}^{(0)}(\DY,l,\nu,\mu_R) = \bar \alpha_s(\mu_R) \DY \chi(l,\nu)
\end{equation}
standing for the first-order expansion of the resummation kernel.

The factorization and renormalization scales are chosen according to the characteristic hardness of the final state. 
Throughout this work, we adopt the reference choice $\mu_F = \mu_R = \mu_N$, where the natural scale is defined as $\mu_N = m_{\Omega_{3Q}\perp} + |\vec q_{T_2}|$. 
Here, $m_{\Omega_{3Q}\perp} = \sqrt{m_{\Omega_{3Q}}^2 + |\vec q_{T_1}|^2}$ denotes the transverse mass of the produced baryon. 
This prescription, based on the sum of the transverse scales associated with the tagged final-state objects, is widely employed in precision-QCD calculations and Monte Carlo implementations (see, \emph{e.g.}, Refs.~\cite{Alioli:2010xd,Campbell:2012am}). 
Residual missing-higher-order uncertainties (MHOUs) are estimated through the standard scale-variation procedure, varying $\mu_F$ and $\mu_R$ around $\mu_N$ within the range $\mu_N/2 < \mu_{F,R} < 2\mu_N$, as controlled by the parameter $C_\mu$.

\section{Precision phenomenology at hadron colliders}
\label{sec:phenomenology}

All numerical results presented in this work are obtained within the \textsc{Python}+\textsc{Fortran} {\Jethad} multimodular framework~\cite{Celiberto:2020wpk,Celiberto:2022rfj,Celiberto:2023fzz,Celiberto:2024mrq,Celiberto:2024swu,Celiberto:2025csa,Celiberto:2026ooh}. 
Proton PDFs are described by the {\tt NNPDF4.0} NLO set~\cite{NNPDF:2021uiq}, interfaced through {\tt LHAPDF6}~\cite{Buckley:2014ana}.

\subsection{Fiducial selection and uncertainty estimates}
\label{ssec:obserbables}

Our phenomenological study is centered on the rapidity-separation distribution, namely the cross section differential in
$\DY = y_1-y_2$, where $y_1$ and $y_2$ denote the rapidities of the produced $\Omega_{3Q}$ baryon and the associated light jet, respectively.

This observable is obtained from the azimuthal coefficient ${\cal C}_0$ introduced in Sec.~\ref{ssec:NLL-cross-section}, after integration over the remaining final-state phase space at fixed $\DY$. 
It can be written as
\\
\begin{equation}
 \label{obs:I}
 \frac{\drv \sigma(\DY, s)}{\drv \DY} =
 \int_{|\vec q_{T_1}|^{\rm min}}^{|\vec q_{T_1}|^{\rm max}} 
 \!\!\drv |\vec q_{T_1}|
 \int_{|\vec q_{T_2}|^{\rm min}}^{|\vec q_{T_2}|^{\rm max}} 
 \!\!\drv |\vec q_{T_2}|
 \int_{\max \, (y_1^{\rm min}, \, \DY + y_2^{\rm min})}^{\min \, (y_1^{\rm max}, \, \DY + y_2^{\rm max})} \drv y_1
 \, \,
 {\cal C}_0^{\rm [res]}
\Bigm \lvert_{y_2 \;=\; y_1 - \DY}
 \;,
\end{equation}
\\
where the label ${\rm [res]}$ collectively denotes $\NLLp$, $\HENLOp$, or $\LL$ accuracy.
The integration over one rapidity variable is removed through the kinematic constraint $\DY = y_1-y_2$.

The quantities $|\vec q_{T_1}|$ and $|\vec q_{T_2}|$ entering
Eq.~\eqref{obs:I} explicitly denote the magnitudes of the transverse
momenta of the tagged $\Omega_{3Q}$ baryon and light jet, respectively.
The notion of ``large transverse momentum'' adopted throughout our
analysis refers to the scale hierarchy underlying the semihard, or
Regge-Gribov regime~\cite{Gribov:1983ivg}: the transverse scales of the tagged objects must
remain well above the infrared QCD domain while being much smaller than
the hadronic center-of-mass energy.
For the identified $\Omega_{3Q}$ baryon, its mass provides an additional
physical scale, and the leading-power VFNS fragmentation description is
expected to become progressively more reliable as
$|\vec q_{T_1}|/m_{\Omega_{3Q}}$ increases and mass-suppressed
contributions become less relevant.
No universal sharp criterion of the form
$|\vec q_{T_1}|/m_{\Omega_{3Q}}>C$, with a process-independent constant
$C$, is implied by the factorization framework.
We therefore assess the practical adequacy of the selected
transverse-momentum domain by studying the stability of the predictions
under progressively harder transverse-momentum cuts.

Our reference selection is the asymmetric range
$50<|\vec q_{T_1}|/{\rm GeV}<120$ for the $\Omega_{3Q}$ baryon and
$30<|\vec q_{T_2}|/{\rm GeV}<120$ for the associated jet.
Asymmetric transverse-momentum cuts are commonly employed in semihard
phenomenological studies~\cite{Khachatryan:2016udy} and are known to
enhance sensitivity to high-energy dynamics.
In particular, they suppress the relative importance of fixed-order
contributions~\cite{Celiberto:2015yba,Celiberto:2020wpk}, reduce
Sudakov-enhanced effects associated with nearly back-to-back
configurations~\cite{Mueller:2013wwa,Marzani:2015oyb}, improve
perturbative stability~\cite{Andersen:2001kta}, and mitigate violations
of energy-momentum conservation~\cite{Ducloue:2014koa}.

To examine whether this reference choice lies within a sufficiently
stable semihard domain, and to assess the impact of the hierarchy
between the two tagged transverse scales, we complement it with three
additional selections.
The first is a symmetric range,
$50<|\vec q_{T_1}|/{\rm GeV}<120$ and
$50<|\vec q_{T_2}|/{\rm GeV}<120$.
We then consider an intermediate asymmetric range,
$80<|\vec q_{T_1}|/{\rm GeV}<120$ and
$50<|\vec q_{T_2}|/{\rm GeV}<120$,
followed by a higher asymmetric range,
$100<|\vec q_{T_1}|/{\rm GeV}<120$ and
$70<|\vec q_{T_2}|/{\rm GeV}<120$.
These progressively more restrictive selections permit us to test the
stability of the resummed predictions as the tagged transverse scales
are moved farther into the perturbative domain, while simultaneously
monitoring the reduction of the available production rate.

Concerning rapidity acceptance, we adopt representative LHC-inspired selections.
The $\Omega_{3Q}$ baryon is assumed to be detected in the central region,
$-2.5 < y_1 < 2.5$,
corresponding to the tracking and central-detector coverage~\cite{Chatrchyan:2012xg},
while the accompanying jet can extend into the forward region,
$-4.7 < y_2 < 4.7$,
consistently with current experimental capabilities~\cite{Khachatryan:2016udy}.

A realistic phenomenological study requires a systematic assessment of theoretical uncertainties.
In the present analysis, the dominant uncertainty sources are identified and treated separately before being consistently combined at the observable level.
The considered contributions are:

\begin{itemize}

 \item[$(i)$]
 \textbf{Perturbative H-MHOUs}.
 These originate from the residual dependence on the renormalization and factorization scales entering the hard-scattering subprocess.
 Their effect is estimated by varying the corresponding scales around their central values by a factor between $1/2$ and $2$, thereby probing the size of missing higher-order contributions in the high-energy resummed cross section.

 \item[$(ii)$]
 \textbf{Perturbative F-MHOUs}.
 These uncertainties are associated with the perturbative input of the FFs at their initial scale.
 As discussed in Sec.~\ref{sec:FFs}, they are modeled through variations of the evolution scales entering the FF construction and are encoded in the replica structure of the {\tt OMG3Q1.1} set.

 \item[$(iii)$]
 \textbf{Nonperturbative F-NPWF uncertainties}.
 These reflect the modeling of the nonperturbative wave function entering the initial-scale FFs.
 In the present framework, they are controlled through variations of the transverse-momentum parameter $\vqTTa$, which governs the effective momentum-space structure of the all-heavy bound state within the Suzuki-inspired approach.

 \item[$(iv)$]
 \textbf{Proton PDFs}.
 An additional source of uncertainty originates from the proton PDFs, which are extracted from global fits and therefore contain an intrinsic nonperturbative component.
 However, previous studies of heavy-flavor and all-heavy hadron production have shown that PDF-induced variations are typically subleading with respect to fragmentation-related uncertainties.
 For this reason, we restrict our analysis to the central member of the {\tt NNPDF4.0} set~\cite{NNPDF:2021uiq}.

 \item[$(v)$]
 \textbf{Phase-space numerical integration}.
 The leading numerical uncertainty originates from the multidimensional integration over the final-state phase space (see Eq.~\eqref{obs:I}) and over the Mellin variable $\nu$ (see Eqs.~\eqref{Cl_NLLp_MSb}, \eqref{Cl_LL_MSb}, and~\eqref{Cl_HENLOp_MSb}).
 These integrals are performed using the native routines of {\Jethad}, with numerical errors systematically kept below the $1\%$ level.
 Subleading contributions arise from the integration over the longitudinal momentum fractions entering the PDF-FF convolution (see Eq.~\eqref{LOPIF}) and are found to be negligible.

\end{itemize}

In practice, F-MHOU and F-NPWF effects are simultaneously encoded within the {\tt OMG3Q1.1} replica ensemble, which provides a unified framework for uncertainty propagation (see Sec.~\ref{sec:FFs} and Appendix~\hyperlink{app:A}{A}).
At the level of rapidity distributions, these contributions are consistently combined with H-MHOUs, allowing for a comprehensive assessment of the total theoretical uncertainty affecting the observable.

\subsection{Numerical results and phenomenological implications}
\label{ssec:results}

\begin{figure*}[!t]
\centering

   \hspace{0.00cm}
   \includegraphics[scale=0.395,clip]{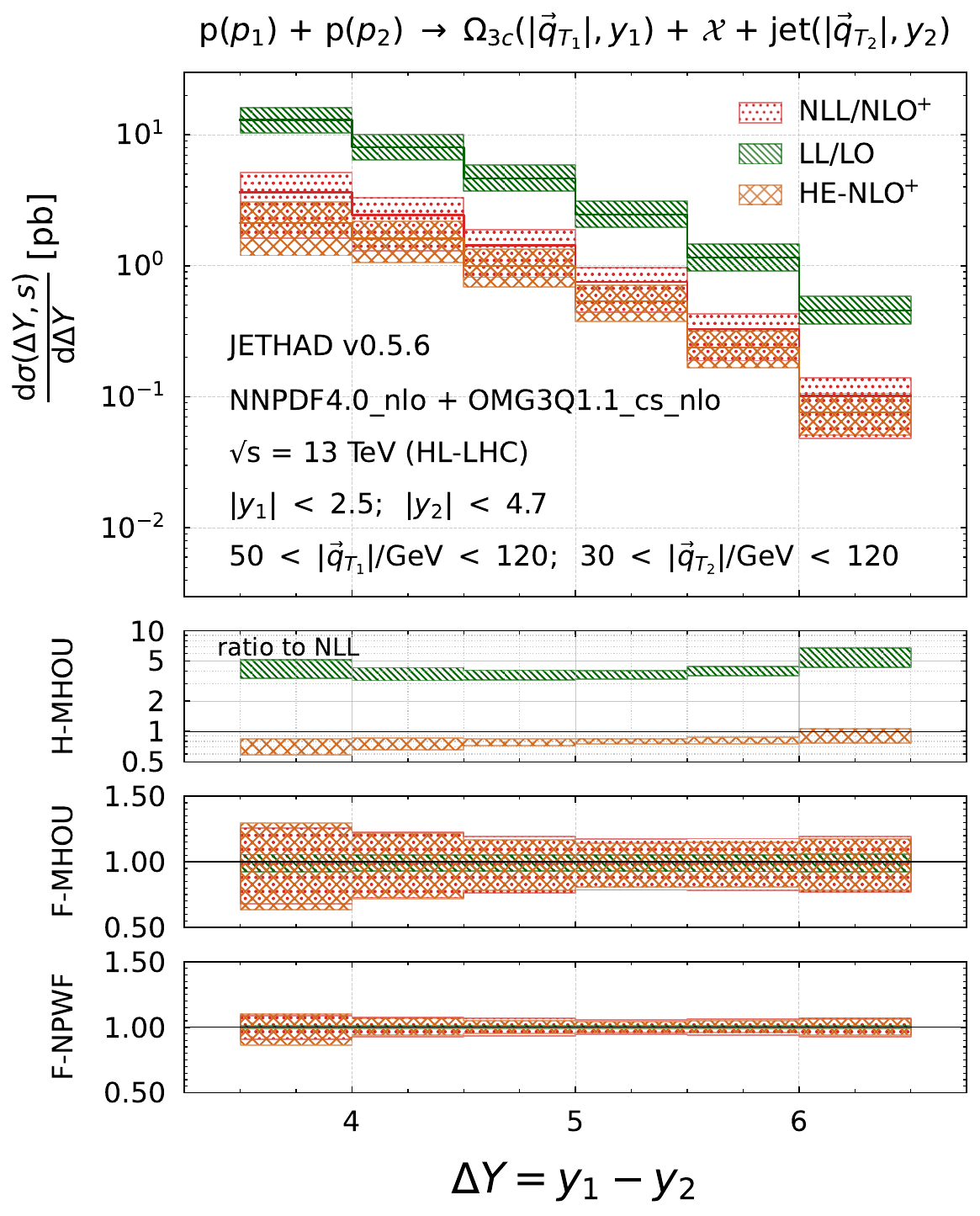}
   \hspace{0.30cm}
   \includegraphics[scale=0.395,clip]{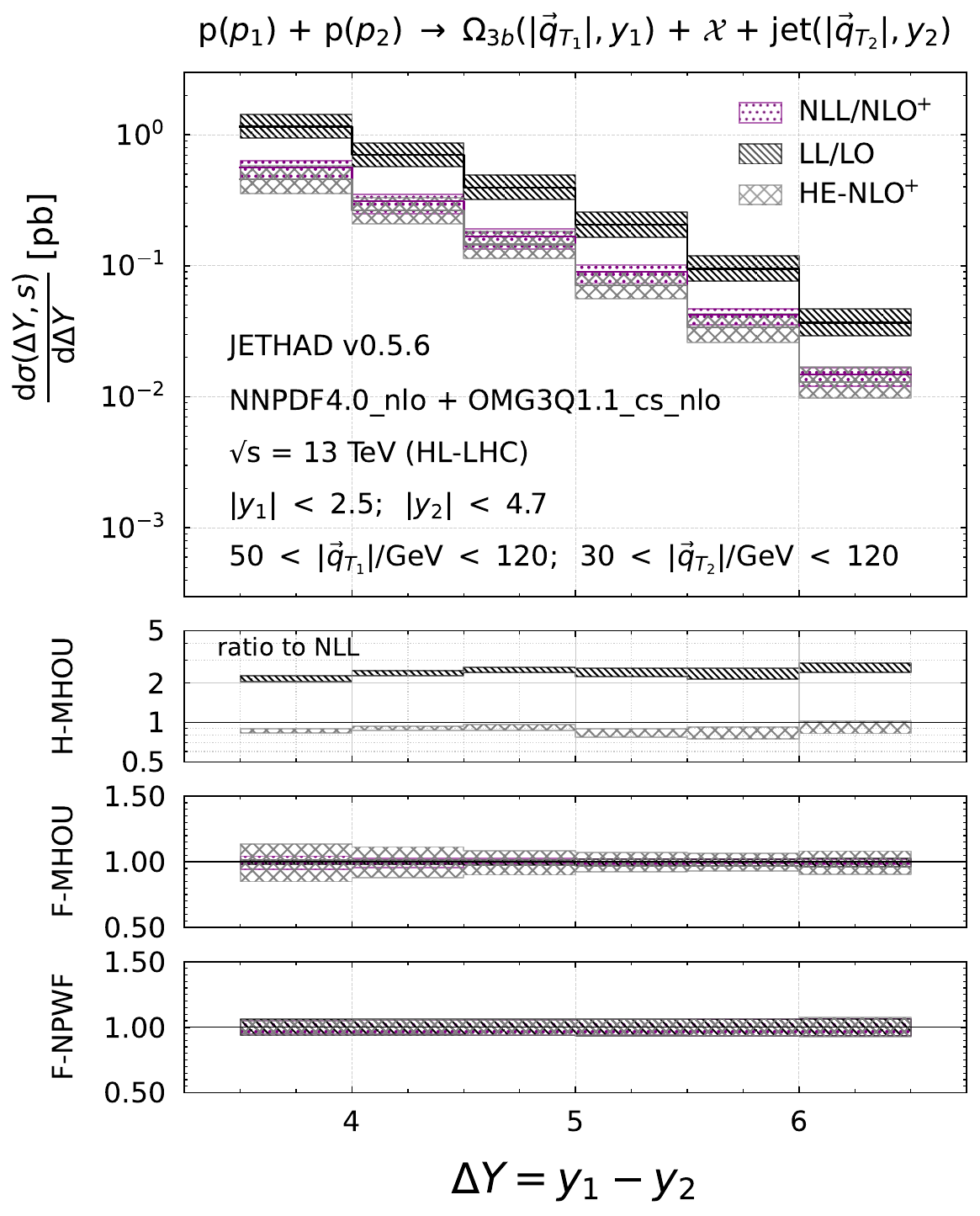}

\caption{Rapidity-separation distributions for semi-inclusive $\Omega_{3Q}$ plus jet production at $\sqrt{s}=13$~TeV (HL-LHC), shown for the triply charmed state $\Omega_{3c}$ (left) and the triply bottom state $\Omega_{3b}$ (right). 
The reference asymmetric transverse-momentum range is defined by $50<|\vec q_{T_1}|<120$~GeV for the $\Omega_{3Q}$ baryon and $30<|\vec q_{T_2}|<120$~GeV for the jet.
Predictions are presented at the $\NLLp$, $\LL$, and $\HENLOp$ accuracy levels. 
The uncertainty envelopes in the upper panels correspond to the total uncertainty obtained by combining H-MHOU, F-MHOU, F-NPWF, and integration effects. 
The lower panels show, from top to bottom: $(i)$ the ratios of $\LL$ and $\HENLOp$ results to the $\NLLp$ prediction, including H-MHOU variations only; $(ii)$ the F-MHOU uncertainty band; and $(iii)$ the F-NPWF uncertainty band, both normalized to the corresponding prediction.}
\label{fig:I_LHC_ref-asym}

\end{figure*}

\begin{figure*}[!t]
\centering

   \hspace{0.00cm}
   \includegraphics[scale=0.395,clip]{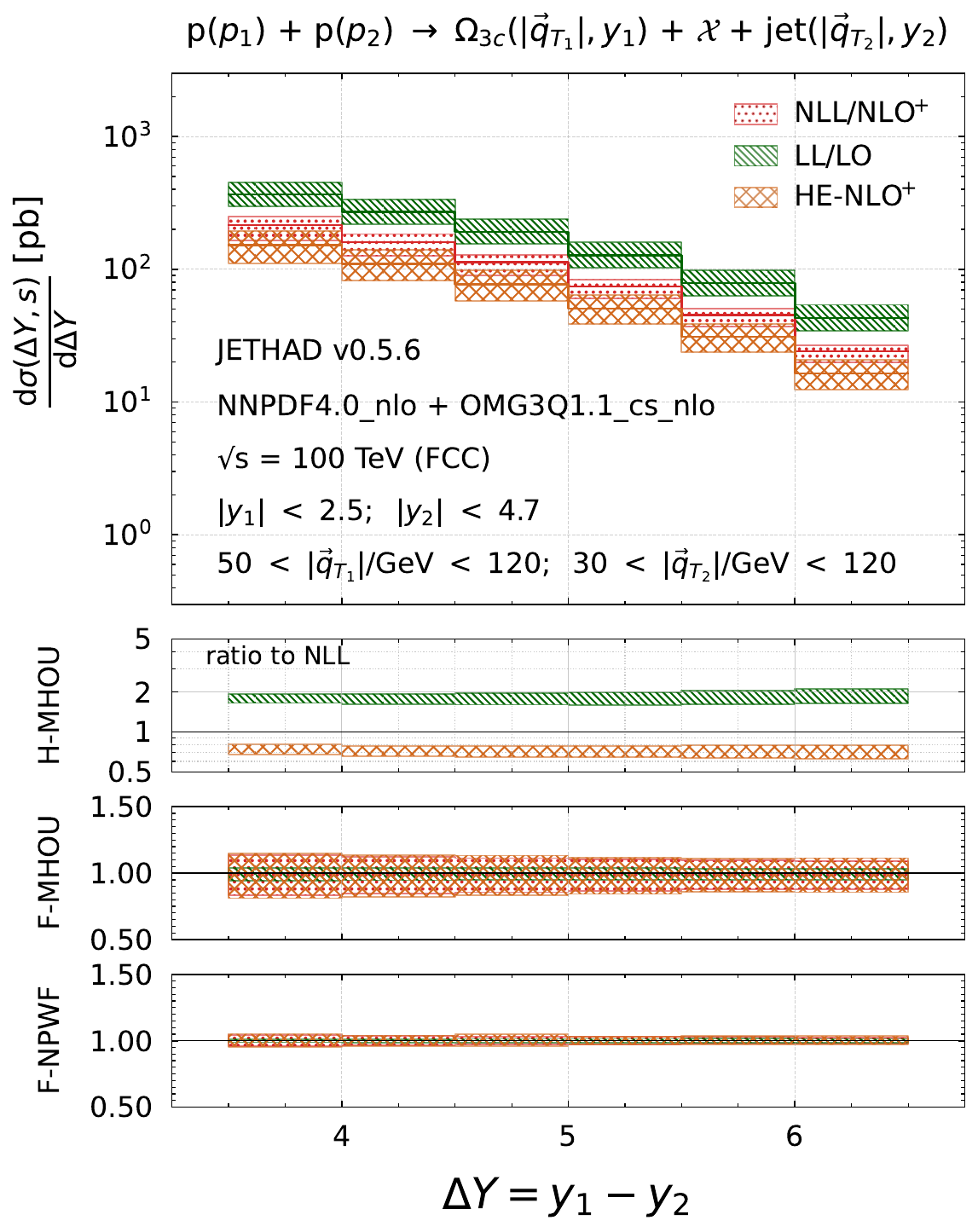}
   \hspace{0.30cm}
   \includegraphics[scale=0.395,clip]{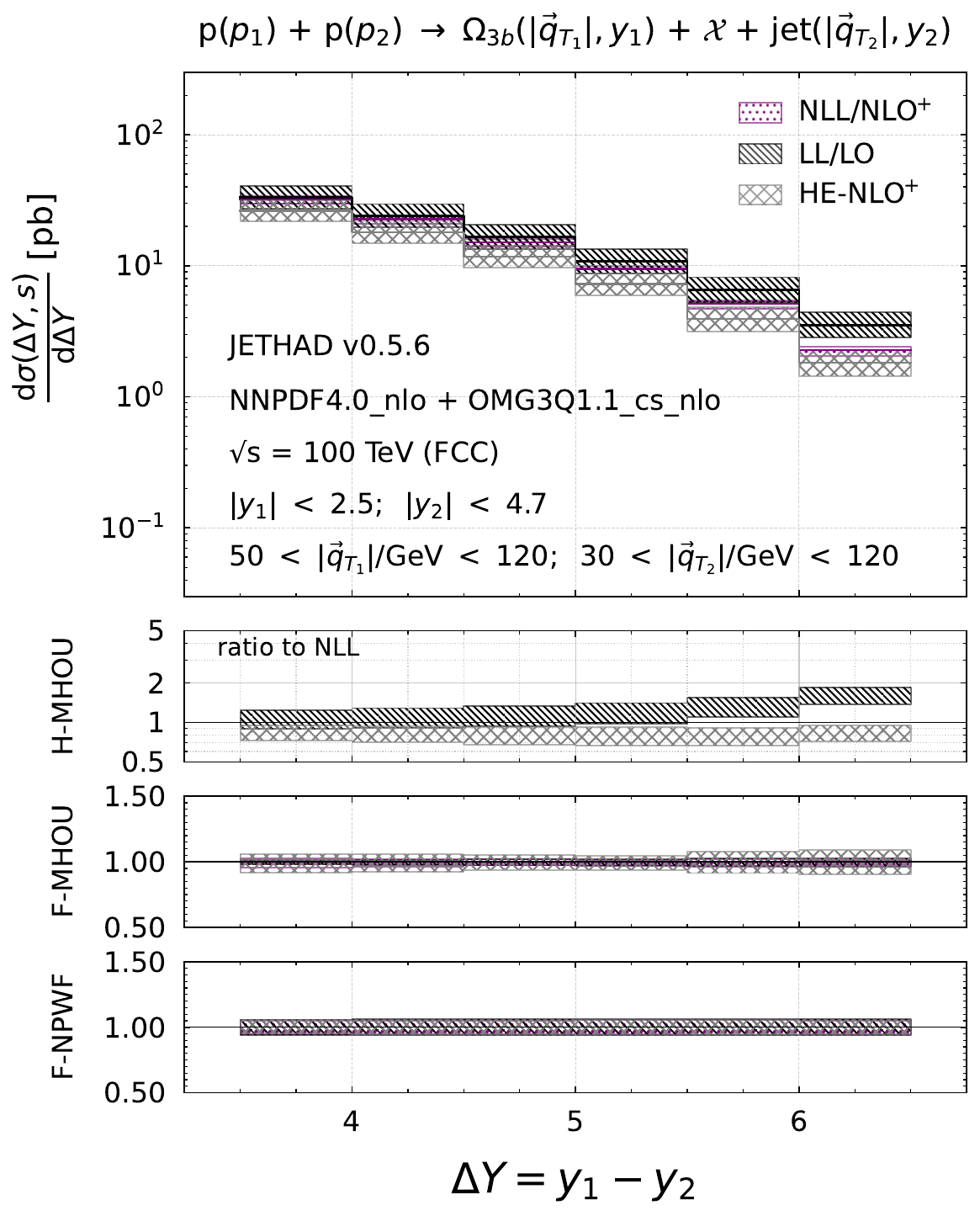}

\caption{Rapidity-separation distributions for semi-inclusive $\Omega_{3Q}$ plus jet production at $\sqrt{s}=100$~TeV (FCC, nominal), shown for the triply charmed state $\Omega_{3c}$ (left) and the triply bottom state $\Omega_{3b}$ (right). 
The reference asymmetric transverse-momentum range is defined by $50<|\vec q_{T_1}|<120$~GeV for the $\Omega_{3Q}$ baryon and $30<|\vec q_{T_2}|<120$~GeV for the jet.
Predictions are presented at the $\NLLp$, $\LL$, and $\HENLOp$ accuracy levels. 
The uncertainty envelopes in the upper panels correspond to the total uncertainty obtained by combining H-MHOU, F-MHOU, F-NPWF, and integration effects. 
The lower panels show, from top to bottom: $(i)$ the ratios of $\LL$ and $\HENLOp$ results to the $\NLLp$ prediction, including H-MHOU variations only; $(ii)$ the F-MHOU uncertainty band; and $(iii)$ the F-NPWF uncertainty band, both normalized to the corresponding prediction.}

\label{fig:I_FCC_ref-asym}
\end{figure*}

Figures~\ref{fig:I_LHC_ref-asym} and~\ref{fig:I_FCC_ref-asym} display our predictions for semi-inclusive $\Omega_{3Q}$ plus light-jet production as a function of the rapidity separation $\DY$ between the baryon and the jet.
Results are shown for the triply charmed state $\Omega_{3c}$ (left panels) and the triply bottom state $\Omega_{3b}$ (right panels), at the HL-LHC energy of $\sqrt{s}=13$~TeV and at the nominal FCC energy of $\sqrt{s}=100$~TeV, respectively.
The rapidity spectra are presented in bins of width $\Delta Y = 0.5$, a choice motivated by future experimental studies and by the need to retain a direct connection with realistic detector-level analyses.

\begin{figure*}[!t]
\centering

   \hspace{0.00cm}
   \includegraphics[scale=0.395,clip]{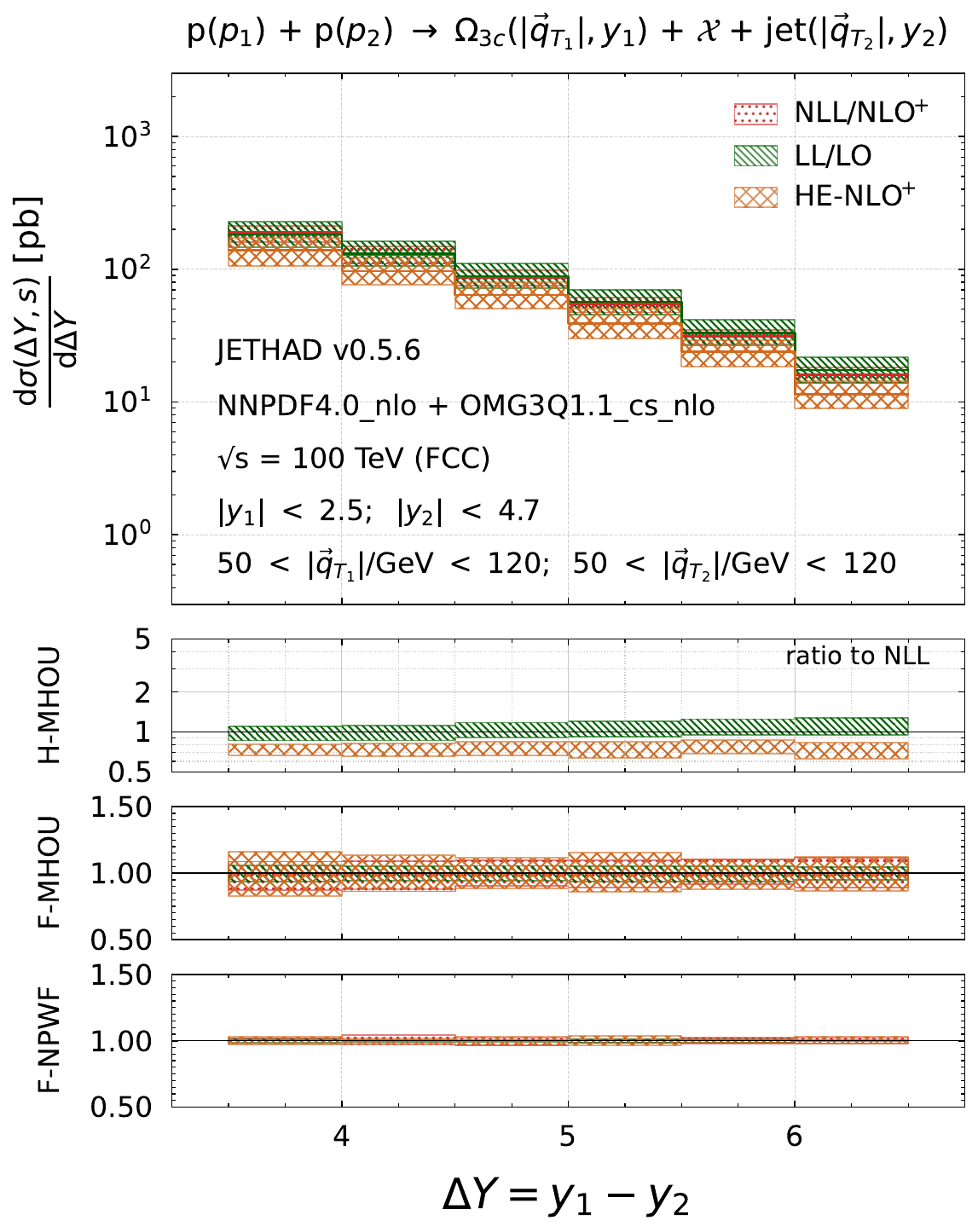}
   \hspace{0.30cm}
   \includegraphics[scale=0.395,clip]{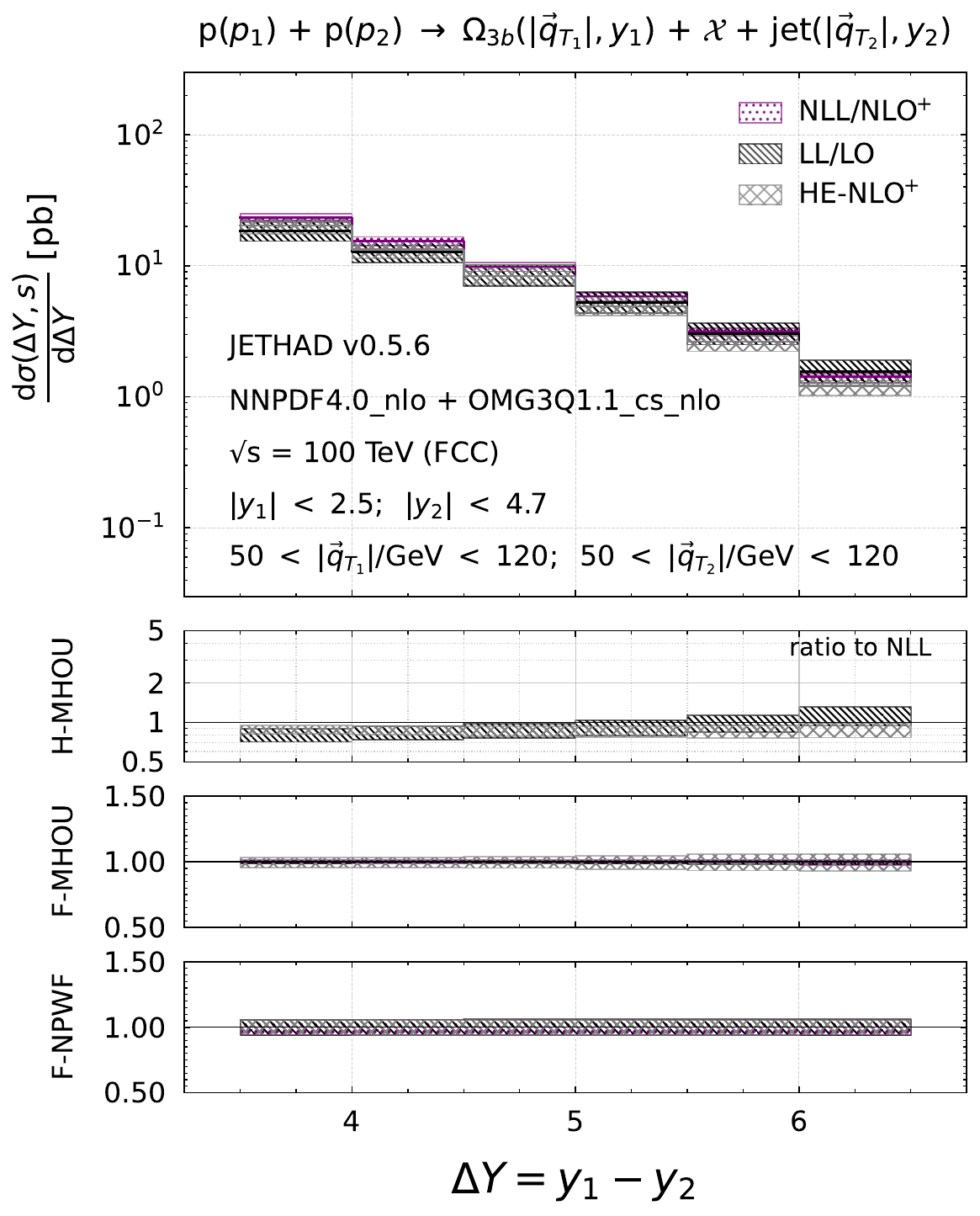}

\caption{
Rapidity-separation distributions for semi-inclusive $\Omega_{3Q}$ plus jet production at $\sqrt{s}=100$~TeV (FCC, nominal), shown for the triply charmed state $\Omega_{3c}$ (left) and the triply bottom state $\Omega_{3b}$ (right). 
The symmetric transverse-momentum range is defined by $50<|\vec q_{T_1}|<120$~GeV for the $\Omega_{3Q}$ baryon and $50<|\vec q_{T_2}|<120$~GeV for the jet.
Predictions are presented at the $\NLLp$, $\LL$, and $\HENLOp$ accuracy levels. 
The uncertainty envelopes in the upper panels correspond to the total uncertainty obtained by combining H-MHOU, F-MHOU, F-NPWF, and integration effects. 
The lower panels show, from top to bottom: $(i)$ the ratios of $\LL$ and $\HENLOp$ results to the $\NLLp$ prediction, including H-MHOU variations only; $(ii)$ the F-MHOU uncertainty band; and $(iii)$ the F-NPWF uncertainty band, both normalized to the corresponding prediction.
}

\label{fig:I_FCC_sym}
\end{figure*}

\begin{figure*}[!t]
\centering

   \hspace{0.00cm}
   \includegraphics[scale=0.395,clip]{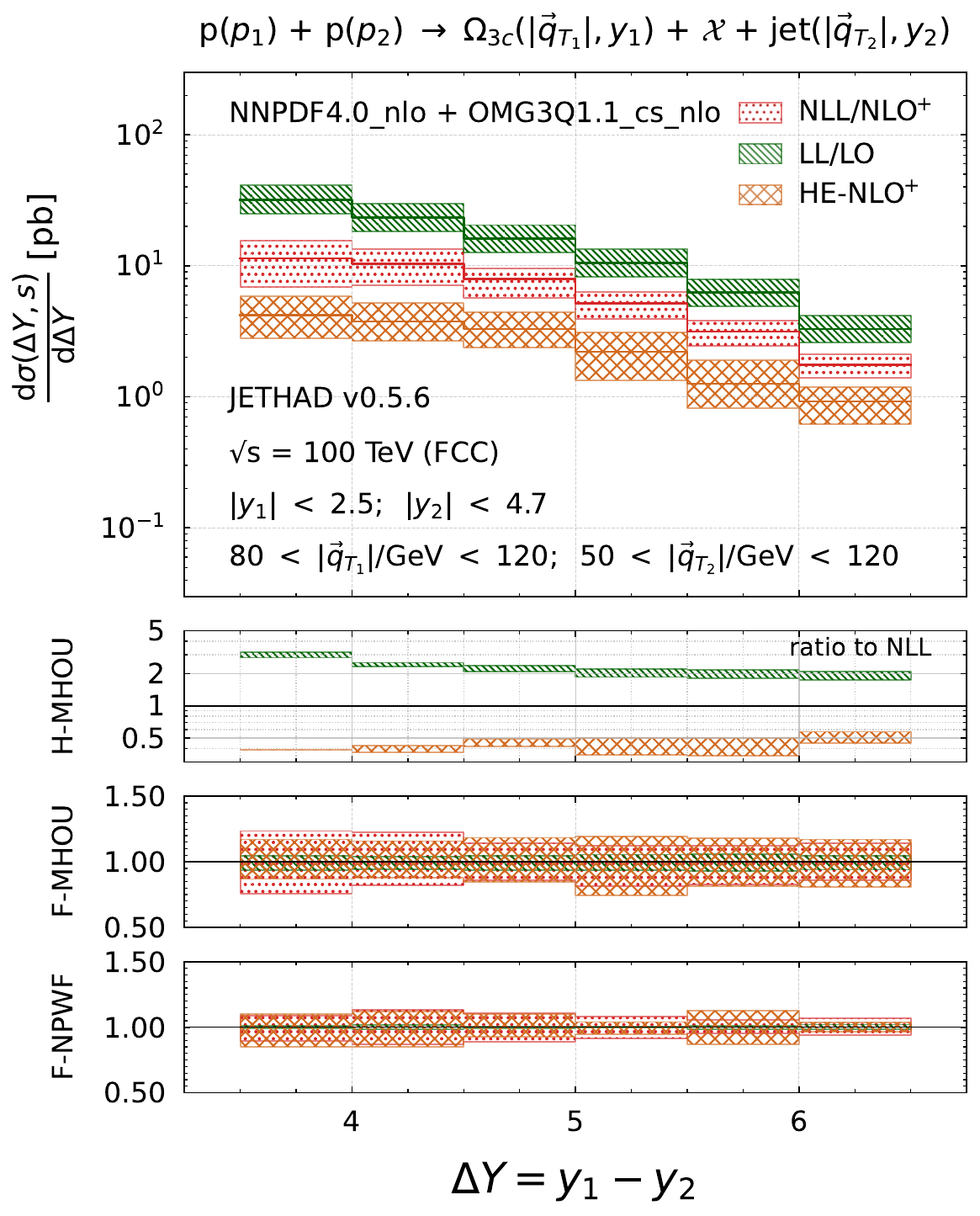}
   \hspace{0.30cm}
   \includegraphics[scale=0.395,clip]{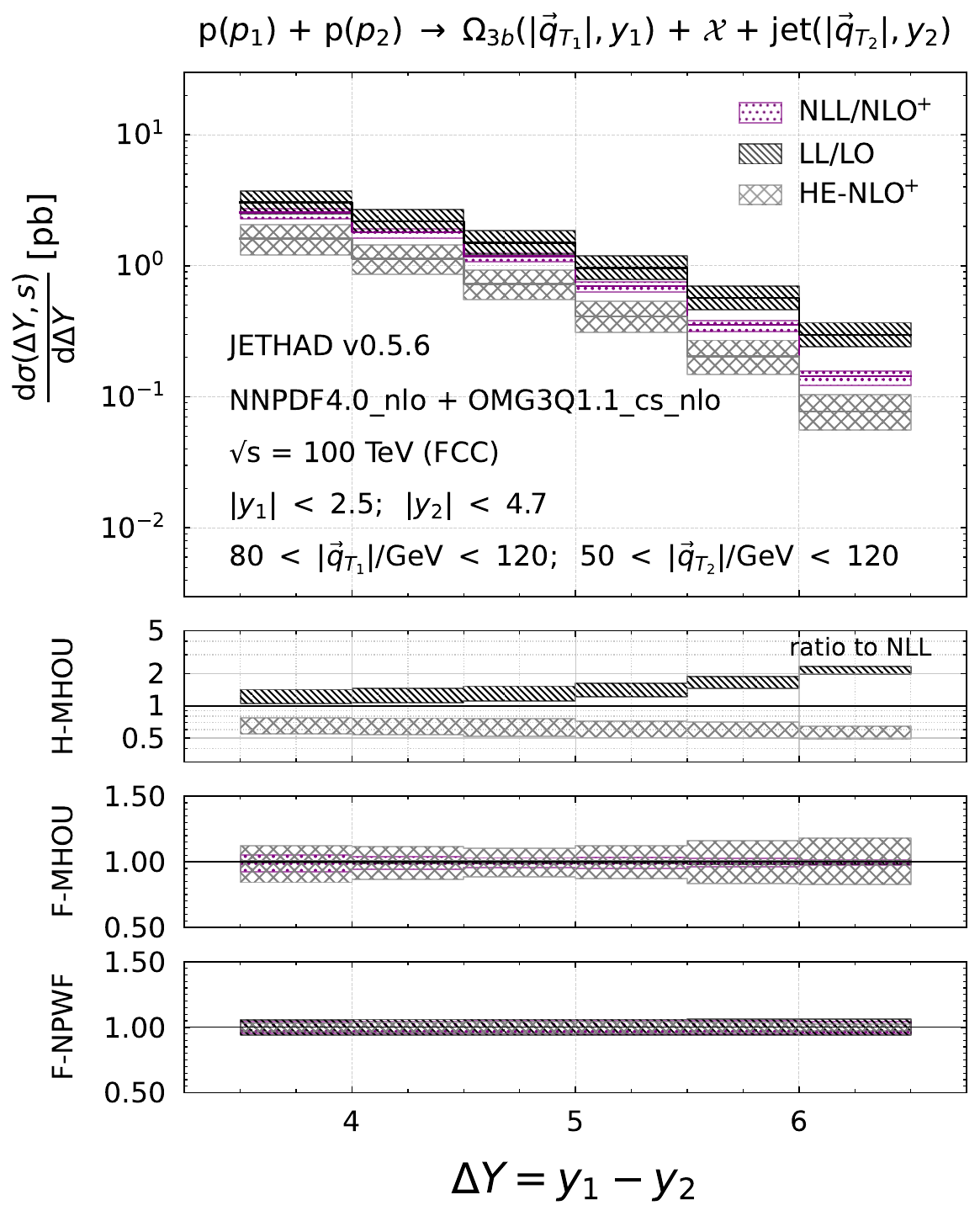}

\caption{
Rapidity-separation distributions for semi-inclusive $\Omega_{3Q}$ plus jet production at $\sqrt{s}=100$~TeV (FCC, nominal), shown for the triply charmed state $\Omega_{3c}$ (left) and the triply bottom state $\Omega_{3b}$ (right). 
The intermediate asymmetric transverse-momentum range is defined by $80<|\vec q_{T_1}|<120$~GeV for the $\Omega_{3Q}$ baryon and $50<|\vec q_{T_2}|<120$~GeV for the jet.
Predictions are presented at the $\NLLp$, $\LL$, and $\HENLOp$ accuracy levels. 
The uncertainty envelopes in the upper panels correspond to the total uncertainty obtained by combining H-MHOU, F-MHOU, F-NPWF, and integration effects. 
The lower panels show, from top to bottom: $(i)$ the ratios of $\LL$ and $\HENLOp$ results to the $\NLLp$ prediction, including H-MHOU variations only; $(ii)$ the F-MHOU uncertainty band; and $(iii)$ the F-NPWF uncertainty band, both normalized to the corresponding prediction.
}
\label{fig:I_FCC_int-asym}
\end{figure*}

\begin{figure*}[!t]
\centering

   \hspace{0.00cm}
   \includegraphics[scale=0.395,clip]{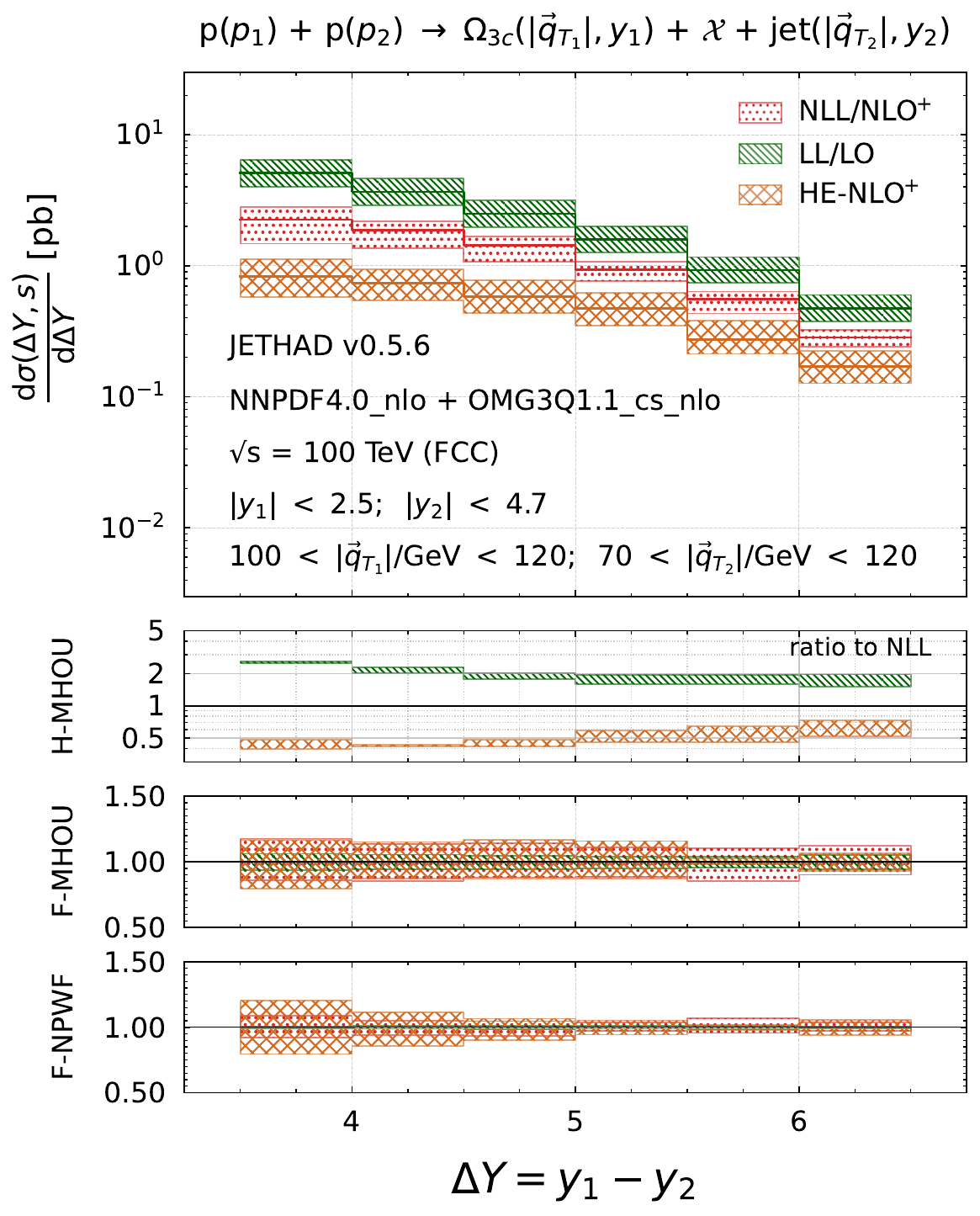}
   \hspace{0.30cm}
   \includegraphics[scale=0.395,clip]{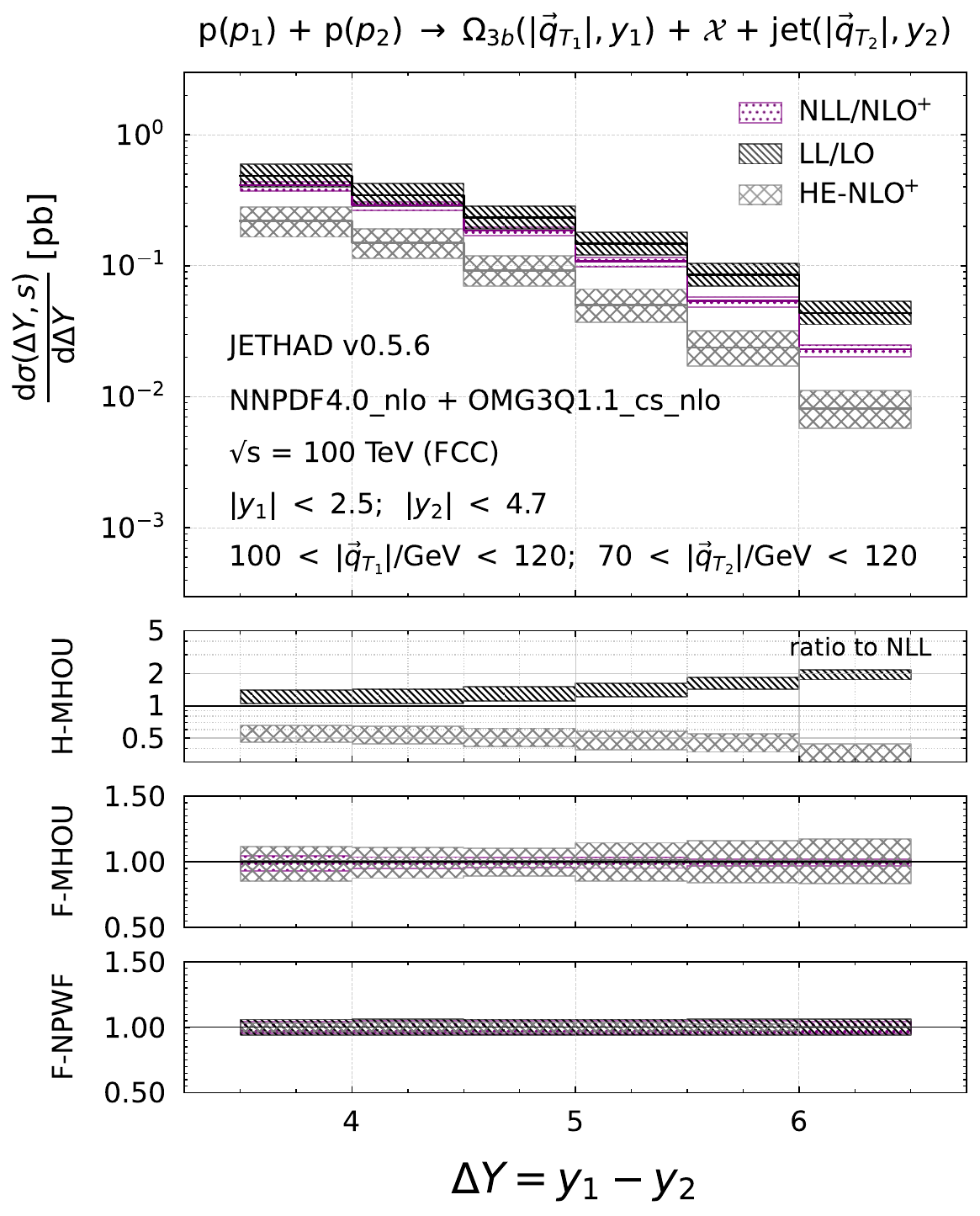}

\caption{
Rapidity-separation distributions for semi-inclusive $\Omega_{3Q}$ plus jet production at $\sqrt{s}=100$~TeV (FCC, nominal), shown for the triply charmed state $\Omega_{3c}$ (left) and the triply bottom state $\Omega_{3b}$ (right). 
The higher asymmetric transverse-momentum range is defined by $100<|\vec q_{T_1}|<120$~GeV for the $\Omega_{3Q}$ baryon and $70<|\vec q_{T_2}|<120$~GeV for the jet.
Predictions are presented at the $\NLLp$, $\LL$, and $\HENLOp$ accuracy levels. 
The uncertainty envelopes in the upper panels correspond to the total uncertainty obtained by combining H-MHOU, F-MHOU, F-NPWF, and integration effects. 
The lower panels show, from top to bottom: $(i)$ the ratios of $\LL$ and $\HENLOp$ results to the $\NLLp$ prediction, including H-MHOU variations only; $(ii)$ the F-MHOU uncertainty band; and $(iii)$ the F-NPWF uncertainty band, both normalized to the corresponding prediction.
}
\label{fig:I_FCC_higher-asym}
\end{figure*}

At the nominal FCC energy, this reference analysis is further complemented by the symmetric, intermediate asymmetric, and higher asymmetric transverse-momentum selections displayed in Figs.~\ref{fig:I_FCC_sym}, \ref{fig:I_FCC_int-asym}, and~\ref{fig:I_FCC_higher-asym}, respectively.
For simplicity and to avoid a redundant proliferation of numerical results, we perform this dedicated cut-dependence study at a single collider energy.
We select the nominal FCC setup since, as the highest-energy configuration considered in this work, it provides the most exploratory environment and the widest kinematic lever arm for testing the semihard regime.
We do not display the corresponding cut-dependence results at the 13~TeV HL-LHC, for which we have verified that the same qualitative pattern is recovered.
These additional FCC configurations are designed to assess the perturbative stability of the resummed predictions against progressively harder transverse-momentum requirements.

A common feature of the reference predictions shown in Figs.~\ref{fig:I_LHC_ref-asym} and~\ref{fig:I_FCC_ref-asym} is the progressive suppression of the cross section as the baryon-jet rapidity interval increases.
This trend originates from the interplay between two competing mechanisms.
While the BFKL ladder enhances the cross section through the growth of high-energy logarithmic contributions at large rapidity separations, the convolution with proton PDFs and $\Omega_{3Q}$ FFs increasingly restricts the available longitudinal phase space.
Within the kinematic region investigated here, the phase-space suppression prevails over the BFKL enhancement, leading to a steadily decreasing $\DY$ spectrum for both $\Omega_{3c}$ and $\Omega_{3b}$ production.

For the reference transverse-momentum selection, a substantial enhancement of the overall normalization is observed when moving from HL-LHC to FCC energies.
For both $\Omega_{3c}$ and $\Omega_{3b}$ production, the cross sections increase by more than one order of magnitude across most of the explored rapidity range, with the largest gains occurring at small and intermediate values of $\DY$.
At $\sqrt{s}=13$~TeV, the $\Omega_{3c}$ distributions lie in the pb-to-tens-of-pb regime, while the corresponding $\Omega_{3b}$ rates are typically one to two orders of magnitude smaller.
At $\sqrt{s}=100$~TeV, the $\Omega_{3c}$ spectra reach the $\mathcal{O}(10^2)$~pb level in the low-$\DY$ region and remain sizable throughout the explored interval, whereas $\Omega_{3b}$ production attains cross sections ranging from a few pb to several tens of pb.
This pronounced energy dependence reflects the rapid growth of partonic luminosities and the much larger high-energy phase space available at FCC energies.
Compared with the previous {\tt OMG3Q1.0} study~\cite{Celiberto:2025ogy}, the lower transverse-momentum thresholds adopted here for both the $\Omega_{3Q}$ baryon and the accompanying jet enlarge the accessible phase space and contribute to the enhancement of the predicted cross sections.

We now examine the dependence of the FCC predictions on the transverse-momentum selection through the additional configurations shown in Figs.~\ref{fig:I_FCC_sym}, \ref{fig:I_FCC_int-asym}, and~\ref{fig:I_FCC_higher-asym}.
Figure~\ref{fig:I_FCC_sym} first replaces the reference asymmetric selection,
$50<|\vec q_{T_1}|<120$~GeV and
$30<|\vec q_{T_2}|<120$~GeV,
with the symmetric range
$50<|\vec q_{T_1}|<120$~GeV and
$50<|\vec q_{T_2}|<120$~GeV.
This comparison directly tests the sensitivity of the predictions to the relative transverse scales of the tagged baryon and jet, without modifying the lower transverse-momentum threshold of the $\Omega_{3Q}$ state.
Figures~\ref{fig:I_FCC_int-asym} and~\ref{fig:I_FCC_higher-asym} then progressively raise the transverse-momentum thresholds to the intermediate asymmetric range,
$80<|\vec q_{T_1}|<120$~GeV and
$50<|\vec q_{T_2}|<120$~GeV,
and to the higher asymmetric range,
$100<|\vec q_{T_1}|<120$~GeV and
$70<|\vec q_{T_2}|<120$~GeV,
respectively.

Taken together, these configurations provide a direct stability test of the resummed predictions as the tagged transverse scales are moved progressively farther into the perturbative domain.
The simultaneous presentation of the $\Omega_{3c}$ and $\Omega_{3b}$ channels is especially useful in this respect, since the same absolute transverse-momentum selection corresponds to substantially different ratios of the baryon transverse momentum to its mass.
The comparison therefore permits us to assess whether the larger mass of the $\Omega_{3b}$ state leads to an enhanced sensitivity to the lower transverse-momentum boundary, while also quantifying the phenomenological cost, in terms of the available production rate, associated with increasingly restrictive selections.

Here and in what follows, stability under variations of the transverse-momentum cuts refers to the perturbative stability of the resummed description, rather than to the absolute normalization of the cross section.
The latter necessarily decreases as the lower transverse-momentum bounds are raised, owing to the corresponding reduction of the available phase space.
Our stability criterion instead concerns the persistence of a controlled hierarchy among the $\LL$, $\NLLp$, and $\HENLOp$ predictions, together with a regular behavior of the associated perturbative uncertainty bands, as the tagged transverse scales are progressively increased.

The comparison between Figs.~\ref{fig:I_FCC_ref-asym} and~\ref{fig:I_FCC_sym} provides a first consistency check of the transverse-momentum dependence anticipated in Sec.~\ref{ssec:obserbables}.
Moving from the reference asymmetric selection to the symmetric one leads to a visibly closer agreement between the $\LL$ and $\NLLp$ predictions, signaling an enhanced perturbative stability when the two tagged transverse scales are more closely balanced.
At the same time, the separation between the $\NLLp$ result and its high-energy fixed-order counterpart, $\HENLOp$, is reduced.
This behavior is consistent with the expected phenomenological role of asymmetric transverse-momentum selections: while symmetric cuts favor a stronger numerical stability among successive resummed accuracies, asymmetric configurations enhance the discriminating power between the resummed prediction and its fixed-order high-energy truncation.
The symmetric-cut result therefore provides a nontrivial consistency check of the high-energy framework and confirms the rationale underlying the reference asymmetric selection.

The intermediate asymmetric configuration of Fig.~\ref{fig:I_FCC_int-asym} provides the first direct test of this stability criterion under simultaneously harder transverse-momentum requirements.
As expected, raising the lower bounds from the reference $(50,30)$~GeV selection to $(80,50)$~GeV produces a substantial reduction of the absolute cross section for both baryon species.
This loss of rate is primarily driven by the reduction of the available phase space and does not signal a deterioration of the resummed description.
Indeed, the characteristic hierarchy among the three perturbative accuracies remains regular throughout the explored $\DY$ interval.
For $\Omega_{3c}$, the $\LL$ prediction continues to lie systematically above the $\NLLp$ result, while the $\HENLOp$ contribution remains below it and approaches the resummed prediction toward larger rapidity separations.
A similarly controlled pattern is observed for $\Omega_{3b}$, with no indication that the larger baryon mass induces an enhanced perturbative instability when the lower transverse-momentum thresholds are raised.

This picture is reinforced by the higher asymmetric configuration of Fig.~\ref{fig:I_FCC_higher-asym}, where the lower transverse-momentum bounds are further increased to $100$~GeV for the $\Omega_{3Q}$ baryon and $70$~GeV for the jet.
Despite the substantial reduction of the available production rate, the perturbative structure of the predictions remains regular over the entire explored rapidity interval.
For $\Omega_{3c}$, the $\LL$ prediction remains systematically above the $\NLLp$ result, while the $\HENLOp$ contribution stays below it and progressively approaches the resummed prediction toward larger $\DY$.
For $\Omega_{3b}$, which provides the more stringent test because of its larger mass, no qualitative rearrangement of the perturbative hierarchy is observed when moving from the reference to the substantially harder transverse-momentum selection.
In particular, the relative behavior of the $\LL$, $\NLLp$, and $\HENLOp$ predictions evolves smoothly, with no evidence of an instability associated with the lower transverse-momentum boundary.

The sequence of transverse-momentum selections therefore indicates that the reference range already lies within a perturbatively controlled semihard regime.
Raising the lower transverse-momentum bounds primarily reduces the available event rate, as expected from the steeply falling transverse-momentum spectrum, rather than inducing a qualitative stabilization that was absent at lower scales.
This conclusion is particularly relevant for $\Omega_{3b}$ production: although the ratio $|\vec q_{T_1}|/m_{\Omega_{3b}}$ is smaller than in the triply charmed channel for the same absolute cut, increasing $|\vec q_{T_1}|^{\rm min}$ from $50$ to $100$~GeV does not reveal a qualitatively different perturbative pattern.
Our results thus support the use of the reference selection for phenomenological studies, while the harder configurations provide increasingly conservative alternatives at the price of a substantial loss in production rate.

Quantitatively, considering the asymmetric selections, the cross section decreases by approximately 1 order of magnitude when moving from the reference $(50,30)$~GeV configuration to the intermediate $(80,50)$~GeV range, and by a further factor of about 5 when the lower transverse-momentum bounds are raised to $(100,70)$~GeV.
The higher asymmetric selection therefore retains only a small fraction of the production rate available in the reference setup, providing a direct measure of the phenomenological cost associated with increasingly conservative transverse-momentum requirements.

The first ancillary panels of Figs.~\ref{fig:I_LHC_ref-asym} and~\ref{fig:I_FCC_ref-asym}, as well as of the additional FCC results in Figs.~\ref{fig:I_FCC_sym}, \ref{fig:I_FCC_int-asym}, and~\ref{fig:I_FCC_higher-asym}, quantify the impact of high-energy resummation by displaying the ratios of the $\LL$ and $\HENLOp$ predictions to the $\NLLp$ reference, including H-MHOU variations only.  
A clear hierarchy emerges at both collider energies and persists across the different transverse-momentum selections. 
The $\LL$ predictions systematically exceed the $\NLLp$ ones, indicating that NLL corrections provide a sizable stabilization of the resummed series.  
This effect is particularly pronounced for $\Omega_{3c}$ production at the HL-LHC, where the $\LL$ over $\NLLp$ ratio reaches several units over a broad region in $\DY$. 
By contrast, the enhancement is significantly milder for $\Omega_{3b}$ production, corresponding to a reduced separation between the $\LL$ and $\NLLp$ predictions in the heavier system.

The $\HENLOp$ predictions remain considerably closer to the $\NLLp$ baseline across the explored transverse-momentum configurations. 
Depending on the collider energy, baryon species, and transverse-momentum selection, the $\HENLOp$ over $\NLLp$ ratio stays below or moderately close to unity, supporting the consistency of our hybrid factorization and confirming that the dominant effects are already captured by the $\NLLp$ description.

A further improvement is observed at FCC energies. 
In particular, the separation between the $\LL$ and $\NLLp$ predictions becomes noticeably smaller, while the $\HENLOp$ curves remain well controlled throughout the explored rapidity range. 
This pattern suggests that the resummed description becomes increasingly natural as the available center-of-mass energy grows and the semihard kinematics more efficiently probes the high-energy regime. 
Consequently, the FCC environment offers not only substantially larger event rates, but also a more stable framework for precision studies of high-energy QCD dynamics.

The second and third ancillary panels isolate the two fragmentation-related uncertainty sources introduced in the {\tt OMG3Q1.1} framework, for the reference setup as well as for the additional transverse-momentum selections.
Both F-MHOUs and F-NPWF uncertainties remain under good control over the entire rapidity range and across the explored transverse-momentum configurations, typically inducing only modest deviations from the central prediction and exhibiting a weak dependence on $\DY$.
The F-NPWF bands are particularly smooth, as variations of the nonperturbative FF parameters primarily affect the normalization and detailed shape of the fragmentation input while leaving the underlying high-energy dynamics unchanged. 
Their effect is therefore propagated through the fragmentation convolution without generating significant distortions of the rapidity spectrum. 
No sign of instability or anomalous enhancement is observed across the explored kinematic domain, either at HL-LHC or FCC energies, nor when progressively harder transverse-momentum selections are imposed.

A particularly noteworthy outcome of the {\tt OMG3Q1.1} analysis is the clear hierarchy among uncertainty sources, which remains qualitatively preserved across the different transverse-momen\-tum choices.  
Fragmentation-related uncertainties remain systematically smaller than the variations induced by changes in the high-energy perturbative description, indicating that both the construction of the $\Omega_{3Q}$ FFs and their DGLAP evolution are under good theoretical control. 
The uncertainty envelopes shown in the main panels combine H-MHOUs, F-MHOUs, F-NPWF effects, and numerical integration uncertainties. 
Beyond providing a realistic estimate of the overall theoretical accuracy, this decomposition allows one to identify the physical origin of the uncertainty budget. 

Unlike earlier analyses based exclusively on central predictions, the {\tt OMG3Q1.1} framework disentangles the relative impact of high-energy resummation, perturbative fragmentation, and nonperturbative FF modeling. 
The resulting picture is remarkably consistent: the dominant variations originate from the perturbative high-energy sector, while the fragmentation component contributes only a limited fraction of the total uncertainty. 
This observation further supports the robustness of the {\tt OMG3Q1.1} predictions and strengthens their suitability for future precision studies of triply heavy baryon production at hadron colliders.

From the perspective of high-energy phenomenology, the results reveal a remarkably stable pattern despite the large rapidity intervals explored. 
Although the $\LL$ predictions can differ substantially from the $\NLLp$ baseline, the matched resummed framework remains well behaved and all fragmentation-related uncertainty sources stay under quantitative control. 
The resulting rapidity spectra are therefore characterized by a clear hierarchy of theoretical effects, where high-energy resummation drives the dominant variations while fragmentation uncertainties provide comparatively small corrections. 
This makes semi-inclusive $\Omega_{3Q}$ plus jet production a particularly attractive channel for investigating the interplay between heavy-hadron fragmentation and high-energy QCD dynamics.

Overall, Figs.~\ref{fig:I_LHC_ref-asym} and~\ref{fig:I_FCC_ref-asym}, complemented by the transverse-momentum-cut study of Figs.~\ref{fig:I_FCC_sym}, \ref{fig:I_FCC_int-asym}, and~\ref{fig:I_FCC_higher-asym}, demonstrate that the {\tt OMG3Q1.1} functions lead to numerically stable and physically transparent predictions for triply heavy baryon production at hadron colliders.
The uncertainty decomposition implemented in the framework provides direct access to the relative impact of perturbative high-energy effects, perturbative fragmentation dynamics, and nonperturbative FF modeling. 
To our knowledge, this represents the first systematic uncertainty-aware study of leading-power $\Omega_{3Q}$ fragmentation and one of the first applications of a fully decomposed fragmentation-uncertainty framework to rare heavy-baryon production in semi-inclusive collider observables.

\section{Summary and Outlook}
\label{sec:conclusions}

In this work we have presented the {\tt OMG3Q1.1} sets, a new generation of collinear FFs for triply heavy baryons within the {\HFNRevo} framework. 
These determinations extend and supersede the previous {\tt OMG3Q1.0} release~\cite{Celiberto:2025ogy} by introducing, for the first time, a systematic uncertainty-aware treatment of heavy-baryon fragmentation. 
Perturbative uncertainties are quantified through replica-based missing-higher-order variations (F-MHOUs) inspired by modern PDF methodologies~\cite{Kassabov:2022orn,Harland-Lang:2018bxd,Ball:2021icz,NNPDF:2024dpb}, while nonperturbative effects are assessed through controlled variations of the wave-function sector (F-NPWF) entering the fragmentation input. 
Both ingredients are embedded into a unified replicalike structure, yielding a flexible and systematically improvable framework for precision studies of rare-baryon production.

Using the {\Jethad} numerical interface and its symbolic-manipulation plugin {\psymJethad}~\cite{Celiberto:2020wpk,Celiberto:2022rfj,Celiberto:2023fzz,Celiberto:2024mrq,Celiberto:2024swu,Celiberto:2025csa,Celiberto:2026ooh}, we investigated semi-inclusive $\Omega_{3Q}$ plus jet production within the $\NLLp$ hybrid collinear/high-energy factorization setup. 
Our phenomenological analysis covers collider energies ranging from the HL-LHC regime at $\sqrt{s}=13$~TeV up to the nominal FCC energy of $\sqrt{s}=100$~TeV. 
The resulting rapidity distributions exhibit a clear and physically interpretable pattern, characterized by a substantial enhancement of production rates at FCC energies together with an increasingly stable perturbative behavior.

A central outcome of the present study is the identification of a well-defined hierarchy among theoretical uncertainties. 
The dominant theoretical sensitivity originates from the high-energy perturbative sector, including hard-factor missing higher-order uncertainties (H-MHOUs), while fragmentation-related effects associated with fragmentation missing higher-order uncertainties (F-MHOUs) and fragmentation nonperturbative wave-function (F-NPWF) variations remain comparatively small throughout the explored kinematic range.
This result demonstrates that both the construction of the {\tt OMG3Q1.1} FFs and their subsequent DGLAP evolution are under good theoretical control. 
More broadly, it establishes triply heavy baryons as theoretically robust probes for precision investigations of high-energy QCD dynamics.

The observed stabilization of the resummed predictions further supports the concept of \emph{natural stability} previously identified in heavy-flavor fragmentation studies~\cite{Celiberto:2022grc}. Although LL predictions can differ substantially from the $\NLLp$ baseline, the resummed high-energy framework remains well behaved and increasingly stable as the collider energy grows. 
This behavior is particularly evident in the FCC regime, where the semihard kinematics more efficiently activates high-energy logarithmic dynamics and provides a favorable environment for precision tests of resummation effects.

From a broader perspective, heavy-baryon fragmentation constitutes a privileged interface between hadron structure and precision QCD. 
The production of triply heavy states requires the simultaneous control of perturbative short-distance dynamics, threshold-sensitive parton evolution, and nonperturbative hadron formation. 
The {\tt OMG3Q1.1} framework provides a quantitative tool to investigate this interplay and establishes a benchmark for future studies of heavy-baryon production across different collider environments.

An additional avenue of investigation concerns the sensitivity of multicharm observables to intrinsic-charm components of the proton wave function~\cite{Brodsky:1980pb,Ball:2022qks,Guzzi:2022rca,NNPDF:2023tyk}. 
Although the production of triply heavy baryons is expected to be largely dominated by gluon-initiated mechanisms, specific kinematic regions may enhance the contribution of charm-initiated channels. 
Recent studies have shown that axial-vector all-charm tetraquarks provide particularly sensitive probes of intrinsic charm through their production dynamics~\cite{Celiberto:2025vra}. 
In this context, uncertainty-aware fragmentation frameworks such as {\tt OMG3Q1.1} offer a natural foundation for future investigations aimed at connecting heavy-flavor production, hadron structure, and exotic spectroscopy.

Beyond their immediate application to the still-unobserved triply heavy baryons, the {\tt OMG3Q1.1} functions should be regarded as part of a broader effort toward a unified description of heavy-baryon formation across flavor sectors. 
Recent years have witnessed remarkable progress in multicharm baryon spectroscopy, culminating in the observation of the doubly charmed states $\Xi_{cc}^{++}$\cite{LHCb:2017iph} and $\Xi_{cc}^{+}$\cite{LHCb:2026pxn}, together with the recent report of the $\Omega_{cc}^{+}$ baryon~\cite{Wang:2026OmegaCC}. 
These discoveries complete the ground-state doubly charmed sector and demonstrate that increasingly complex heavy-baryon systems can now be reconstructed with high precision. 
In this evolving experimental landscape, triply heavy baryons naturally emerge as the next frontier.

The theoretical infrastructure developed in the {\tt OMG3Q1.1} program provides a quantitative baseline for this new generation of studies. 
By combining uncertainty-aware fragmentation with high-energy phenomenology, it enables a systematic investigation of heavy-baryon production from the HL-LHC to future collider facilities, including the FCC~\cite{FCC:2018byv,FCC:2018evy,FCC:2018vvp,FCC:2018bvk} and other next-generation machines~\cite{Chapon:2020heu,Anchordoqui:2021ghd,AlexanderAryshev:2022pkx,InternationalMuonCollider:2024jyv,MuCoL:2024oxj,Black:2022cth,Accardi:2023chb}. 
More broadly, it contributes to a long-term program aimed at establishing a unified and systematically improvable description of heavy-baryon formation, in which recently discovered doubly charmed baryons and future triply heavy states can be interpreted within a common QCD framework.

\section*{Data availability}
\label{sec:data_availability}
\addcontentsline{toc}{section}{\nameref{sec:data_availability}}

The {\tt OMG3Q1.1} collinear FFs~\cite{Celiberto:2026_OMG3Q11} for same-flavor all-heavy $\Omega_{3Q}$ baryons are publicly available at \url{https://github.com/FGCeliberto/Collinear_FFs/}. 
The corresponding {\tt LHAPDF6} package contains the full replica ensemble introduced in this work.

\section*{Acknowledgments}
\label{sec:acknowledgments}
\addcontentsline{toc}{section}{\nameref{sec:acknowledgments}}

The author thanks Marco Bonvini, Matteo Cacciari, Terry Generet, Felix Hekhorn, and Alessandro Pilloni for valuable discussions and insightful exchanges.
This work was supported by the Atracción de Talento Grant No.~2022-T1/TIC-24176 of the Comunidad de Madrid (Spain).

\appendix

\counterwithin*{equation}{section}
\renewcommand\theequation{\thesection\arabic{equation}}

\counterwithin*{table}{section}
\renewcommand\thetable{\thesection\arabic{table}}

\hypertarget{app:A}{
\section{Replica organization of the {\tt OMG3Q1.1} sets}
}
\label{app:A}

The {\tt OMG3Q1.1} release is built upon a structured ensemble of nine replicas designed to quantify the impact of perturbative and nonperturbative uncertainties on the fragmentation process. 
The replica construction follows a two-dimensional variation strategy in which factorization-scale modifications and wave-function-profile deformations are treated independently and subsequently combined.

Perturbative uncertainties are estimated through the variation of the characteristic fragmentation scale by means of the multiplicative factor $K_{\mu}=0.5,\,1,\,2$, providing the F-MHOU component of the uncertainty budget. Nonperturbative effects are instead probed through variations of the transverse-momentum parameter $\vqTTa$, which controls the shape of the initial fragmentation input and enters the F-NPWF uncertainty estimate. 
Three representative values are considered,
\begin{equation}
 \vqTTa \;=\; 55,\;60,\;65~{\rm GeV}^2 \;.
 \label{eq:app:vqTTA}
\end{equation}
The complete replica ensemble is obtained by combining all perturbative and nonperturbative configurations. 
The central prediction corresponds to replica 0, defined by the reference choice $K_{\mu}=1$ and $\vqTTa=60~{\rm GeV}^2$.

The correspondence between replica labels and parameter configurations is reported in Table~\ref{tab:OMG3Q11_replicas}.

\begin{table}[t]
\centering
\begin{tabular}{|c|r|r|}
\toprule
Replica ID & $K_{\mu}$ & $\vqTTa$ [GeV$^2$] \\
\midrule
0 & 1   & 60  \\
1 & 1   & 55  \\
2 & 1   & 65 \\
3 & 0.5 & 60  \\
4 & 0.5 & 55  \\
5 & 0.5 & 65 \\
6 & 2   & 60  \\
7 & 2   & 55  \\
8 & 2   & 65 \\
\bottomrule
\end{tabular}
\caption{Replica structure of the {\tt OMG3Q1.1} set.
Variations of $K_{\mu}$ provide the F-MHOU component, while changes in $\vqTTa$ determine the F-NPWF contribution.}
\label{tab:OMG3Q11_replicas}
\end{table}

\hypertarget{app:B}{
\section{NLO correction to the BFKL kernel}
}
\label{app:B}

The characteristic function associated with the NLO correction to the BFKL kernel in Eq.~\eqref{chi_NLO} takes the form
\begin{equation}
 \label{kernel_NLO}
 \bar \chi(l,\nu)\,=\, - \frac{1}{4}\left\{\frac{\pi^2 - 4}{3}\chi(l,\nu) - 6\zeta(3) - \frac{\drv^2 \chi}{\drv\nu^2} + \,2\,\phi(l,\nu) + \,2\,\phi(l,-\nu)
 \right.
\end{equation}
\[
 \left.
 +\frac{\pi^2\sinh(\pi\nu)}{2\,\nu\, \cosh^2(\pi\nu)}
 \left[
 \left(3+\left(1+\frac{n_f}{N_c^3}\right)\frac{11+12\nu^2}{16(1+\nu^2)}\right)
 \delta_{n0}
 -\left(1+\frac{n_f}{N_c^3}\right)\frac{1+4\nu^2}{32(1+\nu^2)}\delta_{n2}
\right]\right\} \, ,
\]
\vskip 0.25cm
where $\delta_{nm}$ denotes the usual Kronecker delta, and
\begin{equation}
\label{kernel_NLO_phi}
 \phi(l,\nu)\,=\,-\int\limits_0^1 \drv x\,\frac{x^{-1/2+i\nu+l/2}}{1+x}\left\{\frac{1}{2}\left(\psi^\prime\left(\frac{l+1}{2}\right)-\zeta(2)\right)+\mbox{Li}_2(x)+\mbox{Li}_2(-x)\right.
\end{equation}
\[
\left.
 +\ln x\left[\psi(n+1)-\psi(1)+\ln(1+x)+\sum_{k=1}^\infty\frac{(-x)^k}{k+l}\right]+\sum_{k=1}^\infty\frac{x^k}{(k+l)^2}\left[1-(-1)^k\right]\right\}
\]
\[
 =\sum_{k=0}^\infty\frac{(-1)^{k+1}}{k+(l+1)/2+i\nu}\left\{\psi^\prime(k+l+1)-\psi^\prime(k+1)\right.
\]
\[
 \left.
 +(-1)^{k+1}\left[\beta_{\psi}(k+l+1)+\beta_{\psi}(k+1)\right]-\frac{\psi(k+l+1)-\psi(k+1)}{k+(l+1)/2+i\nu}\right\} \; ,
\]
\begin{equation}
\label{kernel_NLO_phi_beta_psi}
 \beta_{\psi}(z)=\frac{1}{4}\left[\psi^\prime\left(\frac{z+1}{2}\right)
 -\psi^\prime\left(\frac{z}{2}\right)\right] \; .
\end{equation}
Furthermore,
\begin{equation}
\label{dilog}
\mbox{Li}_2(x) = \int\limits_0^x \drv y \,\frac{\ln(1-y)}{-y}
\end{equation}
stands for the dilogarithm function.

\clearpage
\bibliographystyle{apsrev}
\bibliography{bibliography}

\end{document}